# Progress of Quantum Molecular Dynamics model and its applications in Heavy Ion Collisions

Yingxun Zhang,[1, 2, *] Ning Wang,[3, 2, †] Qingfeng Li,[4, 5, ‡] Li Ou,[3, 2, §] Junlong Tian,[6, 2, ¶] Min Liu,[3, 2, **] Kai Zhao,[1, ††] Xizhen Wu,[1, 2, ‡‡] and Zhuxia Li[1, 2, §§]

[1]*China Institute of Atomic Energy, Beijing, 102413, China*
[2]*Guangxi Key Laboratory of Nuclear Physics and Technology,*
*Guangxi Normal University, Guilin, 541004, China*
[3]*Guangxi Normal University, Guilin, 541004, China*
[4]*School of Science, Huzhou University, Huzhou, 313000, China*
[5]*Institute of Modern Physics, Chinese Academy of Sciences, Lanzhou, 730000, China*
[6]*School of Physics and Electrical Engineering, Anyang Normal University, Anyang, 455002, China*
(Dated: May 14, 2020)

In this review article, we first briefly introduce the transport theory and quantum molecular dynamics model applied in the study of the heavy ion collisions from low to intermediate energies. The developments of improved quantum molecular dynamics model (ImQMD) and ultra-relativistic quantum molecular dynamics model (UrQMD), are reviewed. The reaction mechanism and phenomena related to the fusion, multinucleon transfer, fragmentation, collective flow and particle production are reviewed and discussed within the framework of the two models. The constraints on the isospin asymmetric nuclear equation of state and in-medium nucleon-nucleon cross sections by comparing the heavy ion collision data with transport models calculations in last decades are also discussed, and the uncertainties of these constraints are analyzed as well. Finally, we discuss the future direction of the development of the transport models for improving the understanding of the reaction mechanism, the descriptions of various observables, the constraint on the nuclear equation of state, as well as for the constraint on in-medium nucleon-nucleon cross sections.

## Contents

*Electronic address: zhyx@ciae.ac.cn
†Electronic address: wangning@gxnu.edu.cn
‡Electronic address: liqf@zjhu.edu.cn
§Electronic address: liou@gxnu.edu.cn
¶Electronic address: tianjl@163.com
**Electronic address: lium`816@hotmail.com
††Electronic address: zhaokai@ciae.ac.cn
‡‡Electronic address: lizwux9@ciae.ac.cn
§§Electronic address: lizwux@ciae.ac.cn

## I. INTRODUCTION

Heavy ion collisions (HICs) provide an unique way to investigate the basic nuclear physics problems in laboratory, such as where is the end of nuclear chart? how can we reach it? what is the properties of neutron-rich matter in cosmos and on the earth? and the origin of elements heavier than Fe? In the past 50 years, a large number of experimental data for HICs have been accumulated and lots of interesting results have been obtained [1–24]. It has been found that the phenomena and the physics processes in HICs at various energy domains are very rich and complicated. Thus, understanding the mechanism behind them is very helpful for us to obtain the knowledge of related nuclear phenomena and physics.

For low energy heavy ion reactions (in this review paper, we consider this energy domain to be $E_{beam} \leq 10$ MeV/nucleon), the synthesis of superheavy nuclei and new neutron (proton)-rich isotopes are the hot topics [25–61]. Related to them, there are also a lot of interesting problems in the reactions, such as the role of the dynamical effect, the nuclear structure effect and the interplay between them, the dependence of the barrier on the reaction systems and incident energies, etc., need to be clarified. Multi-nucleon transfer (MNT) reactions between very heavy nuclei at near or above barrier energies, attracted a lot of attention as well [52–54] since it could lead to production of new heavy and possibly superheavy neutron-rich isotopes, especially those nuclei at waiting point for $r$-process which are important to the nuclear astrophysics. In addition, the study of MNT reaction is highly required due to the new facilities of High Intensity heavy ion Accelerator Facility (HIAF) and Argonne Tandem Linac Accelerator System (ATLAS).

As the beam energy increases, the interplay of one-body mean field and two-body scattering makes the phenomena in this energy domain fascinating and complicated. The ternary and four-fragment breakup process may take place accompanying the binary process. At the low-intermediate energies region, i.e. ∼20 MeV/nucleon to ∼300 MeV/nucleon (the definition of energy domain of intermediate energies, usually differs among authors), multifragmentation process which was thought as a signal of liquid-gas phase transition in finite systems, ap-pears and provides the opportunity to study the equation of state at sub-saturated densities [22, 23, 62–65] and mechanism of liquid-gas phase transition in the finite system. The phenomena relating to the multifragmentation, such as isospin distillation, neutron-rich neck emission, bimodality, ..., evolved with the isospin asymmetry and the size of reaction system and the beam energy as well, are very rich and complicated, and it provides us with the hints of equation of state (EOS), mechanism of fragmentation and information of liquid-gas phase transition for two components finite system.

When the beam energies increase from around 100 MeV per nucleon up to few GeV per nucleon, the collective motion of emitted nucleons and light particles appears, and it was named as collective flow [3, 9, 11–14, 16]. The measurement of collective flow also provides a possibility for the study of the nuclear EOS at supersaturated densities as well as the information of in-medium nucleon-nucleon (NN) cross sections [24], because the changing of the momentum of emitted nucleons are related to the gradient of pressure between the participant and spectator of the reaction system, which is caused by the nucleonic mean field and nucleon-nucleon collisions. The sub-threshold and above threshold production of mesons take place in addition to the multifragmentation and the collective flow effects. The yields of mesons and their yield ratios between different charge states as well as the flow effects evolve with energy and isospin asymmetry of the system [66–71, 73–84, 135], contain a lot information of the isospin asymmetric EOS of nuclear matter, the isospin dependent in-medium NN cross sections and the reaction dynamics. The study of the phenomena at the beam energy from low to intermediate energies, such as the production of new isotopes, multifragmentation, and flow effects closely relating to equation of state stimulate the building of next generation of rare isotope facilities, such as the Intensity Heavy ion Accelerator Facility (HIAF) at Institute of Modern Physics (IMP) in China [85–87], Facility for rare isotope beams (FRIB)[88] in USA, Rare isotope Accelerator complex for ON-line experiment (RAON)[89] in Korea, RI beam factory (RIBF/RIKEN) in Japan[90], SPIRAL2/GANIL in France[91], Facility for Antiproton and Ion Research (FAIR) at GSI in Germany[92], Selective Production of Exotic Species (SPES/LNL) in Italy[93], Nuclotron-based Ion Collider facility at the Joint Institute for Nuclear Research(NICA/JINR) in Russia[94], and the proposed Beijing Isotope-Separation-On-Line Neutron-Rich Beam Facility (BISOL)[95] in China.

In order to describe the phenomena and processes in HICs at each energy domain, kinds of successful models have been developed for explaining and investigating the related physics problems. For instance, for low beam energy reactions, both the diffusion model by using Langevin equation and the di-nuclear model can describe fusion cross sections [46, 52, 53, 53, 54, 96–117]; for the intermediate energy region, the statistical multifragmentation model (SMM) is quite successful in de-scribing the charge distribution of multifragmentations at intermediate energy reactions [118–125], etc. However, heavy ion reaction or collision process is essentially a non-equilibrium process. It is highly demanded to develop a model or theory which can describe the time evolution of reaction, so that one can understand the reaction dynamics and the transient states in the compressing and expanding stages which are important for studying the equation of state of nuclear matter. Also, the theories (models) describing heavy ion collisions in a more unified way from low energies to higher energies are considerably demanded for insight in depth of the properties of nuclear many-body system and reaction dynamics. Thus, the microscopic dynamics models, namely, the microscopic transport models for heavy ion reactions were developed and have been used to study HICs with a great success [63–65, 73, 74, 126–172].

Generally, the transport models adopt different philosophy to solve the transport equations, which can be roughly divided into two types, the Boltzmann-Uehling-Uhlenbeck (BUU) model and the quantum molecular dynamics (QMD) model.

In the BUU approach, the goal is to describe the evolution of the one-body phase space occupation probability $f(\mathbf{r}, \mathbf{p}; t)$ as a function of the time under the action of a mean field potential $U[f]$, usually derived from a density functional, and two-body collisions specified by an in-medium cross section $d\sigma^{\mathrm{med}}/d\Omega$. The non-relativistic BUU equation reads

$$\left(\frac{\partial}{\partial t} + \frac{\mathbf{p}}{m} \cdot \nabla_r - \nabla_r U \cdot \nabla_p\right) f(\mathbf{r}, \mathbf{p}; t) = I_{\mathrm{coll}}(\mathbf{r}, \mathbf{p}; t), \quad (1)$$

$I_{coll}$ is the collision term, which accounts for the particles entering and leaving the phase space $d^3 r d^3 p$. The integro-differential non-linear BUU equation is solved numerically by using the test particle method (TP) [173], as

$$f(\mathbf{r}, \mathbf{p}; t) = \frac{(2\pi\hbar)^3}{g N_{\mathrm{TP}}} \sum_{i=1}^{A N_{\mathrm{TP}}} G(\mathbf{r} - \mathbf{r}_i(t)) \tilde{G}(\mathbf{p} - \mathbf{p}_i(t)), \quad (2)$$

where $N_{\mathrm{TP}}$ is the number of test particles (TP) per nucleon, $\mathbf{r}_i$ and $\mathbf{p}_i$ are the time-dependent coordinate and momentum of the test particle $i$, $G$ and $\tilde{G}$ are the shape functions in coordinate and momentum space, respectively, with a unit norm (e.g. $\delta$-functions or normalized Gaussians). For nucleons, the degeneracy factor $g = 4$ is to define $f(\mathbf{r}, \mathbf{p}, t)$ as the spin-isospin averaged phase space occupation probability. In the BUU approach the distribution function for each isospin (or spin) state in a similar way. In the BUU approach the phase space distribution function is seen as a one-body quantity and a smooth function of coordinates and momenta and it can be approximated better by increasing the number of test particles in the solution. In the limit of $N_{\mathrm{TP}} \to \infty$, the BUU equation is solved exactly. In this limit the solution is deterministic and does not contain fluctuations. In case of studying the cluster and fragment production, it will need to introduce the fluctuation which actually exist in low-intermediate energy heavy ion collisions. There are some effort to do it through the Boltzmann-Langevin equation (BLE) which adds a fluctuation term on the right hand side of Eq. (1),

$$\left(\frac{\partial}{\partial t} + \frac{\mathbf{p}}{m} \cdot \nabla_r - \nabla_r U \cdot \nabla_p\right) f(\mathbf{r}, \mathbf{p}; t) = I_{\mathrm{coll}}(\mathbf{r}, \mathbf{p}; t) + \delta I_{\mathrm{coll}}. \quad (3)$$

This equation is solved approximately in the stochastic mean field (SMF), Boltzmann-Langevin one body (BLOB) [126, 139–141, 141, 174–179].

In the QMD approach, the evolution of $N$-body phase space density distribution $f_N(\mathbf{r}_1, ..., \mathbf{r}_N, \mathbf{p}_1, ..., \mathbf{p}_N; t)$ is formulated. Thus, in the philosophy of QMD, the effect of go-beyond the mean field approach is realized by including correlations and fluctuations from the beginning, and it is at the expense of more rapidly destroying the fermionic character of the system and of reverting to a classical system. The QMD model can be seen as an event generator, where the time evolution of different events is solved independently. The fluctuations in QMD-type codes are regulated and smoothed by choosing the parameter $\sigma_r$, the width of the wave packet. The mean field part in the QMD approach can also be viewed as derived from the Time-Dependent Hartree method with a product of trial wave function of single particle states in Gaussian form that we will mention in section II. Also the collision term, which relocates nucleon wave packets in momentum space, introduces more fluctuations than those for the collision term in BUU. The fluctuations among events are not suppressed in QMD approach even in the limit of infinite number of events. Taking into account all fluctuations and correlations has basically two advantages: i) many-body processes, in particular the formation of complex fragments are explicitly treated, and ii) the model allows for an event-by-event analysis of heavy ion reactions similar to the methods which are used for the analysis of exclusive high acceptance data [146–148].

Stimulated by studying the dynamical effects on the heavy ion reactions near the barrier, we have made a series of improvements on nucleon propagation in the mean field part, nucleon-nucleon collision in two-body collision part, the initialization and the cluster recognization [153, 154, 157, 160] based on the original QMD model during last 20 years. It was named as improved quantum molecular dynamics model (also known as ImQMD). One of the important changes is that we adopt the potential energy density functional to determine the nucleon propagation, which was stimulated by a steady transition taken place during the past several years from the mean-field description of nuclear properties in terms of effective forces to an density functional approach [180–184]. The basic idea of density functional is that the ground-state energy of a stationary many-body system can be represented in terms of the ground state density alone, and thus, energy density functional theory calculations are comparatively simple to imple-ment yet often very accurate and computationally feasible even for systems with large particle numbers. Following this transition, we replaced the mean filed potential part in terms of the effective nuclear interaction by the potential energy density functional, and we will mention it in the following chapter.

Up to now, we have developed three versions of ImQMD designed for different purposes. The details will be given in section II. A series applications of the ImQMD model to the fusion reaction, deep inelastic scattering (multi-nucleon transfer reaction) at near Coulomb barriers and to multifragmentation, collective flow effects and other important phenomena at intermediate and high energies will be presented in this review paper. The study of liquid-gas phase transition in finite nuclear systems connecting to multifragmentation, and the constraints on symmetry energy for asymmetric nuclear matter by comparison of the model calculations with the experimental data, are also presented. Further more, the phenomena at intermediate and high energies related to particle production and collective flow are presented and discussed within the framework of UrQMD model.

This article is organized as follows. In Section II, we will briefly review the many-body transport theory and its solution in the quantum molecular dynamics approaches. In Section III, we will focus on the study of phenomena in heavy ion reaction (collision), and its reaction mechanism. Section IV will present the investigation on the in-medium NN cross sections by using the closed time-path green function method and one-boson exchange model, and the efforts of extracting the in-medium NN cross section by comparing the QMD type model calculations to heavy ion collisions data. Section V will present the investigation on the isospin asymmetric nuclear equation of state, symmetry energy and the uncertainties of these constraints. Section VI will give discussions and prospect on the development of transport theory and its applications in the nuclear physics in the future.

## II. TRANSPORT THEORY AND THE QUANTUM MOLECULAR DYNAMICS MODEL

Utility of the quantum-mechanical phase space distributions for the formulation and solution of scattering and production problems is an important approach in heavy ion collisions. In 1932, Wigner [185] discovered an interesting version of the density matrix which allows the expression of quantum dynamics in a form directly comparable with the classical analog while maintaining the quantum integrity of the conjugacy of the variable $p$ and $x$. This approach can also be extended to the particle production problem by expanding the field operator $\phi$ in terms of the annilation operator $a(p,t)$ [186], and the $N$-particle covariant distribution functions are directly connected with the inclusive differential cross sections. The advantages of this technique are clearly exhibited in application to quantum transport theory [187–189].

The transport equation can be derived by means of the nonequilibrium Green's function technique, i.e. the closed time-path Green's function(CTPGF) technique which is based on the theoretical concepts for a proper many-body description in terms of a real time nonequilibrium field theory initiated by Schwinger in the early sixties. By using the CTPGF technique, the transport equation and the analytical expressions of in-medium two-body scattering cross sections applicable for heavy ion collisions are simultaneously obtained. As the whole theory is complicated but very useful, here we refer to the literatures (see Refs. [190–198]).

In this chapter, we will shortly mention the transport theory which is used for QMD approach.

### A. Transport theory for $N$-body system

The derivation of Boltzmann's $N$-body phase space distribution $f_N(\mathbf{r}_1, ...\mathbf{r}_N; ...\mathbf{p}_1, ..., \mathbf{p}_N)$ provide an intuitive picture of complex collisions process based on the quantum mechanics. As is known that the simultaneous probability for position $\mathbf{r}$ and momentum $\mathbf{p}$ is forbidden in quantum mechanics by Heisenberg's uncertainty principle, i.e., $\Delta\mathbf{r}_i\Delta\mathbf{p}_i \geq \hbar/2$. One way to map a quantum variable to a classical one is to use coarse-grain method, such as Wigner transformation [185] in which both respects the rules of quantum mechanics and recaptures most of the desired features of Boltzmann function. If a wave function $\psi(\mathbf{r}_1, ...\mathbf{r}_N)$ is given, one may build the density matrix with the representation as following,

$$f_N(\mathbf{r}_1, ..., \mathbf{r}_N; \mathbf{p}_1, ..., \mathbf{p}_N) = \tag{4}$$
$$(\frac{1}{2\pi\hbar})^N \int_{-\infty}^{\infty} ... \int d\mathbf{y}_1...d\mathbf{y}_N \psi^*(\mathbf{r}_1 - \mathbf{y}_1, ..., \mathbf{r}_N - \mathbf{y}_N)$$
$$\psi(\mathbf{r}_1 + \mathbf{y}_1, ..., \mathbf{r}_N + \mathbf{y}_N)e^{-2i(\mathbf{p}_1\cdot\mathbf{y}_1 + ... + \mathbf{p}_N\cdot\mathbf{y}_N)/\hbar},$$

to express the probability-function of the simultaneous values of $\mathbf{r}_1, ..., \mathbf{r}_N$ and $\mathbf{p}_1, ..., \mathbf{p}_N$. Eq. (4) is real, but not everywhere positive and it means that the $f_N(\mathbf{r}_1, ..., \mathbf{r}_N; \mathbf{p}_1, ..., \mathbf{p}_N)$ can not be really interpreted as the simultaneous probability for coordinates and momenta. However, the lack of positivity will not hinder the use of it since we are mainly concerned with computation of positive definite asymptotic quantities. When Eq. (4) is integrated with respect to $\mathbf{p}$, the correct probabilities $|\psi(\mathbf{r}_1, ..., \mathbf{r}_N)|^2$ is given; if we integrate Eq. (4) with respect to $\mathbf{r}$, the correct probabilities $|C(\mathbf{p}_1, ..., \mathbf{p}_N)|^2$ can also be verified. Hence, one may get the correct expectation values of mechanical quantities $O$ as a function of

coordinate or momenta for the state $\psi$ as,

$$
\begin{aligned}
< O > &= \int_{-\infty}^{\infty} ... \int d\mathbf{r}_1 ... d\mathbf{r}_N d\mathbf{p}_1 ... d\mathbf{p}_N O(\mathbf{r}_i, \mathbf{p}_i) \quad (5) \\
&\quad f_N(\mathbf{r}_1, ..., \mathbf{r}_N; \mathbf{p}_1, ..., \mathbf{p}_N) \\
&= \int_{-\infty}^{\infty} ... \int d\mathbf{r}_1 ... d\mathbf{r}_N \psi^*(\mathbf{r}_1, ..., \mathbf{r}_N) \\
&\quad O(\mathbf{r}_i, -i\hbar\partial_{\mathbf{r}_i})\psi(\mathbf{r}_1, ..., \mathbf{r}_N),
\end{aligned}
$$

where $f_N(\mathbf{r}_1, ..., \mathbf{r}_N; \mathbf{p}_1, ..., \mathbf{p}_N)$ is the probability-function described above.

For the equation of $N$-body system phase space distribution $f_N(\mathbf{r}_1, \mathbf{r}_2, ..., \mathbf{r}_N; \mathbf{p}_1, \mathbf{p}_2, ..., \mathbf{p}_N)$, it has been derived by Aichline in Ref. [148] and is written as follows,

$$
\begin{aligned}
(\frac{\partial}{\partial t} &+ \sum_i \frac{\mathbf{p}_i}{m} \cdot \nabla_{\mathbf{r}_i}) f_N(\mathbf{r}_1, ... \mathbf{r}_N, \mathbf{p}_1, ... \mathbf{p}_N, t) \quad (6) \\
&= \int \Pi_i d^3 p_i d^3 Q_i d^3 q_i e^{i\mathbf{r}_i \cdot \mathbf{p}_i} \\
&\quad f_0^{(n)}(Q_1, ..., Q_N, q_1, ..., q_N, t) \\
&\quad (I_1(T) + I_2(T) + I_3(T)),
\end{aligned}
$$

where $Q_i$ and $q_i$ are the momenta of final states. $T = \sum t_i + \sum_{k \neq m} \sum_m t_{im} G_0^\dagger t_{ik} + ....$, $t_{im}$ is the sum of all possible transition matrix combinations, and $G_0^\dagger$ is the on shell propagators. $f_0^{(n)}$ is the time evolved free wave packet. The definition of terms $I_1$, $I_2$, and $I_3$ can be found in Ref. [148]. The real part of $I_1 + I_2$ acting as an effective potential has been replaced by two-body potential in this case, and it can be easily related to the nuclear equation of state. The $I_3$ term can be reduced to a sum of terms which contains only absolute squares of transition matrices, and it is assumed to be proportional to products of the cross sections [148].

For the Vlasov equation of Eq. (6), i.e. without $I_3$ term, the time evolution of the phase space density of particles which moves on classical orbits is specified by the Hamilton equation. The equation of motion for the expectation values of $< \mathbf{r}_i >$ and $< \mathbf{p}_i >$ are,

$$
\begin{aligned}
\frac{\partial < \mathbf{p}_i >}{\partial t} &= - < \nabla_i V(\mathbf{r}_1, ... \mathbf{r}_N) >, \quad (7) \\
\frac{\partial < \mathbf{r}_i >}{\partial t} &= \frac{< \mathbf{p}_i >}{m}.
\end{aligned}
$$

Supposing the potential $V(\mathbf{r}_1, ... \mathbf{r}_N)$ is slowly varying, and we can expand $\frac{\partial V}{\partial \mathbf{r}_i}$ as a Taylor series about $< \mathbf{r}_i >$ as following,

$$
\begin{aligned}
&\frac{\partial V(\mathbf{r}_1, ..., \mathbf{r}_N)}{\partial \mathbf{r}_i} \quad (8) \\
&= \frac{\partial V(\mathbf{r}_1, ..., < \mathbf{r}_i >, ..., \mathbf{r}_N)}{\partial < \mathbf{r}_i >} \\
&+ \frac{\partial^2 V(\mathbf{r}_1, ..., < \mathbf{r}_i >, ..., \mathbf{r}_N)}{\partial < \mathbf{r}_i >^2}(\mathbf{r}_i - < \mathbf{r}_i >) \\
&+ \frac{1}{2}\frac{\partial^3 V(\mathbf{r}_1, ..., < \mathbf{r}_i >, ..., \mathbf{r}_N)}{\partial < \mathbf{r}_i >^3}(\mathbf{r}_i - < \mathbf{r}_i >)^2 \\
&+ ....
\end{aligned}
$$

Substitution of the above expansion into Eq.(7), and we have

$$
\begin{aligned}
&\frac{\partial < \mathbf{p}_i >}{\partial t} \quad (9) \\
&= \frac{\partial < V(\mathbf{r}_1, ..., < \mathbf{r}_i >, ..., \mathbf{r}_N) >}{\partial < \mathbf{r}_i >} \\
&+ \frac{\sigma^2}{2}\frac{\partial^3 < V(\mathbf{r}_1, ..., < \mathbf{r}_i >, ..., \mathbf{r}_N) >}{\partial < \mathbf{r}_i >^3},
\end{aligned}
$$

with $< (\mathbf{r}_i - < \mathbf{r}_i >) >= 0$ and $\sigma^2 = < (\mathbf{r}_i - < \mathbf{r}_i >)^2 >$. If the

$$
\begin{aligned}
&|\frac{1}{2}\nabla_{< \mathbf{r}_i >}^3 V(\mathbf{r}_1, ..., < \mathbf{r}_i >, ..., \mathbf{r}_N)\sigma^2| \quad (10) \\
&\ll |\nabla_{< \mathbf{r}_i >} V(\mathbf{r}_1, ..., < \mathbf{r}_i >, ..., \mathbf{r}_N)|,
\end{aligned}
$$

the second term in the right hand of Eq. (9) can ne neglected. While, $< \nabla_{< \mathbf{r}_i >} V(\mathbf{r}_1, ..., < \mathbf{r}_i >, ..., \mathbf{r}_N) > \approx \nabla_{< \mathbf{r}_i >} U(< \mathbf{r}_i >, ..., < \mathbf{r}_N >)$ and the equations in Eq.(7) will be the same as it in the classical equation. Here, the potential energy $U$ is a function of parameters $\{< \mathbf{r}_i >, < \mathbf{p}_i >\}$ and thus $U$ can also be thought as the potential at $\{< \mathbf{r}_i >, < \mathbf{p}_i >\}$ in practical calculations. If the gradients of the potential are strong, the high order terms in Eq. (7) can not be neglected and it causes not only strong force but also large fluctuations around the mean trajectories.

For the collision term, in the actual calculations, the particle collisions are simulated when the particles are sufficiently close [199] with the Monte-Carlo method, and the scattering angle is chosen randomly according to the differential cross section. The outgoing states of collisions also need to be checked whether the states of outgoing particles have been occupied or how much is the probability of the occupation by other particles. If the outgoing states have been fully occupied, the collision will not happen and it is named as Pauli blocking. Otherwise, the collision will happen with certain probability. More details will be given in following sections.

### B. Quantum Molecular Dynamics approach

In the quantum molecular dynamics approach, each nucleon is represented by a Gaussian wave packet,

$$
\psi_i(\mathbf{r}_i) = \frac{1}{(2\pi\sigma_r^2)^{3/4}} e^{-\frac{(\mathbf{r}_i - \mathbf{r}_{i0})^2}{2\sigma_r^2} + i(\mathbf{r}_i - \mathbf{r}_{i0}) \cdot \mathbf{p}_{i0}/\hbar}, \quad (11)
$$

here, $\sigma_r$ and $\mathbf{r}_{i0}$ are the width and centroid of wave packet, respectively. Its Wigner density reads,

$$
\begin{aligned}
f_i(\mathbf{r}, \mathbf{p}, t) &= \frac{1}{(2\pi\sigma_r^2)^{3/2}} e^{-(\mathbf{r}-\mathbf{r}_{i0})^2/2\sigma_r^2} \\
&\quad \frac{1}{(2\pi\sigma_p^2)^{3/2}} e^{-(\mathbf{p}-\mathbf{p}_{i0})^2/2\sigma_p^2} \\
&= \frac{1}{(\pi\hbar)^3} \exp[-\frac{(\mathbf{r}_i-\mathbf{r}_{i0})^2}{2\sigma_r^2} - \frac{(\mathbf{p}_i-\mathbf{p}_{i0})^2}{2\sigma_p^2}],
\end{aligned}
\tag{12}
$$

where $\sigma_r\sigma_p = \hbar/2$. The features of $f_i$ from Gaussian wave function are as follows: (a) $f_i$ does not spread with time for fixed $\mathbf{p}$ or $\mathbf{r}$, even though the underlying wave function does. But it can reproduce the fine structure of nucleon distance in fragmentation. (b) An equation $f(\mathbf{r}, \mathbf{p}, t) = \delta(\mathbf{r}-\mathbf{r}_{i0})\delta(\mathbf{p}-\mathbf{p}_{i0})$ is out of reach, because the uncertainty relation $\Delta x \Delta p_x \geq \frac{\hbar}{2}$. (c) As $\sigma_r$ tends to infinity, $f_i$ spread uniformly over space, while the momentum factor is sharp in momentum.

Since the nuclei of colliding is a $N$-body system, its wave function should be $\psi(\mathbf{r}_1, ... \mathbf{r}_N)$. In the QMD approach, the system wave function is assumed as a direct product of $N$ coherent states, which is in the Hartree approximation,

$$
\psi(\mathbf{r}_1, ..., \mathbf{r}_N) = \phi_{k_1}(\mathbf{r}_1)\phi_{k_2}(\mathbf{r}_2)...\phi_{k_n}(\mathbf{r}_N), \tag{13}
$$
$$
\phi_{k_i}(\mathbf{r}_i) = \tag{14}
$$
$$
\frac{1}{(2\pi\sigma_r^2)^{3/4}} \exp[-\frac{(\mathbf{r}_i-\mathbf{r}_{i0})^2}{2\sigma_r^2} + \frac{i\mathbf{p}_{i0}\cdot(\mathbf{r}_i-\mathbf{r}_{i0})}{\hbar}],
$$

where $\phi_{k_i}(\mathbf{r}_i)$ is the wave function of the $i$th particle at state $k_i$ ($p_i = \hbar k_i$ form). $\phi_{k_i}$ is chosen as Gaussian wave packet to avoid the negative values of phase space distribution ($f_N$).

For the $N$-body Wigner function within the QMD assumption, it reads

$$
f_N(\mathbf{r}_1, ..., \mathbf{r}_N; \mathbf{p}_1, ..., \mathbf{p}_N) = \prod_{i=1}^{N} f(\mathbf{r}_i, \mathbf{p}_i) \tag{15}
$$
$$
= \prod_{i=1}^{N} \frac{1}{(\pi\hbar)^3} \exp[-\frac{(\mathbf{r}_i-\mathbf{r}_{i0})^2}{2\sigma_r^2} - \frac{(\mathbf{p}_i-\mathbf{p}_{i0})^2}{2\sigma_p^2}],
$$

$\mathbf{r}_{i0} = <\mathbf{r}_i>$ and $\mathbf{p}_{i0} = <\mathbf{p}_i>$ are the centroids of wave packets in coordinate and momentum space. The time evolution of the $f_N$, i.e. Eq. (6), is a highly nonlinear integral-differential equation and is difficult to be solved exactly.

Since the width of wave packet is fixed during the time evolution in the QMD approach, the time evolution of the phase space density can be determined from the time evolution of the centroids of the wavepacket in the coordinate and momentum spaces, which are driven by the mean field potential and nucleon-nucleon collisions. Two ingredients in Eq. (6), such as the mean field part and the collision part, are solved separately rather than self consistently in practical calculations. The initialization is also very important for simulating the heavy ion collisions, and we will mention it in the next section.

In the Vlasov model (i.e. only the mean field and without collision part), the time evolution of the centroid of wave packets in the coordinate and momentum space is derived in previous section and it is written as,

$$
\frac{\partial <\mathbf{p}_i>}{\partial t} \approx -\frac{\partial U(\mathbf{r}_{10}, ..., \mathbf{r}_{N0})}{\partial \mathbf{r}_{i0}}, \tag{16}
$$
$$
\frac{\partial <\mathbf{r}_i>}{\partial t} = \frac{\mathbf{p}_{i0}}{m}. \tag{17}
$$

The time evolution of the centroids of $\mathbf{p}_{i0}$ and $\mathbf{r}_{i0}$ have also been derived by using the Euler-Lagrange equations as in Ref. [148],

$$
\frac{d}{dt}\frac{\partial \mathcal{L}}{\partial \dot{\mathbf{p}}_{i0}} - \frac{\partial \mathcal{L}}{\partial \mathbf{p}_{i0}} = 0 \rightarrow \dot{\mathbf{r}}_{i0} = \frac{\mathbf{p}_{i0}}{m} + \nabla_p U_i, \tag{18}
$$
$$
\frac{d}{dt}\frac{\partial \mathcal{L}}{\partial \dot{\mathbf{r}}_{i0}} - \frac{\partial \mathcal{L}}{\partial \mathbf{r}_{i0}} = 0 \rightarrow \dot{\mathbf{p}}_{i0} = -\nabla_p U_i. \tag{19}
$$

The potential energy $U$ in the quantum molecular dynamics model can be directly calculated from the potential operator $\hat{V} = v_{ij} + v_{ijk} + ...$ as follows based on the above assumption,

$$
\begin{aligned}
U &= \sum_{i<j} \int d\Gamma_i d\Gamma_j v_{ij} f_i(\mathbf{r}_i, \mathbf{p}_i) f_j(\mathbf{r}_j, \mathbf{p}_j) \\
&\quad + \sum_{i<j<k} \int d\Gamma_i d\Gamma_j d\Gamma_k v_{ijk} \\
&\quad f_i(\mathbf{r}_i, \mathbf{p}_i) f_j(\mathbf{r}_j, \mathbf{p}_j) f_k(\mathbf{r}_k, \mathbf{p}_k) + ... \\
&\equiv \sum_{i<j} <r_i, r_j|v_{ij}|r_i, r_j> \\
&\quad + \sum_{i<j<k} <r_i, r_j, r_k|v_{ijk}|r_i, r_j, r_k> + ..., \\
&= \sum_{i<j} U_{ij} + \sum_{i<j<k} U_{ijk} + ...,
\end{aligned}
\tag{20}
$$

$d\Gamma_i = d^3r_i d^3p_i$, $v_{ij}$, $v_{ijk}$ are two-body interaction, three-body interaction, respectively. As in Ref. [148], the interactions may consist of local interaction, Yukawa and Coulomb interactions.

The local interaction has the form,

$$
v_{ij} = t_1\delta(\mathbf{r}_1 - \mathbf{r}_2), v_{ijk} = t_2\delta(\mathbf{r}_1 - \mathbf{r}_2)\delta(\mathbf{r}_1 - \mathbf{r}_3), \tag{21}
$$

and one have

$$
U_{ij} = t_1\tilde{\rho}(\mathbf{r}_{i0}, \mathbf{r}_{j0}) = \frac{t_1}{(4\pi\sigma_r^2)^{3/2}} e^{-(\mathbf{r}_{i0}-\mathbf{r}_{j0})^2/4\sigma_r^2}, \tag{22}
$$

$$
\begin{aligned}
U_{ijk} &= \frac{t_2}{(2\pi\sigma_r^2)^3 \cdot 3^{3/2}} \\
&\quad e^{-((\mathbf{r}_{i0}-\mathbf{r}_{j0})^2+(\mathbf{r}_{i0}-\mathbf{r}_{k0})^2+(\mathbf{r}_{k0}-\mathbf{r}_{j0})^2)/6\sigma_r^2} \\
&\approx \frac{t_2}{(2\pi\sigma_r^2)^3 \cdot 3^{3/2}} e^{-((\mathbf{r}_{i0}-\mathbf{r}_{j0})^2+(\mathbf{r}_{i0}-\mathbf{r}_{k0})^2)/4\sigma_r^2}.
\end{aligned}
\tag{23}
$$

In the spin saturated nuclear matter, the three-body interaction can be viewed as the density dependent two-body interaction due to the hard core, and one can replace the three-body interaction to its effective form

$$t_2 \delta(\mathbf{r}_1 - \mathbf{r}_2)\delta(\mathbf{r}_1 - \mathbf{r}_3) = \frac{t_2}{6}\delta(\mathbf{r}_1 - \mathbf{r}_2)\rho((\mathbf{r}_1 + \mathbf{r}_2)/2). \quad (24)$$

The Yukawa interaction is as

$$V^{Yuk} = t_3 \frac{e^{-|\mathbf{r}_1 - \mathbf{r}_2|/\mu}}{|\mathbf{r}_1 - \mathbf{r}_2|/\mu}, \quad (25)$$

with $\mu = 1.5$ fm and $t_3 = -6.66$ MeV which can give the best preservation of nuclear surface for certain parameter sets, and it also gives the contributions to two-body terms.

The parameter $t_1$ and $t_2$ can be determined by fitting the nuclear matter potential,

$$V = \alpha(\frac{\rho}{\rho_0}) + \beta(\frac{\rho}{\rho_0})^2, \quad (26)$$

$\alpha$ term is related to the two-body interaction term, and $\beta$ term is related to the three-body interaction. The two free parameters $\alpha$ and $\beta$ can be determined by the requirement of average binding energy ($E_0$) and compressibility ($K_0 = 9\rho_0^2(\frac{\partial^2 E/A}{\partial \rho^2})_{\rho_0}$) at the normal density. In order to study the effects of different incompatibilities, one generalize the potential to be,

$$V = \alpha(\frac{\rho}{\rho_0}) + \beta(\frac{\rho}{\rho_0})^\gamma. \quad (27)$$

The additional third parameter $\gamma$ can allow us to fix the compressibility independent of other quantities.

The momentum dependent interaction is also important for describing flows over a wide incident energy range with reasonable compressibility. There are the Logarithm-type [200, 201] momentum dependent interaction,

$$U(\Delta \mathbf{p})\delta(\mathbf{r} - \mathbf{r}') = \quad (28)$$
$$1.57[ln(1 + 5 \times 10^{-4}\Delta p^2)]^2 \rho/\rho_0 \delta(\mathbf{r} - \mathbf{r}'),$$

with $\Delta p = |\mathbf{p}_1 - \mathbf{p}_2|$ in units of MeV/c and $U$ in MeV. Another form is the Lorentzian-type momentum-dependent nucleonic mean field $V_{md}$, which was used in RQMD/S and JAM [202, 203]

$$V_{md} = \sum_{k=1,2} \frac{C_{ex}^{(k)}}{\rho_0} \int d\mathbf{p}' \frac{f(\mathbf{r}, \mathbf{p}')}{1 + [(\mathbf{p} - \mathbf{p}')/\mu_k]^2}. \quad (29)$$

$C_{ex}^k$ and $\mu_k$ are the parameters of momentum dependent interaction. The mean field potential (Eq.(29)) leads to the following potential energy,

$$U_{md} = \sum_{k=1,2} \frac{C_{ex}^{(k)}}{\rho_0} \int d\mathbf{r} d\mathbf{p} d\mathbf{p}' \frac{f(\mathbf{r}, \mathbf{p})f(\mathbf{r}, \mathbf{p}')}{1 + [(\mathbf{p} - \mathbf{p}')/\mu_k]^2}. \quad (30)$$

Exact calculation of integral of Eq.(30) is time consuming in the QMD-type model calculations. Thus, in the actual calculations [203], the momentum dependent potential which used in the relativistic QMD framework [172] is

$$\sum_{k=1,2} \frac{C_{ex}^{(k)}}{2\rho_0} \sum_{j(\neq i)} \frac{1}{1 + [\tilde{p}_{ij}/\mu_k]^2} \rho_{ij}, \quad (31)$$

where $\rho_{ij} = \int d^3 r \rho_i(\mathbf{r})\rho_j(\mathbf{r})$. In their formulas, the relative distance $\mathbf{r}_{ij} = \mathbf{r}_i - \mathbf{r}_j$ and $\mathbf{p}_{ij} = \mathbf{p}_i - \mathbf{p}_j$ (for convenience, we use $\mathbf{r}_i$ and $\mathbf{p}_i$ to represent $\mathbf{r}_{i0}$ and $\mathbf{p}_{i0}$ in the following description) in the potentials were replaced by the squared four-vector distance with a Lorentz scalar,

$$\tilde{\mathbf{r}}_{ij}^2 = \mathbf{r}_{ij}^2 + \gamma_{ij}^2(\mathbf{r}_{ij} \cdot \beta_{ij})^2, \quad (32)$$
$$\tilde{\mathbf{p}}_{ij}^2 = \mathbf{p}_{ij}^2 - (p_i^0 - p_j^0)^2 + \gamma_{ij}^2(\frac{m_i^2 - m_j^2}{p_i^0 + p_j^0})^2.$$

The parameters in Eqs. (28) and (31) are determined by reproducing the real part of the global Dirac optical potential (Schrödinger equivalent potential) of Hama et al. [204], in which angular distribution and polarization quantities in proton-nucleus elastic scatterings are analyzed in the range of 10 MeV to 1 GeV. One should note, $\alpha$, $\beta$, $\gamma$ should be readjusted by fitting the EOS in the uniform nuclear matter after including the momentum dependent interaction term.

In the collision part, only binary collisions (two-body level) are considered. The collisions are performed in a point-particle sense with a similar way as in cascade [205] without considering the shape of nucleons. Each pair of particles within an same event is tested for a collision at every time step. In details, there is possible collision between particles 1 and 2, if their minimum distance $d_{12}$ in center of mass of colliding pair satisfy,

$$d_{12} \leq \sqrt{\sigma_{tot}(\sqrt{s})/\pi}, \quad (33)$$

where $\sigma_{tot}(\sqrt{s})$ is the total cross section of incoming particle 1 and 2 at the center of mass energy $\sqrt{s}$. $\sigma_{tot} = \sum_i^{N_c} \sigma_i$, and $i$ represents the outgoing channel and $N_c$ is the maximum number of outgoing channels. Another condition is that the particle 1 and 2 can move long enough for colliding in the time interval $-\delta t/2$ to $\delta t/2$. The channel of outgoing particles ($ic$) is chosen randomly according to the relative weights of the different reaction cross sections, such as

$$\sum_{i=1}^{ic-1} \sigma_i/\sigma_{tot} < \xi \leq \sum_{i=1}^{ic} \sigma_i/\sigma_{tot}, \quad (34)$$

$\xi$ is a random number. The momenta of the outgoing particles are generated randomly according to the angular differential cross sections and in agreement with the energy-momentum conservation laws. Thus, the cross sections constitute another major part of the model. In the original QMD model, the experimental values of nucleon-nucleon (baryon-baryon) cross sections in free space are used and the medium correction on the cross section is based on it. The Pauli blocking is considered as that in VUU [146, 206–209].

In addition, fragmentation is an important mechanism for intermediate energy heavy ion collisions. Thus, one also needs to identify the fragments at the end of simulations. A reasonable method for identifying the fragments from the simulation results of QMD calculations is needed for obtaining the reaction observables. It is found that the analyzing code also contains important physical contents and influence the final results to a certain extent.

## C. Improved Quantum Molecular Dynamics Model

The original version of QMD code we used was developed in the Frankfurt [210], and there are lots of developments accompanied with appearance of the new generation facilities along the beam energy, as well as isospin degree of freedom. Here, we briefly introduce the improvements we have made in past decades, and it was also known as improved quantum molecular dynamics model (ImQMD).

In our following description, for convenience, we use $\mathbf{r}_i$ and $\mathbf{p}_i$ to represent the $\mathbf{r}_{i0}$ and $\mathbf{p}_{i0}$. The adjustable parameters in the model can be divided into two types. One is related to the numerical calculations, and another is physics parameters. The adjustable parameter related to the numerical calculations is the width of Guassian wavepacket, the time step in computation, et ac. The adjustable parameters related to the physics are the mean field parameters and in-medium correction parameters, which totally have about 5-15 parameters. The exact number depends on the physics we study.

### 1. Nucleonic mean field

With the beam energy decreasing to the Coulomb barrier, the fermion properties of nucleons become more and more important and it naturally requires to develop the original transport model for including or mimicking these effects, such as including the effects from antisymmetric wave function and time evolution of width of wave function, which has led the antisymmetric molecular dynamics model (AMD) [142], Fermionic molecular dynamics model (FMD) [211, 212], extended quantum molecular dynamics model (EQMD) [213]. However, these treatments face a extremely large cost in calculation and thus hardly apply to heavy nuclear systems. One has to find a way to balance the efficiency of computation and physics before the revolution of computation ability.

In the framework of QMD approach, it can also be refined by improving the mean-filed part in two sides. One is to adopt a reasonable energy density functional, which can well reproduce the properties of finite nuclei, in the Hamiltonian equation. For example, one can include the Pauli potential in Hamiltonian [147, 214–217], which is a phenomenological repulsive momentum dependent potential, for mimicking the fermions' properties. Thus, the equation of motion of centroid of wave packet for particle $i$ reads,

$$\dot{\mathbf{r}}_i = \frac{\partial H}{\partial \mathbf{p}_i} + \frac{\partial H^{Pau}}{\partial \mathbf{p}_i}, \tag{35}$$

$$\dot{\mathbf{p}}_i = -\frac{\partial H}{\partial \mathbf{r}_i} - \frac{\partial H^{Pau}}{\partial \mathbf{r}_i}. \tag{36}$$

Another *ad hoc* method to mimic this effects is to adopt the phase space constraints [150–152]. This method requires to check whether the phase space occupation number of each particle ($\bar{f}_i$) violates the fermi-dirac distributions, i.e. $\bar{f}_i > 1$, during the evolution. If the phase space occupation number is greater than 1, the momentum direction of nucleons will be rearranged to let the phase space occupation number less than or equal to 1. More details can be found in Refs. [150–152].

In the ImQMD model, we mainly improve the mean field part based on the concept of energy density functional. It comes from the approximation used in the QMD approach, where the potential energy as a function of centroid of wave function are used (i.e., $< \nabla_{\mathbf{r}_i} V(\mathbf{r}_1, ..., \mathbf{r}_N) > \approx \nabla_{<\mathbf{r}_i>} U(< \mathbf{r}_1 >, ..., < \mathbf{r}_N >)$). Thus, one can also directly calculate the potential energy $U$ from its energy density functional $U = \int u[\rho] d^3 r$, by using the $\rho(\mathbf{r}) = \sum \rho_i = \frac{1}{(2\pi\sigma_r^2)^{3/2}} e^{-(\mathbf{r}-\mathbf{r}_{i0})/2\sigma_r^2}$. In this case, the self contribution to density is considered.

Till now, there are three kinds of Skyrme-type energy density functional used in the ImQMD which depends on the energy region we used.

1), In order to study the heavy ion reaction at low beam energy, we adopt a reasonable energy density functional (EDF) derived from the Skyrme EDF and the Fermi constraints (similar concept as phase space constraints) is used. It is usually named as ImQMD-$v_2$ [153, 160, 218–223], and mainly used at the beam energy above Coulomb barriers and less than Fermi energy. In the simulations, calculations are stopped when the dynamical processes are finished, for example, whether the composite system reach equilibrium. The corresponding stop time is later 1000-5000 fm/$c$ than the projectile and target contacting time.

The Hamiltonian reads

$$H = T + U = \sum \frac{p_i^2}{2m} + \int u_\rho d^3 r + U_{Coul}. \tag{37}$$

The potential energy density functional $u_\rho$ used in the ImQMD-$v_2$ for low energy heavy ion collisions (near and above the Coulomb barrier energy domain) reads

$$u_\rho = \frac{\alpha}{2} \frac{\rho^2}{\rho_0} + \frac{\beta}{\gamma+1} \frac{\rho^{\gamma+1}}{\rho_0^\gamma} + \frac{g_{sur}}{2\rho_0}(\nabla\rho)^2$$
$$+ \frac{C_s}{2\rho_0}[\rho^2 - \kappa_s(\nabla\rho)^2]\delta^2 + g_{\rho\tau}\frac{\rho^{\eta+1}}{\rho_0^\eta}, \tag{38}$$

where the asymmetry $\delta = (\rho_n - \rho_p)/(\rho_n + \rho_p)$, $\rho_n$ and $\rho_p$ are the neutron and proton densities. The $g_{sur}$ is the coefficient related to the density gradient. $C_s$ is the symmetry potential coefficient, and $\kappa_s$ is the parameter related to isospin dependent density gradient term. $g_{\rho\tau}$ term is obtained from the contribution of the $\rho\tau$ term in Skyrme energy density functional, where the $\tau$ is the kinetic energy density and expressed with the density $\rho$. The details can be found in Refs. [153, 160, 218, 219, 221]. The parameters in the $u_\rho$ as well as the width of the wave-packet,

$$\sigma_r = \sigma_0 + \sigma_1 A^{1/3}, \qquad (39)$$

in the coordinate space, here $A$ is the number of nucleons in projectile or target, are given in Table I which are determined by fitting the properties (including the stability) of nuclei at ground state, the fusion excitation functions of a number of heavy ion fusion reactions at energies around the Coulomb barrier and the charge distributions in multifragmentation process at Fermi energies.

For the lower beam energies, the excitation of system is low and the momentum distribution of the reaction system is not far from the Fermi-Dirac distribution at zero temperature. Thus, we roughly use $\eta = 5/3$ in the $g_{\rho\tau}$ term to approximately describe the contribution from the $\rho\tau$ term. Actually, the exact calculations of $\rho\tau$ terms in transport models should be directly based on the relative momentum of nucleons, and it will give the obviously momentum dependent interaction and effective mass. It will be interesting to check how the momentum dependent interaction influence the low energy reactions.

2), When the beam energies are high enough to trigger the multifragmentation, which has close relation with the EOS in a wide density range, investigation on the nuclear EOS becomes possible, especially for isospin asymmetric nuclear equation of state with the building of new generation rare facility in last couple decades. It requires to develop the transport codes which can incorporate kinds of effective interactions such as the widely used Skyrme interactions (Skyrme potential energy density functionals). At this energy region, the calculations are stopped after 100-200 fm/$c$ when the projectile and target contact, and the exact values depends on the beam energy we studied.

In the version of ImQMD05 [154–156], it is mainly applied in the energy ranging from 20 MeV/nucleon to 300 MeV/nucleon. The Coulomb interaction is as same as the previous treatments, but the nucleonic potential energy density functional, i.e. $u = u_\rho + u_{md}$, reads as:

$$
\begin{aligned}
u_\rho = {} & \frac{\alpha}{2}\frac{\rho^2}{\rho_0} + \frac{\beta}{\gamma+1}\frac{\rho^{\gamma+1}}{\rho_0^\gamma} + \frac{g_{sur}}{2\rho_0}(\nabla\rho)^2 \\
& + \frac{g_{sur,iso}}{\rho_0}(\nabla(\rho_n-\rho_p))^2 + g_{\rho\tau}\frac{\rho^{8/3}}{\rho_0^{5/3}} \\
& + u_\rho^{sym}.
\end{aligned} \qquad (40)
$$

The last term in Eq. (40) is the symmetry potential energy density functional,

$$u_\rho^{sym} = [A_{sym}\frac{\rho}{\rho_0} + B_{sym}(\frac{\rho}{\rho_0})^\gamma + C_{sym}(\frac{\rho}{\rho_0})^{5/3}]\delta^2\rho, \quad (41)$$

which makes it possible to investigate the different Skyrme forms of density dependence of symmetry energy. For the Skyrme interactions, symmetry potential energy terms come from two-body, three-body and momentum dependent interaction terms, and its related parameters are $A_{sym}$, $B_{sym}$, and $C_{sym}$, respectively. All the parameters in ImQMD, such as $\alpha$, $\beta$, $\eta$, $g_{sur}$, $g_{sur,iso}$, $g_{\rho\tau}$ and $A_{sym}$, $B_{sym}$, $C_{sym}$ can be derived from the standard Skyrme parameters, $x_0$, $x_1$, $x_2$, $x_3$, $t_0$, $t_1$, $t_2$, $t_3$, $\sigma$. The relationship between the standard Skyrme parameters and parameters in the ImQMD can be found in Ref. [158].

If one sets $A_{sym} = C_{sym} = 0$ and $B_{sym} = \frac{C_s}{2}$, $u_\rho^{sym}$ becomes

$$u_\rho^{sym} = \frac{C_s}{2}(\frac{\rho}{\rho_0})^{\gamma_i}\rho\delta^2, \qquad (42)$$

and one can easily investigate the power law form of density dependence of symmetry energy. In the following text, we use $\gamma_i$ to denote the symmetry potential parameters in the power law form of density dependence of symmetry energy.

According to the Eq.(41) and Eq.(42), one can write the density dependence of the symmetry energy of cold nuclear matter as

$$
\begin{aligned}
S(\rho) = {} & \frac{\hbar^2}{6m}(\frac{3\pi^2\rho}{2})^{2/3} + A_{sym}\frac{\rho}{\rho_0} \\
& + B_{sym}(\frac{\rho}{\rho_0})^\gamma + C_{sym}(\frac{\rho}{\rho_0})^{5/3},
\end{aligned} \qquad (43)
$$

and

$$S(\rho) = \frac{\hbar^2}{6m}(\frac{3\pi^2\rho}{2})^{2/3} + \frac{C_s}{2}(\frac{\rho}{\rho_0})^{\gamma_i}, \qquad (44)$$

respectively. In Eq.(43) and Eq.(44), the first term is kinetic symmetry energy term which comes from the kinetic energy contributions. The rest terms are the symmetry potential energy.

The energy density associated with the mean-field momentum dependence is represented as

$$
\begin{aligned}
u_{md} = {} & \frac{1}{2\rho_0}\sum_{N_1,N_2}\frac{1}{16\pi^6}\int d^3p_1 d^3p_2 f_{N_1}(\mathbf{p}_1)f_{N_2}(\mathbf{p}_2) \\
& 1.57[\ln(1 + 5\times10^{-4}(\Delta p)^2)]^2,
\end{aligned} \qquad (45)
$$

$f_N$ are nucleon Wigner functions, $\Delta p = |\mathbf{p}_1 - \mathbf{p}_2|$. The energy is in MeV and momenta are in MeV/$c$. With the interaction as in Eq.(45), the Skyrme-EOS will be modified with additional repulsion from momentum dependent interaction. In order to get the reasonable Skyrme EOS under the momentum dependent interaction as in Eq.(45), we use the $u'_{md}$ in the model, which is obtained by substraction the $u_{md}$ at T=0 MeV [158], i.e.,

$$u'_{md} = u_{md} - u_{md}(T=0), \tag{46}$$

for refitting the Skyrme EOS.

TABLE I: Model parameters adopted in ImQMD-v2.

| Para. | $\alpha$ (MeV) | $\beta$ (MeV) | $\gamma$ | $g_{sur}$ (MeVfm$^2$) | $g_\tau$ (MeV) | $\eta$ | $C_s$ (MeV) | $\kappa_s$ (fm$^2$) | $\rho_0$ (fm$^{-3}$) | $\sigma_0$ (fm) | $\sigma_1$ (fm) |
|---|---|---|---|---|---|---|---|---|---|---|---|
| IQ2 | $-356$ | 303 | 7/6 | 7.0 | 12.5 | 2/3 | 32 | 0.08 | 0.165 | 0.88 | 0.090 |
| IQ3 | $-207$ | 138 | 7/6 | 18.0 | 14.0 | 5/3 | 32 | 0.08 | 0.165 | 0.94 | 0.018 |
| IQ3a | $-207$ | 138 | 7/6 | 16.5 | 14.0 | 5/3 | 34 | 0.4 | 0.165 | 0.94 | 0.020 |

3) In the version ImQMD-Sky [159, 224], a standard parametrization of the Skyrme potential energy density functional with only the spin-orbit interaction neglected is used. It is also mainly used in the beam energy ranging from 20 MeV/nucleon to 300 MeV/nucleon.

The nucleonic potential energy density functional $u$ is

$$
\begin{aligned}
u &= \frac{\alpha}{2}\frac{\rho^2}{\rho_0} + \frac{\beta}{\gamma+1}\frac{\rho^{\gamma+1}}{\rho_0^\gamma} + \frac{g_{sur}}{2\rho_0}(\nabla\rho)^2 \\
&+ \frac{g_{sur,iso}}{\rho_0}(\nabla(\rho_n-\rho_p))^2 \\
&+ A_{sym}(\frac{\rho}{\rho_0})\delta^2\rho + B_{sym}(\frac{\rho}{\rho_0})^\gamma\delta^2\rho \\
&+ u_{md}^{sky}.
\end{aligned}
\tag{47}
$$

The difference between Eqs. (40-41) and Eq. (48) is that we replace the $g_{\rho\tau}(\frac{\rho}{\rho_0})^{5/3}\rho$ and $C_{sym}(\frac{\rho}{\rho_0})^{5/3}\rho\delta^2$ in Eq. (40) with the exact Skyrme type momentum dependent interaction terms $u_{sky}^{md}$ as in Eq. (48). The energy density of Skyrme type momentum dependent interaction we used is,

$$
\begin{aligned}
u_{md}^{sky} &= u_{md}(\rho\tau) + u_{md}(\rho_n\tau_n) + u_{md}(\rho_p\tau_p) \tag{48} \\
&= C_0\int d^3p d^3p' f(\mathbf{r},\mathbf{p})f(\mathbf{r},\mathbf{p}')(\mathbf{p}-\mathbf{p}')^2 + \\
&D_0\int d^3p d^3p'[f_n(\mathbf{r},\mathbf{p})f_n(\mathbf{r},\mathbf{p}')(\mathbf{p}-\mathbf{p}')^2 \\
&+ f_p(\mathbf{r},\mathbf{p})f_p(\mathbf{r},\mathbf{p}')(\mathbf{p}-\mathbf{p}')^2].
\end{aligned}
$$

This formula is derived based on the Skyrme-type momentum dependent interaction, $\delta(\mathbf{r}-\mathbf{r}')(\mathbf{p}-\mathbf{p}')^2$, and $f(\mathbf{r},\mathbf{p})$ is the nucleon phase space density. In QMD approaches, $f(\mathbf{r},\mathbf{p}) = \sum_i \frac{1}{(\pi\hbar)^3}\exp[-(\mathbf{r}-\mathbf{r}_i)^2/2\sigma_r^2 - (\mathbf{p}-\mathbf{p}_i)^2/2\sigma_p^2]$. The coefficients $C_0$ and $D_0$ can be determined for fitting

$$
\begin{aligned}
\mathcal{H}_{eff} &= \frac{1}{8}[t_1(2+x_1)+t_2(2+x_2)]\tau\rho \tag{49} \\
&+ \frac{1}{8}[t_2(2x_2+1)-t_1(2x_1+1)](\tau_n\rho_n+\tau_p\rho_p),
\end{aligned}
$$

in nuclear matter at T=0 MeV, where $f$ in Eq. (48) becomes the zero temperature Fermi-Dirac distributions

and $\tau = \frac{3}{5}(\frac{3\pi^2}{2})^{2/3}\rho^{2/3}$ in Eq. (49). Finally, we have

$$C_0 = \frac{1}{16\hbar^2}[t_1(2+x_1)+t_2(2+x_2)], \tag{50}$$

$$D_0 = \frac{1}{16\hbar^2}[t_2(2x_2+1)-t_1(2x_1+1)]. \tag{51}$$

Thus, both the effects of symmetry energy potential and the neutron/proton effective mass splitting on the isospin asymmetric heavy ion collisions can be studied simultaneously with this code.

Furthermore, the same parametrization of the potential energy density makes better relation between the studies of heavy ion collisions and the nuclear structure. One should note that if the form of $u_{md}^{sky}$ is adopted in the ImQMD model, one can not use it to study the reaction at the beam energy above 300 MeV/nucleon. The reason is that the Skyrme type optical potential increases to infinity with the momentum, which violates the optical potential behavior obtained in experiments of nucleon-nucleus reaction.

Correspondingly, the equation of state of cold nuclear matter from the density functional used in the ImQMD can be written as,

$$
\begin{aligned}
E/A &= \frac{3\hbar^2}{10m}(\frac{3\pi^2}{2}\rho)^{2/3} \tag{52} \\
&+ \frac{\alpha}{2}\frac{\rho}{\rho_0} + \frac{\beta}{\gamma+1}\frac{\rho^\gamma}{\rho_0^\gamma} + g_{\rho\tau}\frac{\rho^{5/3}}{\rho_0^{5/3}} \\
&+ S(\rho)\delta^2,
\end{aligned}
$$

with the gradient terms vanish in the uniform nuclear matter. The density dependence of symmetry energy $S(\rho)$ is as same as in Eq.(43) or Eq.(44), depending on which form of density dependence of the symmetry energy is used. The terms $g_{\rho\tau}$ in Eq.(52) and $C_{sym}$ in $S(\rho)$ come from the energy density of Skyrme type momentum dependent interaction as in Eq.(48) when T=0 MeV in nuclear matter.

### 2. Collision part

In transport models for simulating the low-intermediate energy heavy ion collisions, the nucleon-nucleon collisions are determined with the concept of closest approaches. In the ImQMD model, each pair of particles within an same event is tested for a collision at every time step. More explicitly, suppose there is possible collision between particles 1 and 2, specified with $(t_0, \vec{r}_1)$ and $(E_1, \vec{p}_1)$, and $(t_0, \vec{r}_2)$ and $(E_2, \vec{p}_2)$, respectively, at the current time $t_0$ in the reference frame of system. In the center-of-mass frame of the two particles, their trajectories $\vec{\mathcal{R}}_i^*(t^*)$ without a mean field are straight lines pointing along the constant velocities $\vec{v}_1^* = \vec{p}_1^*/E_1^*$ and $\vec{v}_2^* = \vec{p}_2^*/E_2^*$. The asterisks represent quantities in the two-particle center-of-mass frame, while quantities without asterisk are in the calculational reference frame. The transformations from system center of mass reference to the two-particle center of mass reference frame for momentum and coordinate are,

$$\mathbf{p}_i^* = ((\gamma - 1)\mathbf{p}_i \cdot \frac{\boldsymbol{\beta}}{\beta^2} - \gamma E_i)\boldsymbol{\beta} + \mathbf{p}_i, \qquad (53)$$

$$\mathbf{r}_i^* = (\gamma - 1)\mathbf{r}_i \cdot \frac{\boldsymbol{\beta}}{\beta^2}\boldsymbol{\beta} + \mathbf{r}_i, \qquad (54)$$

where,

$$\boldsymbol{\beta} = \frac{\mathbf{p}_i + \mathbf{p}_j}{E_i + E_j}, \gamma = \frac{1}{\sqrt{1 - \beta^2}}. \qquad (55)$$

As the trajectories of 1 and 2 are known exactly, the minimum distance can be calculated as

$$d_\perp^{*2} = (\vec{r}_1^* - \vec{r}_2^*)^2 - \frac{[(\vec{r}_1^* - \vec{r}_2^*) \cdot \vec{v}_{12}^*]^2}{v_{12}^{*2}}, \qquad (56)$$

with $\vec{v}_{12}^* = \vec{v}_1^* - \vec{v}_2^*$. In the ImQMD codes, the distance condition for a collision to occur is

$$\pi d_\perp^{*2} < \sigma. \qquad (57)$$

Here, the frame-independent definition of the cross section (via the impact parameter in the two-particle rest frame) is an important factor in ensuring the approximate reference-frame independence.

Another criteria is to judge whether the collision occurs during the time interval of the current time step. In the ImQMD codes, the time of the closest approach is considered in the two-particle center-of-mass frame, where it may be written as

$$t_{\text{coll}}^* = t_0^* - \frac{(\vec{r}_1^* - \vec{r}_2^*) \cdot \vec{v}_{12}^*}{v_{12}^{*2}}, \qquad (58)$$

corresponding to the minimum distance $d_\perp^*$ given by Eq. (56). Note that $\vec{r}_1^*$ and $\vec{r}_2^*$ are the positions at different times $t_1^*$ and $t_2^*$, respectively, in this frame.

In the Bertsch prescription [199], the condition of the closest approach for this time step is set as $|(\vec{r}_1^* - \vec{r}_2^*) \cdot \vec{v}_{12}^*/v_{12}^{*2}| < \frac{1}{2}\delta t$ which is equivalent to $t_{\text{coll}}^* \in [t_0^* - \frac{1}{2}\delta t, \ t_0^* + \frac{1}{2}\delta t]$. $\delta t$ is the time interval in c.m. of two colliding particles, and it is related to the $\Delta t$ in computation reference is, $\delta t = \alpha \Delta t$ and $\alpha$ is defined as

$$\alpha = \gamma \frac{E_1^* E_2^*}{E_1 E_2} \qquad (59)$$

with Lorentz factor $\gamma = 1/\sqrt{1 - \beta^2}$ where $\beta$ is the velocity of the center-of-mass of the colliding pair. We have the usual time dilation factor $\alpha = 1/\gamma$ in the limit that the two particles have a common velocity. This method can well describe the nucleon-nucleon collision ranging from low-intermediate energy heavy ion collisions. With the beam energy is high enough, the relativistic effects become more and more important, and the time order of nucleon-nucleon collision may depend on the frame. There are some discussions on it [171], but we will not extend this point in detail according to the scope of this paper.

The nucleon-nucleon cross sections and their differential cross sections in free space are taken from Ref. [225]. The in-medium total nucleon-nucleon cross section $\sigma_{total}^*$ is taken as the form, $\sigma_{total}^* = (1 - \eta(E_{beam}))\sigma_{total}^{free}$, where the $\sigma_{total}^{free}$ is taken as that in Ref. [225]. $\sigma_{total}^* = \sigma_{el}^* + \sigma_{inel}^*$, and $\sigma_{el}^*$ and $\sigma_{inel}^*$ are the in-medium elastic and inelastic two-body cross section. The elastic and inelastic channel is determined by the possibility, $P_{el} = \sigma_{el}^*/\sigma_{total}^*$. After generating a random number $\xi$, the collision for elastic channel is determined by $\xi < \sigma_{el}^*/\sigma_{total}^*$. If the inelastic collision happens, for example $NN \rightarrow N\Delta$, the mass of $\Delta$ resonance should be sampled according to a normalized probability distribution for the mass of produced $\Delta$, i.e. $P(m, s)$ [226], where $m$ is the mass of produced $\Delta$ and $s$ is the center of mass energy. The momentum direction of outgoing nucleons is determined by the differential cross section with the Monte-Carlo method by considering the energy-momentum conservation [199].

The decay time of resonance is sampled according to its decay probability in its rest frame, i.e.,

$$P(t_{dec}) = 1 - e^{-\frac{t_{dec}}{\tau}} = 1 - e^{-\Gamma t_{dec}}, \qquad (60)$$

where $\tau$ is $1/\Gamma$, and $\Gamma$ is the decay width. Thus, we have

$$t_{dec} = -\frac{1}{\Gamma}ln(1 - \xi), \qquad (61)$$

$\xi$ is a random number. The decay time in the computational frame ($t_{dec}'$), is related to that in the rest frame of the decaying particle ($t_{dec}$) by a Lorentz factor, e.g.,

$$t_{A \rightarrow BC}' = \gamma t_{A \rightarrow BC} = \frac{E}{m_A} t_{A \rightarrow BC}, \qquad (62)$$

where $E_A = \sqrt{m_A^2 + \mathbf{p}^2}$ is the energy of particle $A$ in computation reference. The momentum of decayed particles can be calculated based on the kinematic relationship.

After the collision or decay, the corresponding momenta will be changed if there is no Pauli blocking, but the positions of them do not change.

### 3. Pauli blocking

The outgoing nucleon or baryon after attempted scattering will be checked whether it is Pauli-blocked. The occupation probability $f_i'$ at the centroid of the scattered wave packet with final momentum $P_i'$ is obtained from the Wigner distribution function corresponding to the QMD wave function given in Eq. (11), with the self-contribution excluded, i.e.,

$$
\begin{aligned}
f_i' &= f_\tau(\vec{R}_i, \vec{P}_i') \\
&= \frac{1}{2/(2\pi\hbar)^3} \frac{1}{(\pi\hbar)^3} \sum_{k \in \tau(k \neq i)} e^{-(\vec{R}_i - \vec{R}_k)^2/2\sigma_r^2} \quad (63) \\
&\times e^{-2(\sigma_r/\hbar)^2(\vec{P}_i' - \vec{P}_k)^2},
\end{aligned}
$$

with $\tau = n$ or $p$, which estimates the probability of finding nucleons in a phase space cell of dimension $(2\pi\hbar)^3$. The factor $2/(2\pi\hbar)^3$ results from consideration of the spin in the phase-space cell. The prefactors combine into a total factor 4. The occupation probability for $i$ is finally taken as $\min\{1, f_i'\}$.

Furthermore, the criteria $\frac{4\pi}{3} r_{i'k}^3 \frac{4\pi}{3} p_{i'k}^3 \geq h^3/8$ is also used in the ImQMD codes. $i'$ denotes the outgoing nucleon, $k$ represents the other surrounding nucleon which should be in the range of all nucleons except itself. It means that the outgoing nucleon should not be too close to others in phase space. Similar method used in our developed UrQMD model.

### 4. Initialization

When solving the equations of motion for nucleons, i.e. Eq. (7), one needs to know the initial coordinate and momentum of each nucleon in projectile and target, which is the staring point of simulating heavy ion collisions. A reasonable initial condition is of vital importance for correctly describing heavy ion collision.

In the initialization, the positions and momenta of the nucleons in reaction system are sampled according to the density and momentum distributions of projectile and target nuclei. Considering the reaction geometric (such as the impact parameters) and the beam energy, the projectile and target nuclei are boosted into the center of mass frame of the reaction system. Since the width of wave packet is fixed and missing some quantum effects, the ground state of nuclei in the QMD approaches may deviate from its ground state in nature. The initial nuclei obtained in QMD approaches usually have certain excitation. In order to obtain the reasonable initial nuclei with less excited, there are two methods to handle it. One is to reduce the excitation energy by solving the damped equation of motion to find the energy minimum of the system, and it was introduced in Ref. [213]. This method sometimes encounters a difficulty of finding the reasonable damping parameters and the evolution time for different systems. Another method is to select the pre-prepared initial nuclei that satisfactorily describe the properties of projectile and target nuclei, such as the binding energy, the root-mean-square radius ($< r^2 >^{1/2}$) and the shape for deformed nuclei. Furthermore, the selected pre-prepared initial nuclei should be stable long enough time. One should note that all these check should be under the same nucleon-nucleon interaction or the same energy density functional used in the dynamical calculations.

In the ImQMD model for low energy heavy ion reactions, the second method is adopted [153, 160]. To better describe the properties of neutron-rich nuclei, the neutron skin thickness of neutron-rich nuclei is taken into account in the initialization of the ImQMD model. Based on the 4-parameter nuclear charge radii formula proposed in Ref. [227]

$$
\begin{aligned}
R_c(fm) &= 1.226 A^{1/3} + 2.86 A^{-2/3} \quad (64) \\
&- 1.09(I - I^2) + 0.99 \Delta E/A,
\end{aligned}
$$

with which the 885 measured charge radii can be reproduced with a rms deviation of 0.022 fm (rms deviation is $\sqrt{\frac{1}{m} \sum (R_{exp}^i - R_{the}^i)^2}$ ), and the linear relationship between the neutron skin thickness $\Delta R_{np} = \langle r_n^2 \rangle^{1/2} - \langle r_p^2 \rangle^{1/2}$ and the isospin asymmetry $I = (N-Z)/A$ [228]

$$
\Delta R_{np} = 0.9 I - 0.03, \quad (65)
$$

one can obtain the proton radii $R_p = \sqrt{\frac{5}{3} \left[ \langle r_c^2 \rangle - 0.64 \right]}$ from the charge radii $\langle r_c^2 \rangle^{1/2} = \sqrt{\frac{3}{5}} R_c$ and the neutron radii $R_n = \sqrt{\frac{5}{3} \left[ \langle r_p^2 \rangle^{1/2} + \Delta R_{np} \right]}$. The nucleon positions are sampled within the hard sphere with a radius $R_p - w_r$ for the protons and $R_n - w_r$ for the neutrons, respectively. Here, $w_r = 0.8$ fm is to take into account the influence of the width of the wave-packet in the coordinate space. Only those initially prepared nuclei kept good ground state properties and stable for a long enough time are finally selected as the initial nuclei applied for the simulation of the reaction.

For study of fusion reaction at near barrier energy, a more fine procedure is adopted. Fig. 1 shows the average density distribution of $^{208}$Pb and $^{132}$Sn with 500 events. The solid symbols denote the results of the ImQMD model at the initial time, in which the value of $\sigma_r$ is obtained based on the Eq.(39), and the curves denote the corresponding results of the Skyrme Hartree-Fock calculations with the force SkM* [229]. One can find the event averaged density distribution is in very good consistency with the Skyrme Hartree-Fock calculation at given $\sigma_r$.

In version ImQMD-05 and ImQMD-Sky, the sampling is treated less complicated than in the lower energy version. The binding energy and nuclear radius are adopted to select the reasonable initial nuclei, but the stability of initial nuclei are not forced to keep very long time due to the fact that the reaction time decreases quickly with beam energy increases.

One should note that the initial fluctuation is involved naturally in the QMD-type models. It comes from the randomness of the position and momentum of each particle at the initial time when each event is initialized under the macroscopic conditions. To illustrate the initial fluctuation in QMD type models, Fig. 2 shows the density distribution sampled in the box with periodic boundary condition. The required density profile is $\rho(r) = \rho_0 + 0.2\rho_0 sin(2\pi z/L)$ with L=20 fm, which is shown as black solid line. The color lines are the sampled density profile in QMD approach from ten different events. One can find that the profiles of the density distribution fluctuate obviously. This initial fluctuation propagates following the evolution of system controlled by the equation of motion Eq. (7) and scattering process. It plays important roles in many processes of heavy ion reactions, for instance, in the MNT reaction, and multi-framentation process. However, one should note it is not as exactly same as the fluctuations due to many-body correlations (i.e., $f_N(\mathbf{r}_1, ..., \mathbf{r}_N, \mathbf{p}_1, ..., \mathbf{p}_N)$, with $N \geq 2$), which is the main venue to go beyond dissipative mean field dynamics. The physical fluctuation contributes part of the random force in terms of macroscopic dynamical model for heavy ion reactions [126, 139–141] and the other part is from the scattering process [158, 230]. It is still quantitatively unclear what is the strength of the random force, so as the strength of fluctuation, in the heavy ion reactions at different energies. How to extract the initial physical fluctuation obviously needs further study.

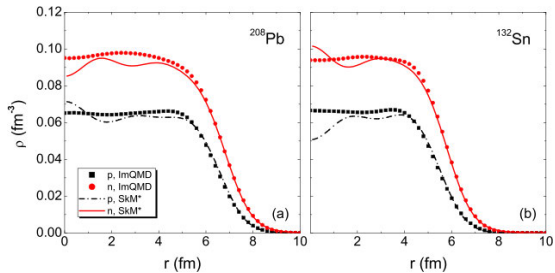

FIG. 1: (Color online) Density distribution of $^{132}$Sn and $^{208}$Pb at the initial time. The red for neutrons and the black for protons. Taken from Ref. [218].

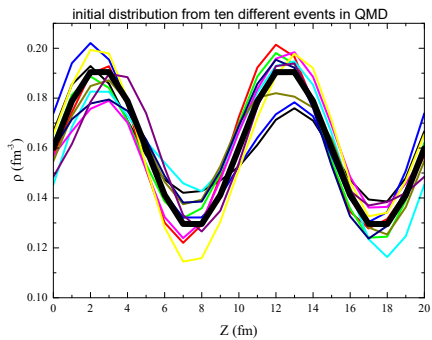

FIG. 2: (Color online) The thick black line is the required density profile, $\rho(r) = \rho_0 + 0.2\rho_0 sin(2\pi/Lz)$. The lines with different colors correspond to the density distribution from 10 different events.

## 5. Cluster recognization

One obvious character of QMD approach is that it can describe the cluster formation due to the $N$-body correlations caused by the overlapping wave packets, initial fluctuation, and nucleon-nucleon scattering process fluctuation. They are identified by the cluster recognition method in the real calculations at the end of simulation time at which the dynamical process is finished. All the fragments are still in excitation, and the secondary decay is allowed. Thus, one can expect that the distance between nucleons in the same fragment should not be changed beyond the criteria of nucleon force range.

Based on this idea, the minimum spanning tree algorithm is adopted to recognize the fragments. In this algorithm, the nucleons with relative distances of coordinate and momentum satisfying the conditions $|\mathbf{r}_i - \mathbf{r}_j| < R_0$ and $|\mathbf{p}_i - \mathbf{p}_j| < P_0$ are supposed to belong to a fragment. Here, $\mathbf{r}_i$ and $\mathbf{p}_i$ are the centroid of the wave packet for $i$th nucleon in their coordinate and momentum space. $R_0$ and $P_0$ are phenomenological parameters determined by fitting the global experimental data, such as the intermediate mass fragments (IMF) multiplicities [148, 154]. They should roughly be in the range of nucleon-nucleon interaction. Typical values of $R_0$ and $P_0$ used in the QMD approaches are about 3.5 fm and 250 MeV/c [148, 154], respectively. This approach has been quite successful in explaining some fragmentation observables, such as the charge distributions of the emitted particles, IMF multiplicities [148, 154], yield ratios of free neutrons to protons (n/p), and the double n/p ratios in heavy ion collisions [156].

On the other hand, the MST method fails to describe other details in the production of nucleons and light charged particles [148, 154, 231]. For example, while the yields of $Z = 1$ particles are overestimated, the yields of $Z = 2$ particles are underestimated partly because the strong binding of $\alpha$ particle can not be well described in transport models. Strong enhancements of the productions of neutron-rich isotopes observed in isoscaling [232], dynamically emitted heavy fragments [233] in neutron-rich HICs, and neutron-rich light charged particles (LCP) at mid-rapidity [234] have not been described well. Furthermore, most transport models predict more transparency than that observed experimentally in central collisions at intermediate energy [158, 235] due to insufficient production of fragments in the mid-rapidity region. Previous studies show that these problems cannot be resolved by changing only the mean field or nucleon-nucleon cross section in transport models.

There have been many attempts to improve the cluster recognition algorithm. More sophisticated algorithms such as the early cluster recognition algorithm (ECRA) [236], the simulated annealing clusterization algorithm (SACA) [237, 238], and the minimum spanning tree procedure with binding energy of fragments (MSTB)[239], have been developed to provide better description of the IMF multiplicities or the average $Z_{max}$ of the fragments. However, these algorithms do not address the lack of isospin dependence in cluster recognition.

By considering the properties of neutron-rich nuclei, such as neutron skin or neutron halo effect and Coulomb effects, in Ref. [157] a method of cluster recognition, namely, the iso-MST was proposed to mimic the isospin dependence in the cluster formation, in which the different values of $R_{nn}^0$, $R_{np}^0$, and $R_{pp}^0$ are chosen. It was found that $R_{nn}^0 = R_{np}^0 = 6$ fm and $R_{pp}^0 = 3$ fm give the suitable description on the isospin related observables in heavy ion collision at intermediate energies.

Our results show that the iso-MST method makes the suppression of $Z = 1$ particles and enhancement of fragments, especially for heavier fragments with $Z \geq 12$. Furthermore, we find enhanced production of neutron-rich isotopes at mid-rapidity. Consequently, isospin-sensitive observables, such as the double ratios, DR(t/$^3$He), and isoscaling parameter $\alpha$ increase to larger values. The widths of the longitudinal and transverse rapidity distributions of $Z = 1-6$ particles also change. In all the observables examined, the effects introduced by the iso-MST algorithm are relatively small but in the direction of better agreement with data [157].

## D. Ultra-relativistic Quantum Molecular Dynamics Model

In relativistic heavy ion collisions, kinds of baryons and mesons can be produced and the relativistic effects should be well considered. It is known that the UrQMD model inherits analogous principles as the QMD model [210] in its mean-field part and the relativistic QMD (RQMD)model[172] in the corresponding two-body collision part. It is successfully extended to describe HICs with beam energy starting from as low as several tens of MeV per nucleon (low SIS) up to the highest one available at CERN Large Hadron Collider (LHC).

In the UrQMD model [84, 171, 240, 241], it contains 55 different baryon species (including nucleon, $\Delta$ and hyperon resonances with masses up to 2.25 GeV/$c^2$) and 32 different meson species (including strange meson resonances), which are supplemented by their corresponding antiparticle and all isospin-projected states. The collision part of UrQMD has a good treatment on the sequence of collision/decay and on the frame dependence issue by using the collision/decay time table in the computational frame, where the collision time of baryons or decay time of reasonances is calculated based on Eq. (58) and Eq. (61), (62) and sorted according to the time order. This method is much better for handling the resonance particles's collision and decay during the propagation time step [226]. When the beam energy is less than 200 GeV/nucleon, the above described algorithm predicts that the particle multiplicities and collision numbers are less than 3% between the laboratory frame and the center of mass frame. Further discussion on the different treatments on the attempted collision are reviewed and compared in Ref. [226, 372].

In addition, based on the cascade mode which is also constantly updated [242], it is also possible to incorporate mean field interactions in the transport calculation. Since the EOS based on the first-principle lattice QCD calculations is still not available in current model investigations, two alternative methods to consider the strong influence of EOS on the dynamics of the expanding system have been tried by our group: the (mean-field) potential updates [73, 74, 243] and the UrQMD+hydrodynamics hybrid mode [244, 245]. In recent years, some improvements we incorporated in the UrQMD especially for HICs at intermediate and low energies are, briefly, as follows: 1) including and enriching the potentials such as the symmetry potential [246], the spin-orbit interaction [247], the magnetic field effect [248], the potentials for produced mesons [249, 250], as well as the Skyrme potential energy density functional [251, 252]; 2) incorporating the medium corrections effects on the NN elastic cross sections [77, 170, 253]; 3) improving the Pauli blocking by introducing the restriction of phase space for scattered particles [170].

# III. STUDY OF THE PHENOMENA AND THE MECHANISM IN LOW-INTERMEDIATE ENERGY HEAVY ION REACTIONS

## A. Heavy ion fusion reactions

Searching for the limits of the existence of nuclei is of fundamental importance for nuclear physics study. It is known that demarcation line for existence of a nucleus is driplines beyond which protons or neutrons leak out of nuclei [25–61]. The proton dripline has been reached for many isotopic chains. However, the neutron dripline is known only up to oxygen ($Z = 8$). The superheavy nuclide with $Z = 118$, $A = 294$ marks the current upper limit of nuclear charge and mass. Thus, to search for the way to produce new isotopes near driplines and superheavy nuclei becomes one of the most important tasks in nuclear research. The heavy ion fusion reactions and multi-nucleon transfer reactions are the most important methods for this aim.

In heavy ion fusion reactions, it is of great importance to explore the nucleus-nucleus potential and the formation process (neck dynamics) of compound nucleus, which are usually affected by dynamic effects, nuclear structure effects, isospin effects and so on. In this part, we try to understand it in the framework of microscopic transport model, ImQMD-v2, from the following points: a) dynamical nucleus-nucleus potential, b) neck dynamics and c) excitation function of fusion reaction, as they are the most important and relevant aspects relating to the reaction dynamics.

### 1. Dynamical nucleus-nucleus potential

. The nucleus-nucleus potential is commonly described as a function of the center-to-center distance between the projectile and target nuclei, and consists of a repulsive Coulomb term and a short-ranged attractive nuclear component. Obviously, it evolves with the time during the reaction process, because the shapes of reaction partners evolve with time due to the rearrangement of particles in the system. The nucleus-nucleus potential depends on the reaction energy and the reaction system (for example, the neutron-richness, the strength of shell effects etc), which is named as the dynamical nucleus-nucleus potential [254].

Generally, the nucleus-nucleus potential can be calculated based on the nucleon-nucleon interaction of reaction system. By using the microscopic transport model, ImQMD, one can calculate the dynamical nucleus-nucleus potential microscopically [220] in which the densities of the system and the relative distance $R$ between the two nuclei are functions of the evolution time.

When the projectile and target nucleus are well separated ($R \gg R_1 + R_2$) ($R_1$ and $R_2$ are the charge radii of the projectile and the target nucleus, respectively), the collective relative motion plays a dominant role and the excitation energy of the reaction partners could be negligible. The nucleus-nucleus potential is thus expressed as

$$V_1 = E_{\text{c.m.}} - T_R, \tag{66}$$

where $T_R$ is the kinetic energy of relative motion of two colliding nuclei, which can be obtained in the ImQMD simulations since the position and momentum of each nucleon can be recorded at every time step in the time evolutions. $E_{\text{c.m.}}$ is incident kinetic energy associated with the motion of center of mass of the system. It is related to the incident kinetic energy in laboratory $E$ as

$$E_{c.m.} = \frac{m_B}{m_A + m_B} E, \tag{67}$$

with A and B are the projectile and target, respectively.

After the di-nuclear system is formed ($R < R_1 + R_2$), the nucleus-nucleus potential is described by a way like the entrance channel potential [255]

$$V_2 = E_{\text{tot}}(R) - \bar{E}_1 - \bar{E}_2, \tag{68}$$

where $E_{\text{tot}}(R)$ is the total intrinsic energy of the composite system which is strongly dependent on the dynamical density distribution of the system. $\bar{E}_1$ and $\bar{E}_2$ are the time average of the energies of the projectile and target nuclei, respectively, which are obtained from the energies of the projectile-like and target-like nuclei in the region $R_T < R < R_T + 8$. $R_T = R_1 + R_2$ denotes the touching point. In the calculations of $E_{\text{tot}}(R)$, $\bar{E}_1$ and $\bar{E}_2$ in Eq. (68), the extended Thomas-Fermi (ETF) approximation for the intrinsic kinetic energy of the reaction system is adopted (see Refs. [220, 256] for details).

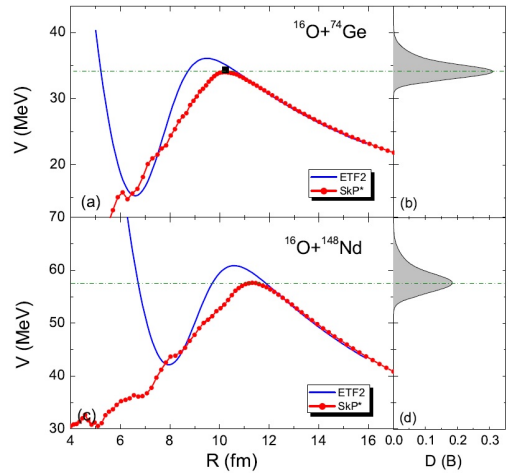

FIG. 3: (Color online) (a) and (c): Nucleus-nucleus potential for $^{16}$O+$^{74}$Ge and $^{16}$O+$^{148}$Nd. The circles and solid curves denote the results of the ImQMD simulations and the entrance-channel potential with the Skyrme-energy density functional plus the ETF2 approach, respectively. The squares denote the extracted most probable barrier height from the measured fusion excitation function. (b) and (d): empirical barrier distribution function proposed in Ref. [43]. Taken from Ref. [221].

The nucleus-nucleus potential is written as a smooth function between $V_1$ and $V_2$,

$$V_b(R) = \frac{1}{2}\text{erfc}(s)V_2 + [1 - \frac{1}{2}\text{erfc}(s)]V_1, \tag{69}$$

and

$$s = \frac{R - R_T + \delta}{\Delta R}, \tag{70}$$

with $\delta = 1$ fm, $\Delta R = 2$ fm. The obtained nucleus-nucleus potential $V_b(R)$ approaches to $V_1$ with the increase of $R$, and approaches to $V_2$ with the decrease of $R$.

Fig. 3 shows the calculated dynamical nucleus-nucleus potentials for fusion reactions $^{16}$O+$^{74}$Ge and $^{16}$O+$^{148}$Nd by using the ImQMD model with the parameter set SkP*. The blue curves denote the corresponding entrance-channel potential with the Skyrme energy-density functional plus the extended ThomasCFermi approximation including the terms up to second order (ETF2) approach in which the sudden approximation for the densities is used. The empirical barrier distribution functions for these two reactions are presented in Fig. 3 (b) and (d). The dashed lines give the positions of the most probable barrier heights $B_{\text{m.p.}}$. The black squares denote the extracted most probable barrier heights from the measured barrier distributions $D(E) = d^2(E\sigma_{\text{fus}})/dE^2$ based on the fusion excitation functions. For $^{16}$O+$^{148}$Nd, the statistics of the measured data for the fusion cross sections are not many enough to extract the most probable barrier height. It was found that the dynamical barrier height from the microscopic dynamics transport model depends on the incident energy in the fusion reactions [220]. Here, the incident energy $E_{c.m.} = 1.1 B_{m.p.}$ in the ImQMD simulations. The obtained dynamical barrier height $B_{dyn} \approx B_{m.p.}$ at this incident energy for the fusion events. The static potential barriers from the sudden approximation for the densities are evidently higher but relatively thinner than the dynamical ones. To reasonably reproduce the fusion excitation functions, the empirical barrier distributions [see the sub-figures (b) and (d) in Fig. 3] were proposed to take into account the nuclear structure effects and the multi-dimensional character of the realistic barrier in the ETF2 approach, and the value of the peak is lower than the corresponding barrier height from the entrance-channel potential which is based on the spherical symmetric Fermi functions for the densities of the two nuclei and the frozen-density approximation.

To understand the energy dependence of the potential barrier from the view point of the dynamical effects, the density distribution of the fusion events in $^{16}O+^{186}W$ at $E_{c.m.} = 66$ MeV was investigated in Refs. [220, 256], and the dynamical deformations of the reaction partners are evident.

In Ref. [254], it was the first time to explore the energy dependence of the potential barrier. It was found that the potential barrier around the Coulomb barrier increases with incident energy increasing, and its up-limits approaching the static one under the sudden approximation. The height of dynamic barrier decreases with decrease of the incident energy, and finally approaches low limit which is close to the adiabatic one.

### 2. Neck Dynamics

. The neck region is defined as connecting the projectile and target nuclei during the reaction, where the density is lowest along the centers of projectile and target nuclei, and it could be neutron-rich warmed, un-compressed region compared with the rest part of the reaction system. It is known that the neck dynamics play important roles in fusion-fission, light charged particle emission, and so on.

Fig. 4 shows the time evolution of the density at neck region in the fusion reaction $^{132}Sn+^{40}Ca$ at an incident energy of $E_{c.m.} = 115$ MeV. Here, the density distribution of 328 fusion events over a total of 1000 simulation events is studied for the head-on collisions with the parameter set IQ3a. At $t = 400$ fm/$c$, the reaction partners begin to touch each other, and the ratio of neutron-to-proton density $\rho_n/\rho_p$ reaches 2.7 which is higher than the $N/Z$ of the compound nucleus by a factor of two. With the increase of the density at neck, the value of $\rho_n/\rho_p$ decreases with some oscillations and gradually approaches the corresponding neutron-to-proton ratio of the compound nucleus ($N/Z = 1.37$). The extremely neutron-rich density at the neck region can significantly suppress the height of the Coulomb barrier for the fusion reactions induced by neutron-rich nuclei at energies around and be-low the barrier. Furthermore, one can see from Fig. 4 (e) that the surface diffuseness of the reaction partners at the neck side is obviously larger than that at the other side due to the transfer of nucleons.

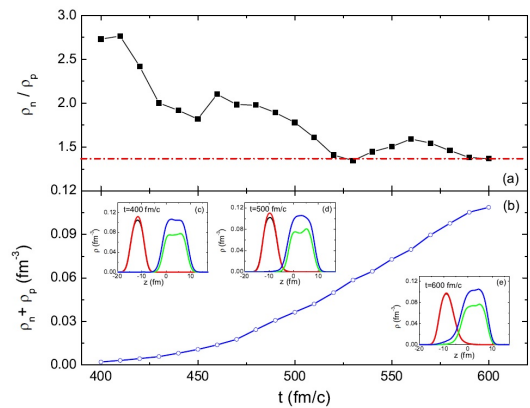

FIG. 4: (Color online) Time evolution of the density at neck region in fusion reaction $^{132}Sn+^{40}Ca$ at an incident energy of $E_{c.m.} = 115$ MeV. (a) Ratio of neutron-to-proton density at neck. The dash-dotted line denotes the corresponding ratio of the compound nuclei. (b) Density of the fusion system at neck. The sub-figures show the density distribution of the fusion system at $t = 400, 500$ and $600$ fm/$c$, with the red and the blue curves for the neutrons and the others for the protons. Taken from Ref. [218].

### 3. Fusion cross sections

. It is known that the fusion potential between two nuclei is closely related to the surface properties of the nuclei. The neutron skin thickness of neutron-rich nuclei in heavy ion fusion reactions should affect the fusion barrier and thus the fusion cross sections. To explore the influence of the symmetry potential on the fusion excitation function, the fusion cross sections of a number of fusion reactions are systematically investigated with the ImQMD model by adopting different parameter sets. Through creating certain bombarding events (hundreds to thousands) at each incident energy $E_{c.m.}$ and each impact parameter $b$, and counting the number of fusion events, we obtain the fusion probability (or capture probability for reactions leading to super-heavy nuclei) $g_{fus}(E_{c.m.}, b)$ of the reaction, by which the fusion (capture) cross section can be calculated:

$$\sigma_{fus}(E_{c.m.}) = 2\pi \int b \, g_{fus} \, db \simeq 2\pi \sum b \, g_{fus} \, \Delta b. \quad (71)$$

To consider the influence of the Coulomb excitation, the initial distance $d_0$ between the projectile and target should be much larger than the fusion radius. The collective boost to the sampled initial nuclei is given by $E_{kin} = E_{c.m.} - Z_1 Z_2 e^2/d_0$ at the initial time, with the center-of-mass energy $E_{c.m.}$, the charge number $Z_1$ and $Z_2$ for the projectile and target nuclei, respectively. Here, the initial distance between the reaction partners at z-direction (beam direction) is taken to be $d_0 = 30$ fm for the intermediate reaction systems and 40 fm for the ones with stronger Coulomb repulsion such as $^{132}$Sn+$^{40}$Ca. In the calculation of the fusion (capture) probability, event will be counted as a fusion (capture) event if the center-to-center distance between the two nuclei is smaller than the nuclear radius of the compound nuclei (which is much smaller than the fusion radius) and the time evolution will be terminated for those events that will not going to fusion in order to save the CPU time.

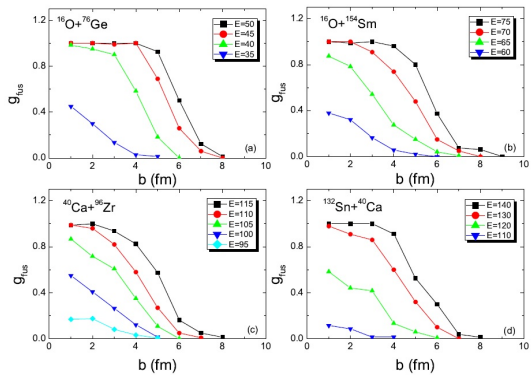

FIG. 5: (Color online) Fusion probability of the fusion reactions with IQ3a as a function of impact parameter $b$ for different beam energy (in MeV). Taken from Ref. [218].

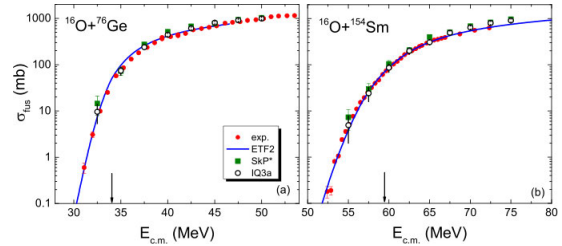

FIG. 6: (Color online) Fusion excitation functions of $^{16}$O+$^{76}$Ge and $^{16}$O+$^{154}$Sm. The solid circles denote the experimental data taken from [257] and [38], respectively. The blue curves denote the results with an empirical barrier distribution in which the fusion barrier is obtained by using the Skyrme energy-density functional together with the extended Thomas-Fermi (ETF2) approximation [43, 44]. The solid squares and open circles denote the results of ImQMD with the parameter set SkP* and IQ3a, respectively. The statistical errors in the ImQMD calculations are given by the error bars. The arrows denote the most probable barrier height based on the barrier distribution function adopted in the ETF2 approach. Taken from Ref. [218].

As an example, Fig. 5 shows the calculated fusion probability as a function of impact parameters. For the fusion reactions at energies above the Coulomb barrier, the fusion probability looks like a Fermi distribution, i.e. the fusion probability is about one for the central and mid-central collisions. At energies around the Coulomb barrier, the fusion probability decreases quickly with the impact parameter, which implies that the centrifugal potential due to the angular momentum affects the results significantly at this energy region.

A number of fusion reactions were studied with the ImQMD [220, 256]. As examples, the fusion excitation functions for $^{16}$O+$^{76}$Ge and $^{16}$O+$^{154}$Sm are shown in Fig. 6. The experimental data can be reproduced very well at energies around the barrier. We also note that in the present version of ImQMD model, the surface diffuseness of heavy nuclei is slightly over-predicted due to the approximate treatment of the Fermionic properties of nuclear system, which causes the over-predicted fusion cross sections at sub-barrier energies for the reactions with heavy target nuclei. In addition, for the fusion reactions with doubly-magic nuclei such as $^{208}$Pb, the fusion cross sections at sub-barrier energies are significantly over-predicted by the ImQMD calculations due to the neglecting of the shell effects. The strong shell effect of nuclei can inhibit the dynamical deformation and nucleon transfer, and therefore inhibit the lowering barrier effect. For some neutron-rich fusion systems such as

$^{40}$Ca+$^{96}$Zr and $^{132}$Sn+$^{40}$Ca, the results of the ImQMD model are relatively better, which could be due to that the neutron-rich effect is more evident than the shell effect in these reactions.

### B. Multi-nucleon transfer reactions

The synthesis of nuclei towards driplines and superheavy nuclei, has been of experimental [258–269] and theoretical interest [222, 223, 270–278]. Fusion reactions with stable beams as the traditional approach to successfully synthesize the superheavy elements (SHEs) with $Z = 107 - 118$ in the last forty years [279–290]. Up to now, the number of observed extremely neutron-rich nuclides is still very limited and about 4000 masses of neutron-rich nuclides in nuclear landscape are unmeasured. For the new nuclei in the 'northeast' area of the nuclear map, it is difficult to be reached in the fission reactions and thus fragmentation processes are widely used nowadays. Due to the 'curvature' of the stability line, it is also difficult for reaching these new more neutron-rich nuclei in fusion reactions with stable projectiles because of the lack of neutron number.

The revival interest of MNT reaction in intermediate systems or between actinide nuclei at low-energy collisions has arisen [55, 115, 291] for synthesis of the unknown nuclei and superheavy nuclei. As one of the efficient method to produce very neutron-rich nuclei, which are important for understanding nuclear structure at the extreme isospin limit of the nuclear landscape, the MNT reaction may overcome the limits that the number of observed neutron-rich nuclides is very limited at mass region $A > 160$, due to that neither traditional fusion reactions with stable beams nor fission of actinides eas-ily produce new neutron-rich heavy nuclei in this region. The MNT process should compete with fusion process as well as elastic and inelastic scattering for intermediate mass nuclear systems. With the increase of system charge, fusion process is gradually suppressed and completely forbidden for actinide nuclear system due to strong Coulomb repulsion. So, it is interesting to investigate how the MNT reaction mechanism evolves with the size of nuclear system. Here we study the MNT reactions for three reaction systems corresponding to different nuclear size region : 1) MNT in $^{86}$Kr+$^{64}$Ni at 25 MeV/nucleon; 2) $^{154}$Sm+$^{160}$Gd at $E_{c.m.}$=440 MeV (=5.6 MeV/nucleon) ; and 3) $^{238}$U+$^{238}$U at 7 MeV/nucleon.

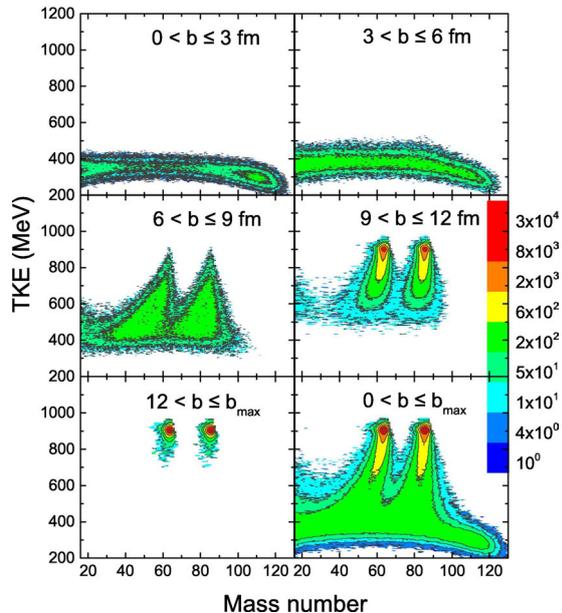

FIG. 7: (Color online) Mass-TKE distribution in the calculation of $^{86}$Kr+$^{64}$Ni at the incident energy of 25 MeV/nucleon. Taken from Ref. [219].

As a $N$-body and microscopic dynamical model, the ImQMD model has been used to study MNT reactions in heavy ion reactions because a large number of degrees of freedom, such as those in the excitation and deformation of projectile and target, neck formation, nucleon transfer, different types of separation of the composite system, and nucleon emission can be considered simultaneously.

Fig. 7 shows the mass vs total kinetic energy (mass-TKE) distribution in the ImQMD-v2.2 calculations for different impact parameters for the reaction $^{86}$Kr+$^{64}$Ni at 25 MeV/nucleon [219]. The incident energy is much higher than the Coulomb barrier, which is very suitable for studying the competition among fusion, quasi-elastic scattering, deep inelastic scattering and multifragmentation. As shown in Fig. 7, at central collisions, fusion and binary scattering (quasi-elastic and deep-inelastic scattering leading to multi-nucleon transfer) play equal important role. And with the increase of impact parameter, the fusion probability approaches to zero gradually, whereas the binary scattering events including the

MNT process becomes dominate. Other processes such as ternary breakup and multifragmentation process at this energy also take place. One can see that in central collisions, the neck of the di-nuclear system can be well formed and quickly broadened at such an incident energy and many nucleons are transferred between projectile and target, and the kinetic energy of the relative motion between two colliding nuclei is significantly dissipated to the excitation energy of the composite system. As a consequence, the masses of fragments are distributed in a much broader region, in which the products come from quite different reaction processes, such as fusion, binary breakup, ternary breakup and as well as multifragmentation. With the increase of impact parameter, the mass distribution becomes narrower due to decrease of nucleon transfer between reaction partners.

The production cross sections of isotopes are calculated by using the ImQMD model (ImQMDv2.2 version) together with a statistical code [292] for describing the secondary decay of fragments. The mass distributions of elements $Z = 30$ to $Z = 35$ are in good agreement with experimental data as shown in Fig. 8.

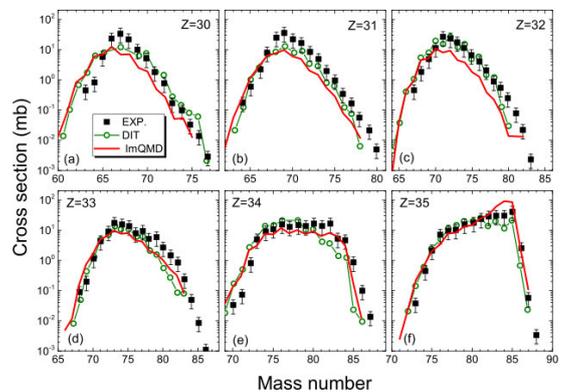

FIG. 8: (Color online) The isotopic distributions for elements from $Z = 30 - 35$ in $^{86}$Kr+$^{64}$Ni at 25 MeV/nucleon. Taken from Ref. [219].

The second reaction system used to produce very neutron-rich nuclei by MNT reaction is the neutron-rich reaction system $^{154}$Sm+$^{160}$Gd. It was investigated by Ning Wang $et~al.$ [293] to study the production of new neutron-rich lanthanides at $E_{c.m.}$=440MeV. The dynamical study by the ImQMD model (ImQMD-v2.2) shows that it is impossible to produce super-heavy nuclei in this reaction due to the disappearance of the capture pocket and the rapid increase of the potential with decreasing of the relative distance, it might produce new neutron-rich nuclide during the deep inelastic scattering process and moreover the fission barriers of lanthanides are relatively high to prevent fission of heavy fragments in the secondary decay process. Therefore this reaction is interesting for the synthesis of unmeasured lanthanides. By using the same approach used in the reaction $^{86}$Kr+$^{64}$Ni at 25 MeV/nucleon where the measured isotope distribution of products were reasonably well reproduced [219], the productions of unknown neutron-rich nuclei for reaction $^{154}$Sm+$^{160}$Gd at $E_{c.m.}$=440 MeV are calculated. More than 40 extremely neutron-rich unknown nuclei with $Z$ between 58 − 76 are observed and the production cross sections are at the order of $\mu b$ to $mb$ as shown in Fig. 9. The contour plots are the production probabilities of fragments in logarithmic scale for $^{154}$Sm+$^{160}$Gd at the energy of $E_{c.m.}$=440MeV. The curves denote the $\beta$-stability line described by Greens formula. The circles denote the positions of known masses in AME2012. The results show that multi-nucleon transfer in the neutron-rich reaction system $^{154}$Sm+$^{160}$Gd is an efficient way to produce new neutron-rich lanthanides. By analyzing the angular distribution of the produced heavy fragments, it is suggested that $20° < \theta_{lab} < 60°$ might be a suitable angular range to detect these extremely neutron-rich heavy nuclei.

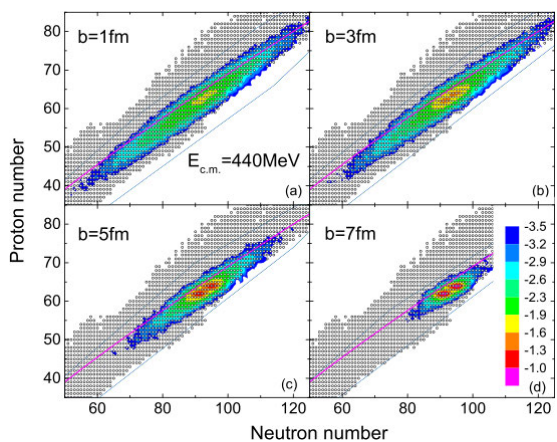

FIG. 9: (Color online) Contour plots are the production probabilities of fragments in logarithmic scale for $^{154}$Sm+$^{160}$Gd at the energy of $E_{c.m.}$=440 MeV. The curves denote the $\beta$-stability line described by Greens formula. The circles denote the positions of known masses in AME2012. Taken from Ref. [294].

The third reaction system we studied is the $^{238}$U+$^{238}$U. For elements with Z>100 synthesized by "cold fusion" with lead and bismuth targets and "hot fusions" with actinide targets, only neutron-deficient isotopes were produced compared with the centers of superheavy elements. At present, the pathway to reach beyond $Z = 118$ is not clear. It is well known that any further experimental extension of the region of SHEs to the center of the predicted first island of stability by means of complete fusion is limited by the neutron number of available projectiles and targets and by the very low production cross section as well. The strongly damped reactions between very heavy nuclei, such as U+U (between actinide nuclei) by MNT, could be one possible approach for those purposes [46, 52–54, 109–112, 270].

The theoretical description of the strongly damped reactions between very heavy nuclei, is one of the most difficult problems in nuclear physics due to the extremely large number of degrees of freedom involved.

This motivates the use of microscopic dynamical approaches. The ImQMD model incorporating the statistical decay model (HIVAP) for describing the decay of primary fragments was used to study the reaction $^{238}$U+$^{238}$U at 7 MeV/nucleon (see references Kai Zhao, et al. [223, 275, 276]). In order to calculate the isotope production cross sections of primary and residual fragments for reactions between two actinide nuclei, an empirical model for describing the mass distribution of fission fragments of fissile nuclei was introduced and incorporated with the statistical decay model (details see Ref. [223]).

Fig. 10 shows the comparison between the calculated results of the mass distribution of the products in reaction of $^{238}$U around 7 MeV/nucleon and the data [266]. The same scattering angle cut as in the experimental data is selected, that is, only fragments with scattering angles of $56° < \theta_{c.m.} < 84°$ and $96° < \theta_{c.m.} < 124°$ in the center-of-mass frame are selected [223, 266]. The open triangles in the figure are the calculation results and the experimental mass spectra from Ref. [266] are indicated by solid squares, open squares, solid circles, open circles, and solid triangles for incident energies of 6.09, 6.49, 6.91, 7.10, and 7.35 MeV/nucleon, respectively. From the figure, we find that the behavior of the calculated mass distribution at 7.0 MeV/nucleon is generally in agreement with the data at the incident energy 7.10 MeV/nucleon, except that the yields at the mass region from 170 to 210 are overestimated compared with the experimental data.

The most important features of mass distribution are considered to be the following: (1) A dominant peak around uranium is observed; this can be attributed to the contribution of the reactions with large impact parameters, (2) A steep decreasing yield above U with increasing mass number appears. The products at this mass region stem from large mass transfer in small-impact-parameter reactions. (3) A small shoulder can be seen in the distribution of the products around Pb, compared with the products with a mass near and smaller than uranium for which the yields decrease exponentially as mass decreases. The appearance of the small shoulder is due to the very high fission barrier around Pb. The central and semicentral collisions, and even reactions with b = 8–10 fm, contribute to the shoulder in the region around Pb. (4) In the region below $A \approx 190$, a double hump distribution is observed. This is clearly due to the fission of actinide and transuranic nuclei, which results in the superposition of symmetric and asymmetric fission.

Fig. 11 presents the isotopic production cross sections $\sigma(Z, A)$ for primary (open symbols) and residual fragments (solid symbols) with charge numbers from $Z = 94$ to 101 in the reaction of $^{238}$U + $^{238}$U at 7.0 MeV/nucleon. An angle cut of $32.5° \sim 44.5°$ in laboratory system as same as that in the experiment [259, 265] is taken. The experimental data and calculation results from Refs. [259, 265] denoted by black solid stars and blue lines, respectively, are also shown in the figure for comparison. One sees that the experimental data are generally reproduced, except the mendelevium isotopes ($Z = 101$), which was not detected in the experiment yet. Comparing the isotope distributions for primary and residual fragments, we can find the following three features, i.e., the widths of the isotope distributions for residual fragments are much smaller than those for primary fragments for the same element; the peaks in the isotope distributions for residual fragments shift to less neutron-rich side compared with those for primary fragments; the production cross sections for the most probable residual transuranium fragments ($Z = 96 − 101$) decrease almost exponentially with the increase of fragment charge number. From the isotopic cross section $\sigma(Z, A)$ for primary fragments and residual fragments, we can calculate the cross section $\sigma(Z) = \sum_A \sigma(Z, A)$, and the most probable mass number $A_Z$ for the isotope distribution through $\sigma(Z, A_Z) = \max[\sigma(Z, A)]$ for each element.

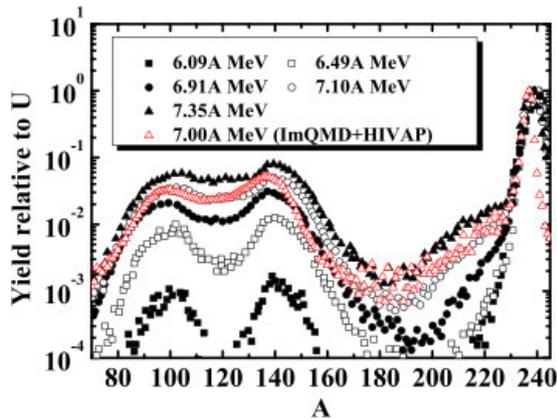

FIG. 10: (Color online) Mass distribution of the products of reaction $^{238}$U+$^{238}$U at different beam energies. In the context, we named the beam energies as MeV/nucleon. Taken from Ref. [223].

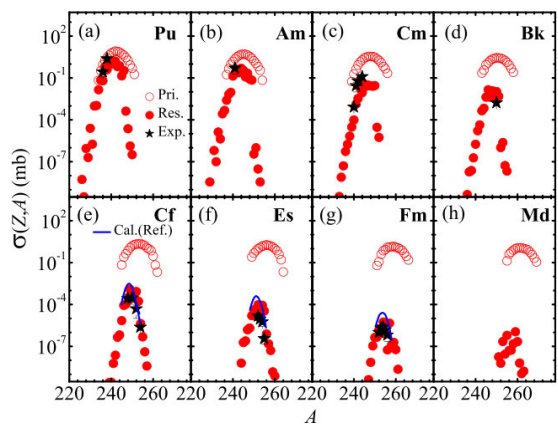

FIG. 11: (Color online) The isotopic distributions for elements from Pu to Md in $^{238}$U+$^{238}$U at 7 MeV/nucleon. Taken from Ref. [275].

Fig. 12 shows the cross sections for primary fragments with $Z \geq 70$ which are plotted by black contour lines for the reaction $^{238}$U+$^{238}$U at 7.0 MeV/nucleon. It shows that a large amount of primary fragments are produced via proton and neutron transfer between projectile and target. And the most probable isotopes of primary fragments are located near the line with the isospin asymmetry close to that of $^{238}$U (the isospin asymmetry is 0.227) on the nuclear map. The superheavy primary fragment (114, 184) (the isospin asymmetry is 0.235) at the center of the first 'island of stability' denoted by cross symbol in red color is not far from this line. The production cross sections for residual fragments are shown by colored rectangles. It can be found that the production cross sections for most of transactinide nuclei are smaller than $10^{-8}$ mb because it is difficult for those primary fragments to survive against fission due to very low fission barrier. For comparison, the area of known nuclei taken from Ref. [295] is presented by the magenta thick line in the figure, and one can find that quite a few unknown neutron-rich isotopes at the 'northeast' area of nuclear map can be produced through multinucleon transfer in this reaction.

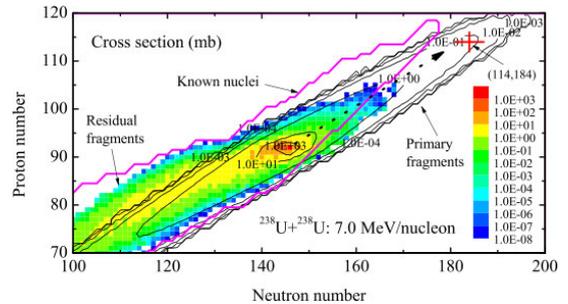

FIG. 12: (Color online) The landscape of the cross sections for primary fragments and residual fragments produced in $^{238}$U+$^{238}$U at 7 MeV/nucleon. Taken from Ref. [276].

For the predicted production of light uranium-like elements with $Z < 92$, we find that they can reach the border of the proton-rich side of known nuclei in the nuclear map. Because of the high fission barrier, the light uranium-like primary fragments can survive against fission more easily and de-excite through neutron evaporation leading to the production of proton-rich nuclei.

### C. Low-intermediate and intermediate-high energy heavy ion collisions

With the beam energy increasing, the reaction systems are compressed much more and are highly excited than low energy heavy ion reactions. It leads the system breaking up into much more fragments, three, four, five and more, i.e. from binary fission, ternary fission, to multifragmentation during its expansion phase. The phenomena appeared in this energy region are very rich. One of the characteristic phenomenon is the multifragmentation, which was considered as a signal of liquid-gas phase transition. The mechanism of mutifragmentation is still an interesting topic now. Another important phenomenon is the collective flow which reflects the collective motion of emitted nucleons or light charged particles due to the pressure gradient between the participant and the spectator, and thus the collective flow can be used to obtain the stiffness of EOS. Above the threshold energy for $\Delta$ production, the resonances appear in the high density region due to the nucleon-nucleon inelastic collisions, and then followed by the production of the mesons. The yields of different kinds of mesons and the ratios between different charged mesons are supposed to be sensitive to the EOS at suprasaturation density. In this chapter, we will make a review of above phenomena appeared in intermediate energy heavy ion collisions and the calculations based on ImQMD and UrQMD models.

### 1. Fragmentation mechansim

As beam energy increases, a ternary breakup was observed experimentally for reaction $^{197}$Au+$^{197}$Au at 15 MeV/nucleon [296]. The ternary breakup observed in this reaction has two characters: (a) the masses of three fragments are in comparable size [296] and (b) the three fragments are nearly aligned along a common reseparation axis [297, 298]. The ternary breakup reaction explored in the experiment is completely different from the commonly known formation of light charged particle accompanied binary fission. The features observed in the ternary breakup reaction between two $^{197}$Au nuclei indicate that the strong dissipation plays important role in the reaction process, and the deep study of ternary breakup can help us to understand the interplay between the one-body or two-body dissipation mechanism.

The ImQMD model is a suitable approach to study the reaction mechanism because in principle, the dissipation from nucleonic potential and nucleon-nucleon collisions, diffusion and correlation effects from classical $N$-body system are all included in the model without introducing any freely adjusting parameter, and thus, it was applied to study the microscopic mechanism of this new type ternary breakup reaction by Tian *et al.* [299]. In Ref. [299] the ternary breakup events were selected according to the condition given by experiment (see Ref. [296]), and the microscopic mechanism of the ternary break was studied by the event tracking with the ImQMD-v2 simulation. Fig. 13 shows the comparison between the calculation results and experimental data for the mass number distributions of the heaviest fragment $A_1$, the middle-mass fragment $A_2$, and the lightest fragment $A_3$ in the ternary breakup reactions of $^{197}$Au+$^{197}$Au at energy of 15 MeV/nucleon. The histograms denote the experimental data of Ref. [296] and the lines with open circles are the calculation results. Clearly, the calculations reproduce the experimental data rather well.

The event tracking in Ref. [299] explored that in the most of ternary breakup events, two $^{197}$Au nuclei had long contact time (about 1000 fm/$c$ or even longer ) and a transient composite system was formed before separating into two parts, i.e. the projectile-like fragment (PLF) and target-like fragment (TLF) parts with large nucleon transfer. The snapshots of the time evolution of the ternary breakup process are showed in Fig. 14, which presented the composite system elongating and forming a neck led to system re-separate and then the PLF (or TLF) further quickly separated into two fragments within a time of about 100 fm/$c$ or little short or little longer.

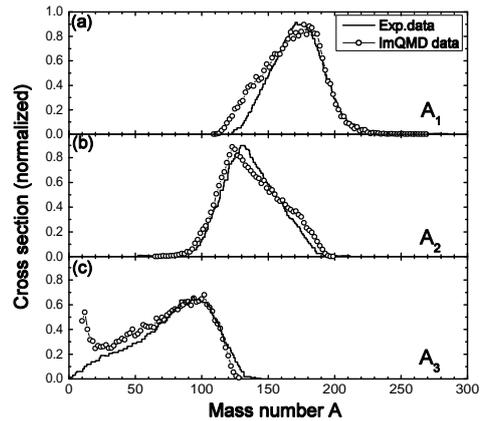

FIG. 13: Mass number distributions of (a) the heaviest $A_1$, (b) middle-mass $A_2$, and (c) the lightest $A_3$ fragments in selected ternary reactions of $^{197}$Au+$^{197}$Au at energy of 15 MeV/nucleon. The histograms denote the experimental data come from Ref. [296], the lines with open circles are the calculation results with the ImQMD model. Taken from Ref. [299].

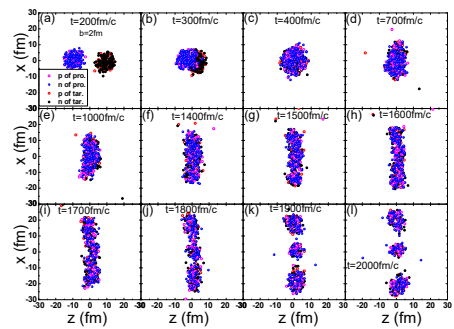

FIG. 14: (Color online) An example of direct ternary events for $^{197}$Au+$^{197}$Au at 15 MeV/nucleon with b= 2 fm. Taken from Ref. [299].

An analysis of the correlation between $A_1$, $A_2$, and $A_3$ was also performed in Ref. [299] in order to further study the mechanism of the ternary breakup process. The distribution of mass asymmetry of $A_3$ with respect to $A_1$ and $A_2$, i.e. $\eta_1 = (A_1 - A_3)/(A_1 + A_3)$ and $\eta_2 = (A_2 - A_3)/(A_2 + A_3)$, for different impact parameters are given in Fig. 15. It can be found that the probability distributions of $\eta_1$ for central and semi-central collisions ($0-6$ fm) are of the Wigner-like distributions while those of $\eta_2$ for all impact parameters are of the Poisson-type ones. From the point of view of probability theory, it implies that there exists a certain correlation between fragments $A_1$ and $A_3$ and no correlation between $A_2$ and $A_3$, which is obviously in consistence with the event tracking analysis discussed previously. From the view point of dynamics, the ternary breaking up in central collisions are in the sequence of $(A_1+A_3)+A_2$ and then $A_1+A_3+A_2$. It means the stronger correlations for $A_1$ and $A_3$ is kept during the evolution, while the correlations between $A_1$ and $A_3$ is weaken as the impact parameter increases.

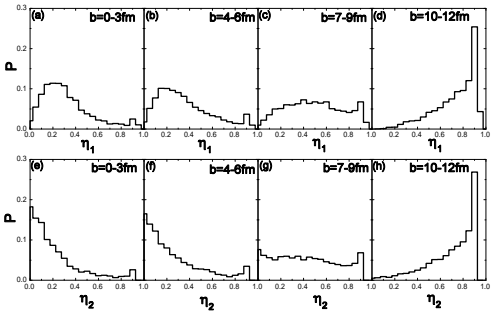

FIG. 15: Mass asymmetry $\eta_1$ (up) and $\eta_2$ (down) probability distribution for impact parameters from b=0 fm to 12 fm for $^{197}$Au+$^{197}$Au at energy 15 MeV/nucleon. Taken from Ref. [299].

More crucial information on the mechanism of the studied ternary processes comes from the features of the PLF breakup in its rest frame in the reaction plane. The basic information can be obtained from velocity vectors of the PLF fragments F1 and F2, and TLF in the laboratory reference frame (see Ref. [299]). In Fig. 16, we present the schematic view of the definition of the out-of-plane angle $\theta$, in-plane angle $\phi$, and the angle of the separation axis with respect to the beam direction $\theta_{cm}$. The reaction plane is defined by the beam direction and the direction of vector of $\vec{V}_{PLF} - \vec{V}_{TLF}$. This definition is as the same as in Refs. [298, 300].

Fig. 17 shows the results of the out-of-plane angle $\theta$ and the azimuthal angle $\phi$ distributions of fragments from PLF → F1 + F2 breakup, and as well as the angle $\theta_{cm}$ distribution obtained from the ImQMD model simulations. The experimental results from Ref. [298] are also shown as red symbols in the figure. Fig. 17 clearly tells us that the most of ternary breakup reaction events are in the reaction plane and three fragments are approximately aligned. The study of the ternary breakup reaction of $^{197}$Au+$^{197}$Au indicates that the fusion reaction of two very heavy nuclei at energies about 15 MeV/nucleon or higher is forbidden because of the strong Coulomb repulsion but a transient composite system may be formed due to the strong dissipation (strong dumped reaction). The formed transient composite system elongates then breaks up into two parts namely, the PLF and the TLF followed by a further breakup of PLF (or TLF) after a short time, leading to a ternary breakup reaction. Similar mechanism can be extended to quaternary reactions.

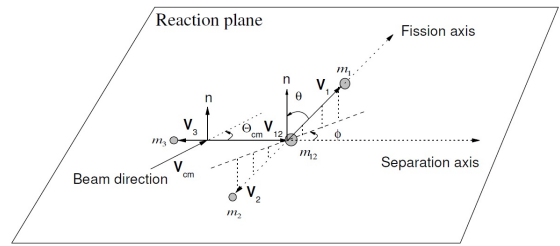

FIG. 16: Schematic view of the reaction in a cascade ternary event, in which the TLF and PLF are formed in the primary deep-inelastic process, and followed by a consecutive break-up from one of the two fragments. Taken from Ref. [300].

Furthermore, one could expect that accompanied light particle emission should have different rapidity distributions for different kinds of breakup mode. Based on this idea, we also analysis the rapidity distributions for light charged particles in binary, ternary and multi- break-up modes.

Fig. 18 presents the time evolution of the density contour plots for typical event of reaction $^{64}$Zn+$^{64}$Zn at $E_{beam}$=35 MeV/nucleon which could be the onset energy of multifragmentation [301]. The results are obtained by the ImQMD05 calculations with soft symmetry energy case, i.e. $\gamma_i$=0.5, where $\gamma_i$ is the symmetry potential energy coefficients in Eq. (42). As shown in Fig. 18, the system breaks via different break-up modes, such as binary, ternary and multifragmentation break-up modes, i.e. (f1), (f2) and (f3), with different probabilities due to the fluctuations which automatically appears in the QMD type models. It is clear that different break-up modes obviously lead to different emission patterns as well as the different angular and rapidity distributions of LCPs and fragments (for example, Fig. 18(f1)-(f3)).

To quantitatively understand the correlation between the break-up modes and LCPs emissions, we plot the rapidity distributions of the yields of LCPs corresponding to three kinds of break-up modes, binary (square symbols), ternary (circle symbols) and multifragmentation (triangle symbols) in Fig. 19. Fig. 19 presents the rapidity distribution of light charged particle $^3$He and $^6$He for $^{64}$Zn+$^{64}$Zn at mid-peripheral collisions. Panel (a) and (b) are for neutron-poor particle, i.e., Y($^3$He), (c) and (d) are for neutron-rich particles, i.e., Y($^6$He). Left panels are for symmetry potential coefficient $\gamma_i$= 2.0 and right panels are for $\gamma_i$=0.5. The rapidity distributions for the yields of $^{3,6}$He are normalized to per event. It is clear that the binary and ternary modes tend to produce more $^3$He and $^6$He at midrapidity than the multifragmentation multifragmentation mode, but the yield of $^6$He at $y_r \approx 0$ from binary mode is 70% larger than that from multifragmentation break-up mode. Consequently, one can expect that the different break up modes lead to different shape of rapidity distribution of LCPs.

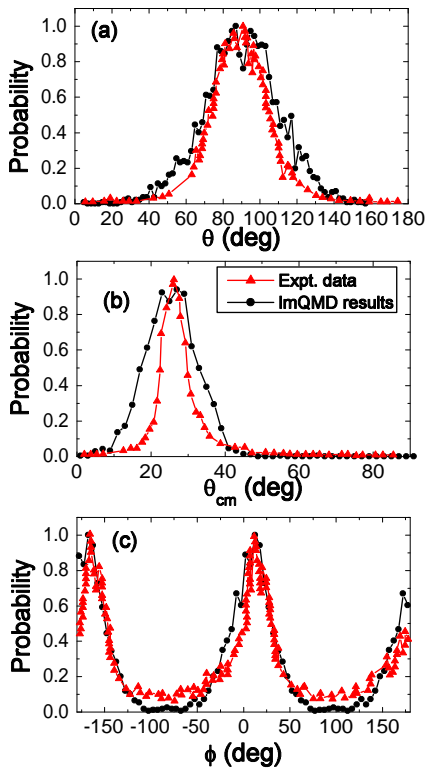

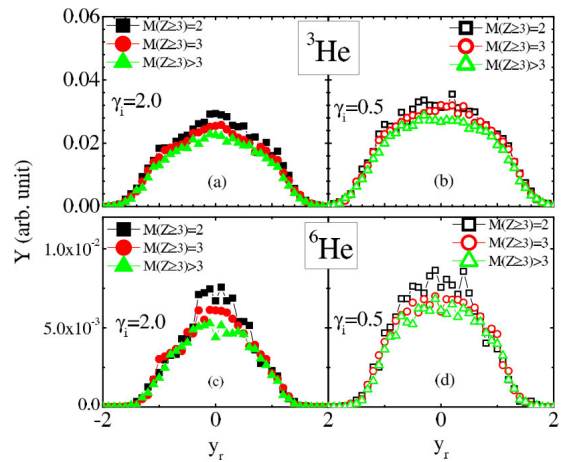

FIG. 19: (Color online)(a) and (b) are the reduced rapidity ($y_r$) distribution for the yield of $^3$He, i.e., Y($^3$He), with binary (square symbols), ternary (circle symbols) and multifragmentation (triangle symbols) break-up modes. (c) and (d) are for Y($^6$He). (a) and (c) are the results with $\gamma_i = 2.0$, (b) and (d) are for $\gamma_i = 0.5$. All of those results are for $^{70}$Zn $+^{70}$Zn at $E_{beam}$=35 MeV/nucleon for $b = 4$ fm. Taken from Ref. [301].

FIG. 17: (Color online) Angular distributions of fragments (a) out-of-plane angle $\theta$, (b) $\theta_{cm}$, and (c) azimuthal angle $\phi$ in cascade ternary reactions. The line with solid circles is the calculated results with the ImQMD model, and the line with triangles denotes experimental data [298]. Taken from Ref. [299].

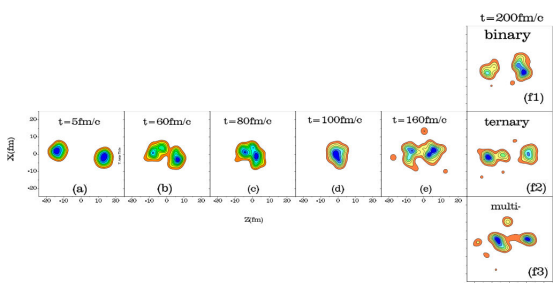

FIG. 18: (Color online) Time evolution of the density contour plots for $^{64}$Ni $+^{64}$Ni at $E_{beam}$=35 MeV/nucleon for $b = 4$ fm from typical events which are calculated with $\gamma_i = 0.5$. Taken from Ref. [301].

breakup mode does. The difference among the yields of neutron-rich light particle Y($^6$He) in the binary, ternary modes and multifragmentation break-up modes is larger compared with that for Y($^3$He) due to stronger isospin migration. For example, for $\gamma_i$=2.0, the yield of $^3$He at $y_r \approx 0$ from binary mode is 35% larger than that from

### 2. Fluctuation and chaoticity in liquid-gas phase transition

The theoretical studies on the properties of nuclear matter predict existing a liquid-gas phase transition at temperature of about 10-20 MeV and subnormal densities. A considerable progress has been made on the theoretical study as well as on the experimental side in order to define and collect a converging ensemble of signals connecting multifragmentation to the nuclear liquid-gas phase transition in last couple decades [3, 302–308]. The fluctuation in spinodal region was thought as one of the origins of liquid-gas phase transition in finite nuclear system. An anomalous increase of fluctuation at a phase transition and a rapid increase of chaoticity at the microscopic level stimulate further study of the dynamics of liquid-gas phase transition in nuclear systems.

A way to characterize the dynamics in the phase transition is to calculate the largest Lyapunov exponent (LLE). The largest Lyapunov exponent is defined as [309–311]

$$\lambda = \lim_{n \to \infty} \frac{1}{n\tau} \ln \frac{||d\vec{X}_n||}{||d\vec{X}_0||}. \quad (72)$$

The quantity $||d\vec{X}_n||$ is the phase space distance between two trajectories corresponding to two concerned events at time $t = n\tau$, more details of definition can be found in Ref. [311]. The LLE is a measurement of the sensitivity of the behavior of a system to initial condition and also gives an idea of the velocity at which the system explores the available phase space, and a positive Lyapunov exponent may be taken as the defining signature of chaos. For the case of nuclear fragmentation, a nucleus at a highly excited state eventually breaks into several fragments with nucleons and light particles, where a given trajectory for the system in the phase space will never come back close to the initial state of the system. So a local-in-time LLE over an ensemble of trajectories whose initial conditions are consistent with the nuclei at a given excitation energy should be used. Thus, the time scales of the inverse LLE compared with density fluctuation become more relevant. If the time scale for density fluctuation is much longer than the inverse largest Lyapunov exponent it indicates that the dynamics during fragmentation of the nuclear system is chaotic enough. In this way the LLE calculated over an ensemble of trajectories can carry the full information of dynamics of the systems in multifragmentation.

Fig. 20 shows the correlations between the LLE, i.e. $\lambda$, and the density fluctuation $\sigma_\rho^2$ at different temperatures from 3 to 19 MeV for systems of $^{124}$Sn of [Fig. 20 (a)] and $^{208}$Pb of [Fig. 20 (b)]. The density fluctuation $\sigma_\rho^2$ is,

$$\sigma_\rho^2 = \frac{<\rho^2(t)> - <\rho(t)>^2}{<\rho(t)>^2}, \qquad (73)$$

with

$$<A> = \int A\rho(\mathbf{r},t)d\mathbf{r}, \qquad (74)$$

and the integration is over whole space. One can see that the maximum values of both the LLE and density fluctuation are located at the same temperature, i.e., the critical temperature. There are two branches in $\lambda \sim \sigma_\rho^2$, one corresponding to the temperature lower than the critical temperature and another corresponding to the temperature higher than the critical temperature. For the low temperature branch, both the $\lambda$ and $\sigma_\rho^2$ increase as the temperature increases, whereas for the high temperature branch they increase as the temperature decreases. Both branches show that the $\lambda$ increases roughly linearly with $\sigma_\rho^2$. This correspondence between the $\lambda$ and $\sigma_\rho^2$ is qualitatively consistent with the discussion based on Refs. [312, 313].

### 3. Collective flow

Collective flow is a motion characterized by space-momentum correlations of dynamic origin, and it carries the information of the pressures generating that motion, the EOS, the in-medium NN cross sections and other properties of the strongly-interacting matter.

As shown in Fig. 21, the flows that have been identified so far are radial, directed, elliptic, triangle flow and high order flow, described by the corresponding coefficients

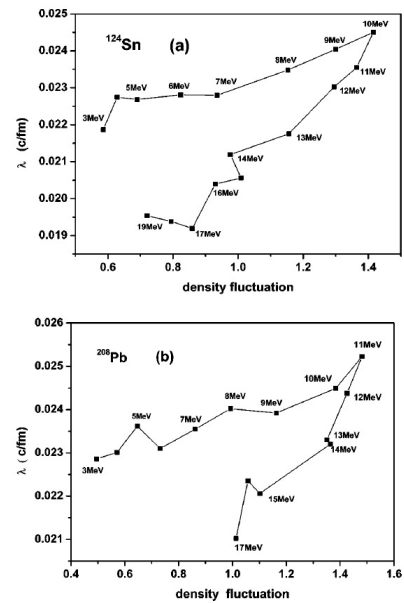

FIG. 20: The relation between the LLE and the density fluctuation at temperatures from 3 MeV to 19 MeV for systems of $^{124}$Sn (a) and $^{208}$Pb (b). Taken from Ref. [311].

of the components of the Fourier decomposition of the azimuth angle distribution of emitted particles

$$\frac{dN}{d\phi} = p_0(1 + 2v_1\cos\phi + 2v_2\cos 2\phi + 2v_3\cos 3\phi + ...), \quad (75)$$

$v_1$ is the directed flow, and $v_2$ the elliptic flow, $v_3$ the triangle flow.

The ImQMD05 model were applied to study the directed and elliptic flows and stopping power in $^{197}$Au+$^{197}$Au reaction at energies lower than 0.4 GeV/nucleon by using SkP, SkM*, SLy7 and SIII Skyrme interaction parameter sets and compared with the experimental data of INDRA, FOPI and Plastic Wall taken from Ref. [314] and it was found that the SkP and SkM* best fit to the data when the energy dependent in-medium NN elastic cross sections were used [154]. As an example, Fig. 22 shows the excitation function of elliptic flow parameters at midrapidity ($|y/y_{beam}^{cm}| < 0.1$) for $Z \leq 2$ particles for $^{197}$Au+$^{197}$Au collisions at $b = 5$ fm [the reduced impact parameter $b/b_{max}$ equals 0.38 and $b_{max} = 1.15(A_P^{1/3} + A_T^{1/3})$]. The general behavior of the excitation functions of elliptic flow parameters $v_2$ calculated with different Skyrme interactions is similar, i.e., the elliptic flow evolves from a preferential in-plane (rotational like) emission ($v_2 > 0$) to out-of-plane (squeeze out) emission ($v_2 < 0$) with an increase of energies. Clearly, one can see that the harder EOS provides stronger pressure which leads to a stronger out-of-plane emission and thus to a smaller transition energy. The transition energies calculated with SkP and SkM* agree with experimental data [314], while those with SIII and SLy7 are too small compared with experimental data.

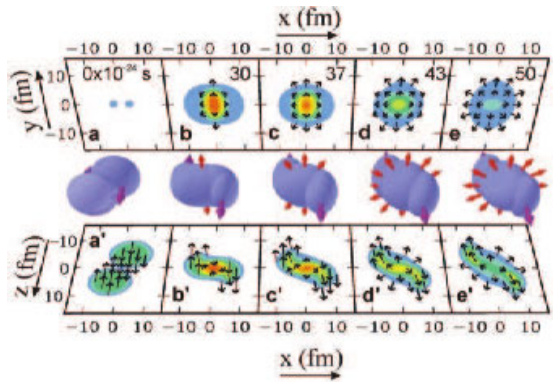

FIG. 21: (Color online) Overview of the dynamics for a Au+Au collision. Time increases from left to right, the center of mass is at r=0, and the orientation of the axes is the same throughout the figure. The trajectories of projectile and target nuclei are displaced relative to a head-on collision by an impact parameter of b=6fm. The three-dimensional surfaces (middle panel) correspond to contours of a constant density $\rho \sim 0.1\rho_0$. The magenta arrows indicate the initial velocities of the projectile and target (left panel) and the velocities of projectile and target remnants following trajectories that avoid the collision (other panels). The bottom panels show contours of constant density in the reaction plane (the x-z plane). The outer edge corresponds to a density of $0.1\rho_0$, and the color changes indicate steps in density of $0.5\rho_0$. The back panels show contours of constant transverse pressure in the x-y plane. The outer edge indicates the edge of the matter distribution, where the pressure is essentially zero, and the color changes indicate steps in pressure of 15 MeV/fm$^3$. The black arrows in both the bottom and the back panels indicate the average velocities of nucleons at selected points in the x-z plane and x-y planes, respectively. Taken from Ref. [24].

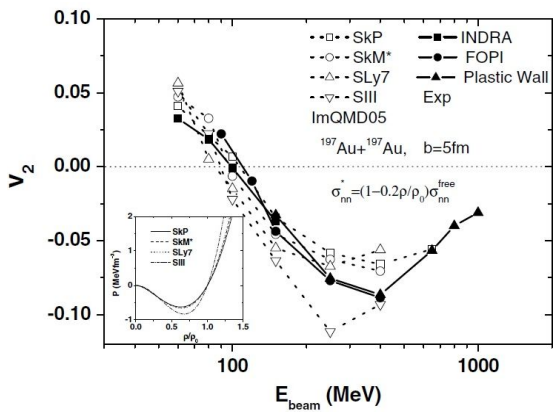

FIG. 22: Excitation functions of elliptic flow parameters at midrapidity for $Z \leq 2$ particles from midcentral collisions of $^{197}$Au+$^{197}$Au calculated with SkP, SkM*, SLy7, and SIII Skyrme interactions. The calculated results are given in the same reference frame as that used for the experimental data, which are taken from [314]. Inset shows pressure as a function of density calculated with the four Skyrme interactions. Taken from Ref. [154].

For the flow effect at further higher energy heavy ion collisions, it is more suitable to use the UrQMD model. The direct and elliptic flow for protons and light charged particles $d$, $t$, and $^4$He at a wide range of beam energies were studied and compared with experimental data [315–318]. In Fig. 23, the excitation function of elliptic flow of protons for reaction $^{197}$Au+$^{197}$Au at energies from SIS to RHIC is presented. The UrQMD model in its cascade (UrQMD2.2) and a mean field mode with the HM-EOS and with the DBHF-like medium modification on nucleon-nuleon elastic cross sections are employed. The rapidity cut of $|y| <0.1$ has been used because this is the appropriate one to compare with the data at lower energies. Due to this cut, the calculation of the cascade mode (without nuclear potential) reaches negative values at low energies. With inclusion of nuclear potential, the model is in line with experimental data at both SIS and AGS energies. In recent years, for a better description of experimental data at SIS energies, the surface and the surface asymmetry energy terms from the Skyrme potential energy density functional are further incorporated into the UrQMD model [252, 319, 320]. It is found that, with a proper parameter set on the in-medium NN cross section, the published collective flow and nuclear stopping data in HICs at intermediate energies can be reproduced well [170, 249, 250, 252, 253, 319–322].

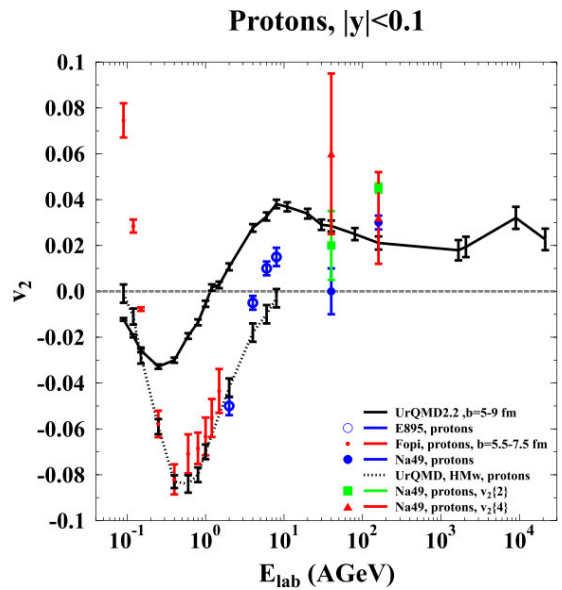

FIG. 23: (Color online) The calculated energy excitation function of elliptic flow of protons in Au+Au/Pb+Pb collisions in mid-central collisions ($b = 5 - 9$ fm) with $|y| < 0.1$ (full line). This curve is compared to data from different experiments for mid-central collisions [315–318]. The dotted line in the low energy regime depicts UrQMD calculations with nuclear potential included. Taken from Ref. [84].

One can find that the elliptic flow calculated with ImQMD05 and UrQMD are coincidence within the common energy region. For experiments, the important issue is how to determine the reaction plane accurately, since the detectors can not cover full solid angle and can not detect all kinds of particles with the same efficiency. It blocks the experimentalists to get the flow accurately, especially for low-intermediate energy heavy ion collisions. Thus, to further improve the accuracy of the experimental measurement of flow is desirable.

### D. Spallation reactions

The neutron/proton induced spallation reactions with the beam energy up to 1 GeV have wide applications in material science [323], biology [324], surgical therapy [325], space engineering [326] and cosmography [327]. Interest in the spallation reactions has recently been renewed because of the importance of intense neutron sources for various applications, such as spallation neutron sources for condensed matter and material science [328–330], accelerator-driven subcritical reactors for nuclear waste transmutation[331, 332] or energy production, as well as for medical therapy. So there is a growing need of nuclear data for spallation reactions at intermediate energies up to 1 GeV for targets not only the neutron production materials such as Pb, Bi, W, but also for surrounding structural materials such as Al, Fe, Ni, Zr and biologic elements such as C, O, Ca.

Experimental data are important for designing spallation sources. However, it is impossible to make measurements for all data that are of importance for the various applications [333]. When experimental data are not available, theoretical model calculations have to be employed to estimate the related data. The moving source (MS) model [334], High Energy Transport Code-Three step model (HETC-3STEP) [335], intra-nuclear cascade evaporation (INC/E) model [336] and quantum molecular dynamics (QMD) model [148, 337–340] have often been utilized in the reactions at energies higher than hundreds MeV.

It is well known that spallation reaction is usually described by three-step processes, i.e. the dynamical non-equilibrium reaction process leading to the emission of fast particles and an excited residue, followed by pre-equilibrium emission, and by the decay of the residue. The first process can be described by microscopic transport theory models, the pre-equilibrium is usually optional in different approaches, and the last one can be described by a statistical decay model. Over the past decade, by applying the ImQMD Model (05 version) merged statistical decay model, a series of studies on the proton-induced spallation reactions at intermediate energies have been made [341–345]. Nuclear data including neutron double differential cross sections (DDXs), proton DDXs, DDXs of light complex particles, i.e. $d$, $t$, $^3$He and $^4$He, mass, charge and isotope distributions can be overall reproduced quite well.

As an example, Fig. 24 shows the results of the neutron and proton DDXs for proton-induced spallation reaction for p+$^{208}$Pb at 256 MeV, 800 MeV and 1 GeV by us-ing the ImQMD model plus statistical decay model(s). In Fig. 25, the mass and charge distributions of productions in the reaction of 500 MeV p+$^{208}$Pb [334, 335] and 800 MeV p+$^{197}$Au [333, 336], respectively, is shown. In Fig. 26, the isotope distributions of residues produced in the reaction of 750 MeV proton on $^{56}$Fe is shown. And Fig. 27 shows comparisons between calculated DDXs and experimental data for light charged particles produced in the reaction of 200 and 392 MeV proton on $^{27}$Al.

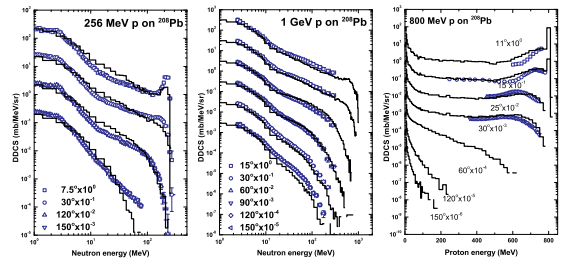

FIG. 24: (Color online) Comparison of ImQMD05+GEM2 calculation results (lines) with experimental data (open symbols) for double differential cross sections of emitted neutrons in 256 [346] and 1000 MeV [347] proton on $^{208}$Pb and emitted protons in 800 MeV proton on $^{208}$Pb [348] respectively. Taken from Ref. [341].

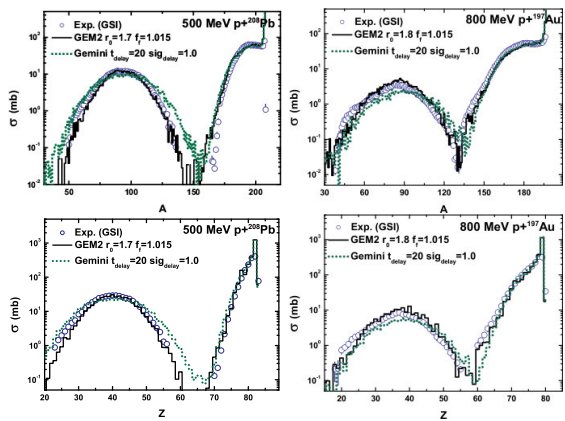

FIG. 25: (Color online) Comparison between the ImQMD05+GEM2 model and ImQMD05+GEMINI model calculation results and experimental data for mass and charge distributions cross sections of products in 500 MeV p+$^{208}$Pb [334, 335] and 800 MeV p+$^{197}$Au [333, 336], respectively. Taken from Ref. [341].

Except the systems shown above, more proton-induced spallation reactions have been analyzed with the ImQMD05 model merging the GEM2 and GEMINI models. The cross sections for products in proton-induced reactions on heavy targets can be reproduced quite well by both models. And it is found that the DDXs of proton and neutron are not sensitive to the parameters of ImQMD and statistical model.

However, to best reproduce the experimental data of mass and charge distributions of productions, two param-eters in each model, i.e., the nucleus radius parameter $r_0$ and the level density parameter modification factor $f_f$ (i.e., $a_f^* = f_f a_f$, $a_f$ is the level density parameter) in GEM2; the delay time for fission $t_{delay}$ and $sig_{delay}$ in GEMINI, need to be tinily readjusted according to the incident energy. For the future, with more precise experimental data becoming available, we expect that the systematic adjustable parameters in GEM2 and GEMINI can be obtained with the present approach. However, there is still some works to be done in order to achieve a universal description for spallation reactions with arbitrary targets and arbitrary incident energy.

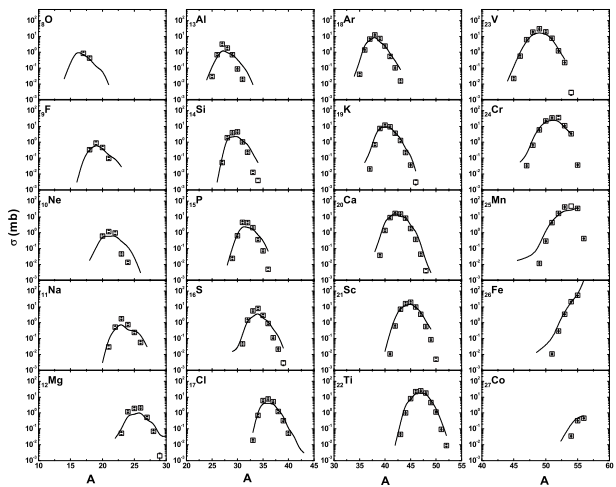

FIG. 26: Comparison of all measured cross sections of products (open squares) from the reaction 750 MeV proton on $^{56}$Fe [349] with ImQMD05+GEMINI calculation results (lines). Taken from Ref. [341].

Another important point is the description on the LCPs produced in the spallation reaction should be refined. The yields of LCPs with high kinetic energy are underestimated in the previous ImQMD model, as the dashed curves shown in Fig. 27, because the description of the LCPs emission in pre-equilibrium process is absent. In order to overcome the limitation of the ImQMD05 model in description of LCPs emission, a phenomenological surface coalescence mechanism is introduced into the ImQMD model. The basic idea of this mechanism is: the leading nucleon ready to leave from compound nuclei can coalesce with other nucleon(s) to form a LCP, and the LCP with enough kinetic energy to overcome Coulomb barrier can be emitted. By systematic comparison between calculation results and experimental data of nucleon-induced reactions, the parameters in the surface coalescence model are fixed. Then with the fixed parameters, chosen once for all, the prediction power of the model is tested by the nucleon-induced reactions on various targets with energies from 62 to 1200 MeV. And it is found that, with surface coalescence mechanism introduced into ImQMD model, the description on the DDXs of LCPs is greatly improved. And Fig. 27 shows comparisons between calculated DDXs and experimental data

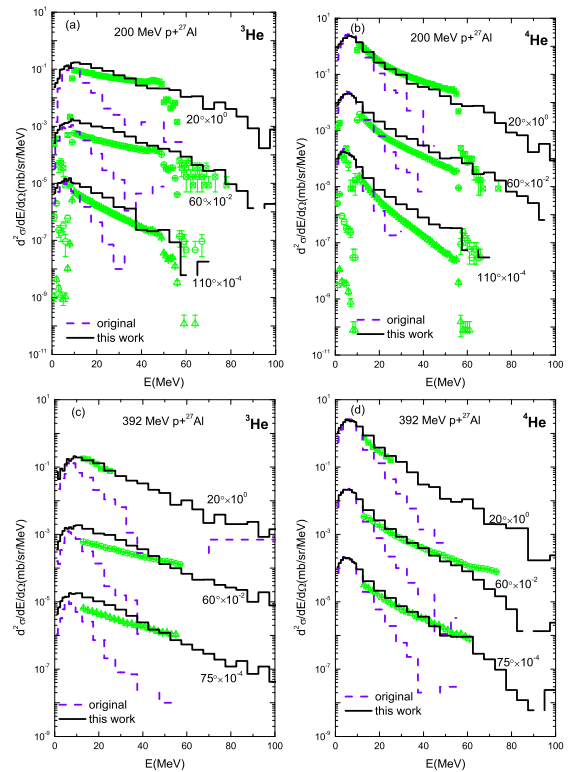

FIG. 27: (Color online) Calculated DDXs of light charged particles produced in the reaction p+$^{27}$Al at 200, and 392 MeV. Experimental data (triangles) are taken from Ref. [350] for 200 MeV p+$^{27}$Al, Ref. [351] for 392 MeV p+$^{27}$Al, respectively. Taken from Ref. [345].

for light charged particles produced in the reaction of 200 and 392 MeV proton on $^{27}$Al.

As we mentioned before, the advantage of the ImQMD model is that the model parameters are directly obtained from the Skyrme interactions. Thus, one can study the impact of the different effective Skyrme interaction on proton-induced spallation reactions. Based on the best Skyrme interaction we obtained, one can also obtain the information of EOS in the mean field level.

## IV. STUDY OF IN-MEDIUM NN CROSS SECTIONS

As we discussed in previous chapter, collision term $I_{coll}$ is another key ingredient of transport equation, and the in-medium effects on the collision cross sections influences the results of the simulation of HICs. Thus, the study of in-medium cross sections and extracting the medium corrections to the cross sections from experimental observables are important, which are the main contents of this chapter.

## A. In-medium NN cross sections from microscopic approach

The in-medium NN cross sections can be calculated by using Bruckner G-matrix method [352–357], one-boson-exchange model (OBEM) [358–360] or the closed-time Green's function approach [361–364] in theory. By using the closed-time path Green's function method, the in-medium elastic and inelastic nucleon-nucleon cross sections can be derived within the framework of self-consistent relativistic Boltzmann-Uehling-Uhlenbeck (RBUU) equations. In the early works [361–363], the in-medium elastic and in-elastic cross sections without considering the isospin dependence were computed based on quantum hadro-dynamical (QHD-1) model. Up to now, by applying the effective Lagrangian such as the QHD-II and its extension in which the couplings to vector-isovector $\rho^\mu$ and scalor-isovector $\delta(a_0(980))$ mesons are considered, the isospin asymmetry $\alpha$, density $\rho$, and center-of-mass energy $\sqrt{s}$ dependence of the in-medium cross sections $\sigma^*$, such as $\sigma^*_{NN\to NN}(\sqrt{s}, \rho, \alpha)$ and $\sigma^*_{NN\to N\Delta}(\sqrt{s}, \rho, \alpha)$ are computed [365].

In Fig. 28 we present both the cross sections and the suppression parameters as a function of c.m. energy $\sqrt{s}$ at different densities, which are obtained with closed-time path Green's function as in Ref. [363]. $\sigma^{*(2)}_{np}$ and $\sigma^{*(2)}_{nn,pp}$ are in panels (a) and (b), and the suppression parameters, $\eta^{(2)}_{np} = \sigma^{*(2)}_{np}/\sigma^{free(2)}_{np}$ and $\eta^{(2)}_{nn(pp)} = \sigma^{*(2)}_{nn(pp)}/\sigma^{free(2)}_{nn(pp)}$ are in panels (c) and (d). The superscript '(2)' means that the cross sections are obtained with closed-time path Green's function method as described in Ref. [155]. One can see that $\sigma^{*(2)}_{np}$ changes little with density and is nearly the same as in free space. On the other hand, the cross section $\sigma^{*(2)}_{nn,pp}$ tends to be suppressed at lower energies, $\sqrt{s} \leq 2.05$ GeV and enhanced at higher energies. Differences in the features of the two cross sections are associated with the differences between the isospin T = 0 and the T = 1 channels and, in particular, presence of a low-energy resonance in the T = 1 channel and effects of $\rho^\mu$ exchange.

Fig. 29 shows the in-medium cross section $\sigma^*_{NN\to N\Delta}$ (the initial isospin averaged one) as a function of $\sqrt{s}$ for different reduced densities and the calculation was performed with the $\Delta$ mass taken to be the pole mass [365]. It is found that at high energies, the $\sigma^*_{NN\to N\Delta}$ monotonously decreases with increasing density. This is similar to previous calculations [362] performed with another parameter set of the equation of state. However, when approaching the threshold energy, the density dependence is somewhat different from that in Ref. [362] due to the neglect of the $\Delta$ mass distribution. This should play a more important role at the lower energies. One might argue that the mass distribution ought to be influenced by the nuclear medium as well, a topic that deserves further investigation.

The results become more interesting for the isospin asymmetries $\alpha = (\rho_n - \rho_p)/(\rho_n + \rho_p) \neq 0$ to be discussed in the following. Owing to the difference in isospin matrices shown in Ref. [365], the cross-section values for the production of $\Delta^{++}$ and $\Delta^-$ are exactly three times larger than those of other channels ($\Delta^+$ and $\Delta^0$) as long as isospin asymmetry $\alpha = 0$. With $\alpha \neq 0$, this relation is destroyed because of the different effective masses for $\Delta^{++}$, $\Delta^+$, $\Delta^0$, $\Delta^-$ and neutron, proton, which make the scattering amplify different as well. It is found that the influence of the mass splitting on $\sigma^*_{pp\to n\Delta^{++}}$ and $\sigma^*_{nn\to p\Delta^-}$ are much stronger than that on the other four channels. Fig. 30 shows the effect of mass splitting in the isospin asymmetric medium on the in-medium cross sections of $\Delta$ production. The ratios $R(\alpha) = \sigma^*(\alpha)/\sigma^*(\alpha = 0)$ of all channels are shown as an example in Fig. 30 as a function of the isospin asymmetry at a typical kinetic energy $E_K = 1$ GeV (correspondingly, $s^{1/2} = 2.326$ GeV) where the maximum of the cross section approaches. The ratio $R(\alpha)$ deviates almost linearly from unity when the value of the isospin asymmetry increases from 0 to 0.5. It occurs with the sequence: R($\alpha$, pp→n$\Delta^{++}$)> R($\alpha$, pp→p$\Delta^+$) > R($\alpha$, pn→n$\Delta^+$) > R($\alpha$, pn→p$\Delta^0$) > R($\alpha$, nn→n$\Delta^0$) > R($\alpha$, nn→p$\Delta^-$). It is further seen that, at $\alpha$ = 0.5, the R($\alpha$) ratio remains within the interval between 0.88 and 1.15 for the $\Delta^+$ and $\Delta^0$ production channels, while it changes more rapidly to 1.74 and 0.73 for $\Delta^{++}$ and $\Delta^-$, respectively.

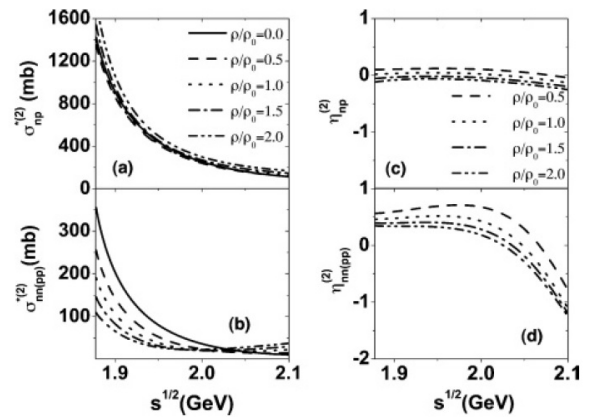

FIG. 28: Energy dependence of in-medium cross sections $\sigma^{*(2)}_{np}$ and $\sigma^{*(2)}_{nn,pp}$ in panels (a) and (b), and of the suppression parameters $\eta^{(2)}_{np}$ and $\eta^{(2)}_{nn,pp}$ in panels (c) and (d), at selected densities. The cross sections have been obtained within the CTPGF approach with QHD-II effective Lagrangian. Taken from Ref. [155].

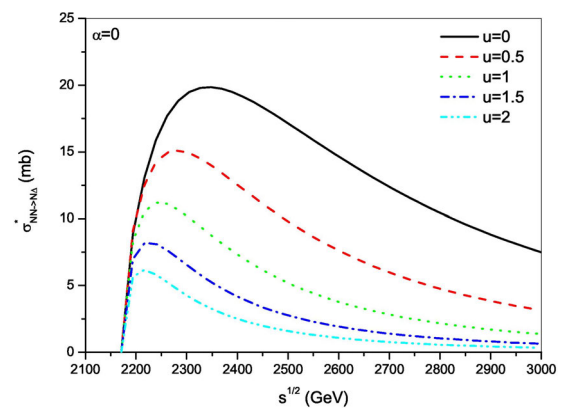

FIG. 29: (Color online) The in-medium cross section $\sigma^*_{NN\to N\Delta}$ as a function of $s^{1/2}$ at several reduced densities. Taken from Ref. [365].

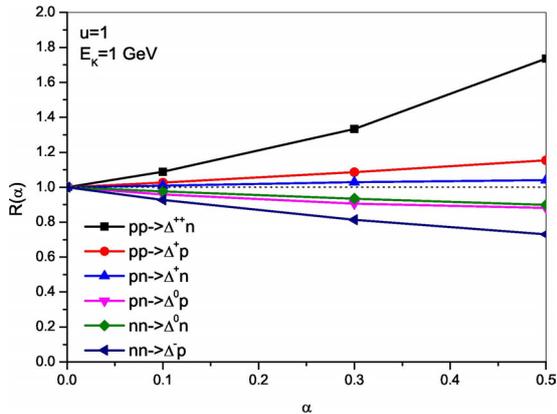

FIG. 30: (Color online) The $R(\alpha) = \sigma^*(\alpha)/\sigma^*(\alpha = 0)$ ratios of all channels (lines with different symbols) as a function of the isospin asymmetry for $u = 1$ and $E_K = 1$ GeV. The horizontal dotted line represents unity. Taken from Ref. [365].

Cui et al. also studied the influence of threshold effects on the in-medium $NN \rightarrow N\Delta$ cross sections [366] by using the one-boson-exchange model (OBEM), where the $NN \rightarrow N\Delta$ cross section was averaged over the mass distribution of $\Delta$ considering the $\Delta$ as short-living resonance. They found that the isospin splitting of medium correction factor $R = \sigma^*_{NN \rightarrow N\Delta}/\sigma^{free}_{NN \rightarrow N\Delta}$ are weaken when one consider the threshold energy effects, i.e., the changes of self-energy when a nucleon turns into a $\Delta$ in process of $NN \rightarrow N\Delta$. In Fig. 31, the $R$ as a function of energy above the threshold, i.e., $Q = \sqrt{s} - \sqrt{s}_{th} = \sqrt{s} - (2m_N + m_\pi)$, for different beam energies are presented. By considering the threshold effects in the calculation of $NN \rightarrow N\Delta$ cross sections, the isospin splitting of $R$ becomes weaker at the beam energy above 0.8 GeV because the changes of scalar and vector self-energies in the process of $NN \rightarrow N\Delta$ become smaller relative to the kinetic energy part.

Constraining those cross sections is essential for reducing the uncertainties of EOS constraints by using the transport models [155, 158, 367]. Also the in-medium cross sections are of interest for their own sake, as they underly the viscosity and other nuclear transport coefficients [368]. One of the powerful HICs observables for extracting the in-medium NN cross sections are the stopping power [369, 370] because it is constructed from the changing of momentum of the emitted particles and reflects the cross sections in the most direct manner. Fur-thermore, their correlation [369] with flow observables which are widely used for determining the stiffness of EOS, is also of interest. We employed the ImQMD and UrQMD models to investigate the effects of in-medium NN cross sections on HICs observables. The cross checking on the results with two models is helpful for us to deeply understand the in-medium NN cross sections and the reliability of transport models, and it will be discussed in the next part.

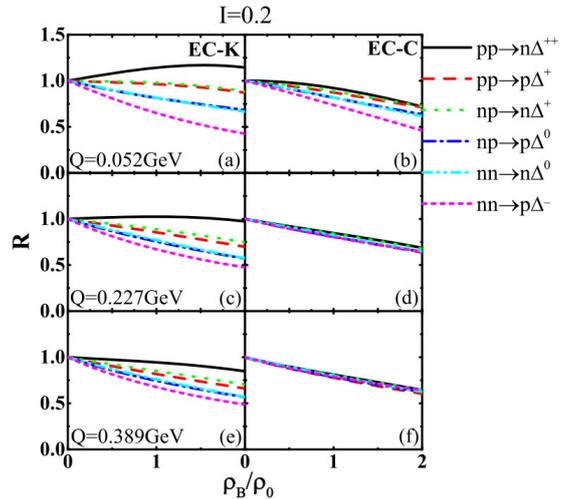

FIG. 31: (Color online) The medium correction factor $R = \sigma^*_{NN \rightarrow N\Delta}/\sigma^{free}_{NN \rightarrow N\Delta}$ as a function of density for different channels (with different colors) for the beam energy at Q = 0.052, 0.227, and 0.389 GeV ($E_{beam} = 0.4$, 0.8 and 1.2 GeV) in isospin asymmetric matter at $I = 0.2$. The left panels are the results without threshold effects, and the right panels are with threshold effects. Taken from Ref. [366].

## B. In-medium NN cross sections from heavy ion collisions

It is quite often in the transport model calculations that the medium corrected elastic cross section adopts a form of $\sigma^*_{NN} = (1 - \eta \rho/\rho_0)\sigma^{free}_{NN}$, with $\eta$ being a parameter and usually set to be 0.2. Compared with that obtained from the theoretical calculations shown in Fig. 28 where the $\sigma^*_{NN}$ depends also on energy in addition to density, the form of the medium corrections $\sigma^*_{NN} = (1 - \eta \rho/\rho_0)\sigma^{free}_{NN}$ are too simple. To obtain the information of the in-medium elastic cross sections from HIC, Y.X. Zhang, et al. [155] proposed ad hoc parametrization inspired by the closed-time path Green's function (CTPGF) results aiming at the description of the excitation function for elliptic flow in the midrapidity region of $|y_{c.m.}/y^{c.m.}_{beam}| < 0.1$ in Au+Au collisions. Fig. 32 shows the ImQMD05 calculation results of excitation functions of elliptic flow parameter $v_2$ and directed flow $P^0_{xdir} = p_{xdir}/u^{beam}_{c.m.}$ with different Skyrme density functionals and in-medium elastic cross sections.

There are three kinds of in-medium NN cross sections, i.e. $\sigma^{*(1)}$, $\sigma^{*(2)}$, and $\sigma^{*(3)}$, where $\sigma^{*(1)}$ is the one given in expression ($\sigma^{*}_{NN} = (1 - \eta\rho/\rho_0)\sigma^{free}_{NN}$ and $\eta = 0.2$), $\sigma^{*(2)}$ is the results calculated by using CTPGF method based on QMD-II effective Lagrangian, $\sigma^{*(3)}$ is the *ad hoc* parameterization given by Ref. [155]. It is seen that the $\sigma^{*(3)}$ shows the best fit to the data and the $\sigma^{*(1)}$ is the worst among three kinds in-medium cross section. It is found that the best fit to the experimental flow data with $\eta=0.2$ for $E_{beam} \leq 150$ MeV/nucleon, $\eta=0$ for $E_{beam}$=150-200 MeV/nucleon, $\eta$=-0.2 for $E_{beam} = 200$-400 MeV/nucleon, can well reproduce the data of flow and stopping power simultaneously.

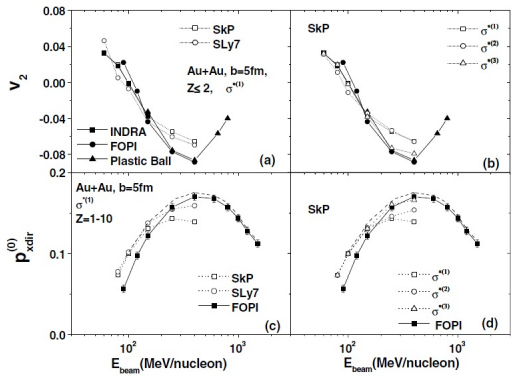

FIG. 32:   The top panels display midrapidity elliptic flow. The bottom panels display maximal scaled directed flow. The data are represented by solid symbols and calculations by open symbols. The left panels in the figure illustrate the sensitivity of calculations to the employed mean field. The right panels illustrate the sensitivity of calculations to in-medium cross sections. Taken from Ref. [155].

The impact of nucleon-nucleon cross sections on nuclear stopping is also investigated. Fig. 33 shows a variety of results pertinent to $vartl$. The $vartl$ is defined as the ratio of the rapidity variance in the transverse direction to the rapidity variance in the longitudinal direction,

$$vartl = \frac{< y_t^2 >}{< y_z^2 >}, \qquad (76)$$

which has been used as a measure of the nuclear stopping. The emitted charged particles with $Z = 1 - 6$ from central collisions of symmetric or near-symmetric systems are selected, with the contributions of different particles weighted with $Z$. The panels (a) and (b) in Fig. 33 show calculated distributions in longitudinal and transverse rapidities in 400 MeV/nucleon Au+Au collisions at $b/b_{max} < 0.15$. The corresponding $vartl$ values for different calculations are quoted in those panels. The panel (c) compares the calculated $vartl$ excitation functions for Au+Au to the data. Finally, the panel (d) compares the calculated dependence of $vartl$ on system charge to data at 400 MeV/nucleon. From Fig. 33, one can learn that there is a good chance to restrict the in-medium cross sections using measured $vartl$ but less chance to restrict the mean field. Our results show both flow and stopping power data favor the $\sigma^{*(3)}$, which is consistent with the recent analysis from INDRA data [371].

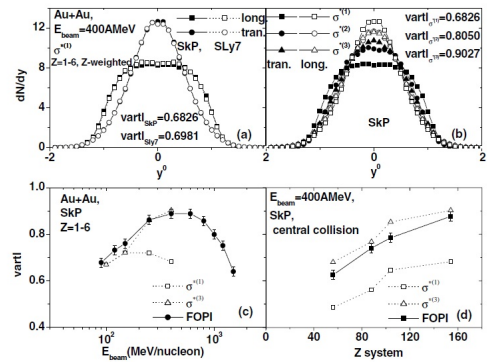

FIG. 33:   Comparison of characteristics of central colliding systems in the longitudinal and transverse rapidities. Panels (a) and (b) display the calculated rapidity distributions of particles from central ($b/b_{max} < 0.15$) 400 MeV/nucleon Au+Au collisions and illustrate, respectively, the sensitivity of calculations to the mean field and to the cross sections. Bottom panels (c) and (d) display the calculated and measured beam energy dependencies of the variance ratio $vartl$, respectively, on energy for the Au+Au collisions and on net system charge at collision energy 400 MeV/nucleon. The data are from FOPI Collaboration [369]. Taken from Ref. [155]

Very recently, P.C. Li *et al.* also discuss the similar things with UrQMD model [253]. They adopted a more complicated medium correction form, which reads

$$\sigma^{*}_{NN} = \mathcal{F}(\rho, p)\sigma^{free}_{NN}, \qquad (77)$$

with

$$\mathcal{F}(\rho, p) = \frac{\lambda + (1-\lambda)e^{-\rho/\rho_0\xi} - f_0}{1 + (p/p_0)^\kappa} + f_0. \qquad (78)$$

$p$ is the momentum of nucleon in center of mass of two colliding nucleons, $\xi$ and $\lambda$ are the parameters which determine the density dependence of the cross sections, $\kappa$ is the parameter which determine the momentum dependence of the cross sections.

As shown in Fig. 34, the degree of nuclear stopping ($R_E = \frac{E_{\perp}}{2E_{\parallel}}$) in central Au + Au collisions as a function of the beam energy are investigated. The results obtained with medium correction factor $\mathcal{F}(\rho, p) = \sigma^{*}_{NN}/\sigma^{free}_{NN} = 0.5$ are the largest and those with $\mathcal{F} = 0.2$ are the smallest of all. Again, this result from the nuclear stopping observables also consistently supports that the medium correction factors of about 0.2 and 0.5 are required for reasonably describing the degree of nuclear stopping in HICs at $E_{lab} = 40$ and 150 MeV/nucleon, respectively. Furthermore, the difference between the results from the FU3FP1 (corresponding to $\lambda = 1/3$, $\xi = 1/3$, $f_0 = 1$, $p_0 = 0.425$ GeV/c, and $\kappa = 5$ in Eq. (78)) parametrization and from $\mathcal{F} = 0.3$ steadily increases with increasing beam energy. Calculations with FU3FP1 fit the experimental data quite well and reproduce the slightly increased stopping power with increasing beam energy, whereas others fail to reproduce the observed beam-energy dependence. In the future, rigorous extracted information of in-medium NN cross sections also require a refined Pauli blocking in the transport models [372].

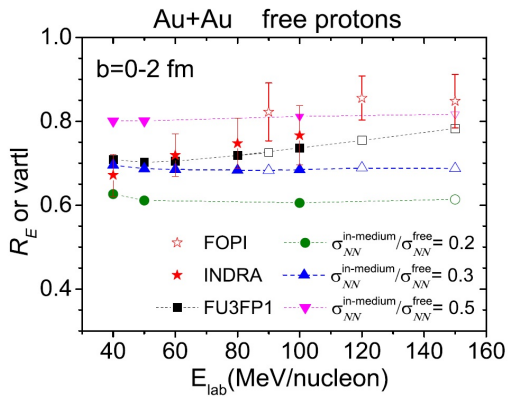

FIG. 34: (Color online) Beam-energy dependence of $R_E$ (the solid symbols) and *vartl* (the open symbols) for free protons from central $^{197}$Au + $^{197}$Au collisions. Calculations with four different medium correction factors (the dashed lines with different symbols) are compared with the FOPI (the open stars) and INDRA (the solid stars) experimental data. Taken from Ref. [253].

### C. In-medium NN cross sections from spallation reactions

Proton induced reactions also provide the information of in-medium NN cross sections below the normal density with a much clearer picture. In the following, the medium correction is taken as,

$$\sigma_{\text{tot}}^* = \sigma_{\text{in}}^{\text{free}} + \sigma_{\text{el}}^* = \sigma_{\text{in}}^{\text{free}} + \mathcal{F}(u, \delta, p)\sigma_{\text{el}}^{\text{free}}. \quad (79)$$

Where the $\sigma_{\text{el}}^{\text{free}}$ and $\sigma_{\text{in}}^{\text{free}}$ are the free isospin dependent elastic and inelastic cross sections, respectively. The form of the medium correction factor is as same as that proposed by Q.F. Li *et al.* in Refs. [77, 170, 252]. The $\mathcal{F}(u, \delta, p)$ depends on the nuclear-reduced density $u = \rho/\rho_0$, the isospin-asymmetry $\delta = (\rho_n - \rho_p)/(\rho_n + \rho_p)$ and the momentum $p_{NN}$. On condition of momentum independent medium correction, $\mathcal{F}(u, \delta, p) = F_\delta \cdot F_u$,

$$F_u = \lambda + (1 - \lambda)\exp(-u/\zeta), \quad (80)$$

$$F_\delta = 1 - \tau_{ij} A(u)\delta, \quad A(u) = \frac{0.85}{1 + 3.25u}, \quad (81)$$

$$i = j = n/p, \ \tau_{ij} = \mp 1; \ i \neq j, \ \tau_{ij} = 0.$$

If the medium correction has a obvious momentum dependence, $\mathcal{F}(u, \delta, p) = F_\delta^p \cdot F_u^p$,

$$F_{\delta/u}^p = \begin{cases} f_0, & p > 1 \text{ GeV/c}, \\ \frac{F_{\delta/u} - f_0}{1 + (p/p_0)^\kappa} + f_0, & p \leq 1 \text{ GeV/c}, \end{cases} \quad (82)$$

with $p$ being the momentum in the NN center-of-mass two colliding nucleons. By varying the parameters $\lambda$, $\zeta$, $f_0$, $p_0$ and $\kappa$ one can obtain different medium correction on nucleon-nucleon elastic cross sections (NNECS). The parameter sets used are listed in Table II and III. Among these parameter sets, FU1-3 and FP1-5 are taken from Refs. [77, 170, 252], FU4 and FP6 are used in Ref. [373].

TABLE II: Parameter sets used for the density-dependent correction factor $F_u$.

| Set | $\lambda$ | $\xi$ |
|-----|-----------|-------|
| FU1 | $\frac{1}{3}$ | 0.54568 |
| FU2 | $\frac{1}{4}$ | 0.54568 |
| FU3 | $\frac{1}{6}$ | 1/3 |
| FU4 | $\frac{1}{5}$ | 0.45 |

TABLE III: Parameter sets used for the momentum dependence of correction factor $F_p$.

| Set | $f_0$ | $p_0$ (GeV/c) | $\kappa$ |
|-----|-------|---------------|----------|
| FP1 | 1 | 0.425 | 5 |
| FP2 | 1 | 0.225 | 3 |
| FP3 | 1 | 0.625 | 8 |
| FP4 | 1 | 0.3 | 8 |
| FP5 | 1 | 0.34 | 12 |
| FP6 | 1 | 0.725 | 10 |

The various in-medium NNECS obtained from combinations by parameterizations FU1, FU2, FU3 and FP1, FP2, FP3 are tested by the excitation function of reaction cross section (RCS) for p+$^{56}$Fe in Fig. 35. By using the in-medium NNECS, the descriptions on the excitation function of RCS are great improved. Especially the combinations of FU2+FP3 and FU3+FP3 give the best two results except a little deviation at low energies. According to the momentum dependence of the in-medium NNECS $F_u^p$, the enhancement effect of momentum correction does not yet appear at such low energies if FP3 is adopted. FU4 provides a reasonable correction effect between ones given by FU2 and FU3.

More experimental data of the RCS of proton induced reaction from light to heavy targets $^{12}$C, $^{27}$Al, $^{40,48}$Ca, $^{90}$Zr, $^{118}$Zn, $^{208}$Pb are used to test the obtained in-medium NNECS FU4FP6. As examples, the excitation functions of RCS for nucleon-induced on various targets, such as n+$^{12}$C, n+$^{63}$Cu, n+$^{208}$Pb, p+$^{12}$C, p+$^{40}$Ca, and p+$^{118}$Sn, calculated by the ImQMD model with FU4FP6 are shown in Fig. 36. One can see that all experimental data are reproduced quite well. In particular, not only are the excitation functions of RCS for the targets along the $\beta$-stable line well described, but also the excitation function of RCS for $^{48}$Ca, which is far away from the $\beta$-stable line, is also well reproduced. In addition to the proton-induced reactions, the good description on the excitation functions of RCS for the neutron-induced reac-tions means that the medium correction of the isospin dependence of NNECS is reasonably reproduced.

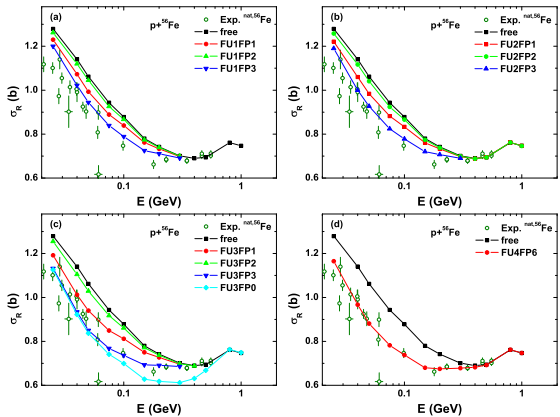

FIG. 35: (Color online) Excitation functions of RCS for p+$^{56}$Fe calculated with free NNECS and various in-medium NNECS compared with experimental data, respectively. Taken from Ref. [373].

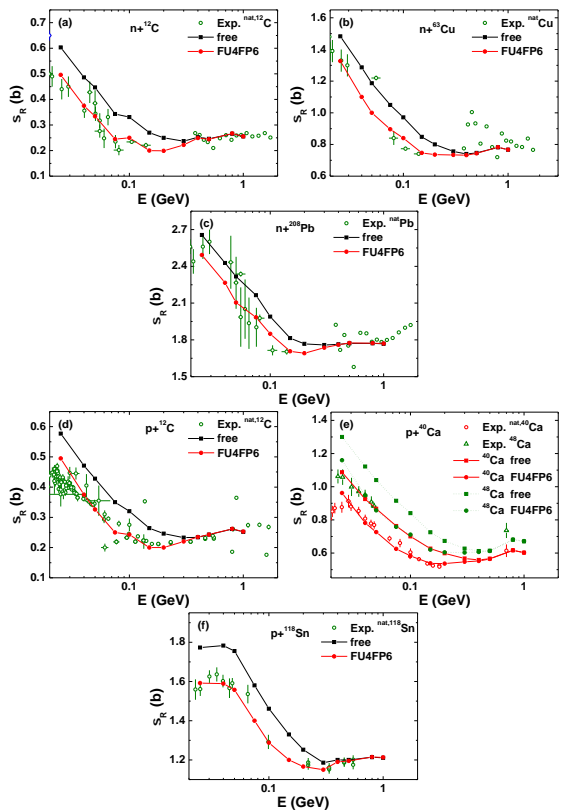

FIG. 36: (Color online) Excitation functions of RCS for nucleon-induced on various targets calculated with free NNECS and FU4FP6 in-medium NNECS compared with experimental data, respectively. Taken from Ref. [373].

## V. SYMMETRY ENERGY AT DIFFERENT DENSITIES AND TEMPERATURES AND THE CONSTRAINTS

The relationship between pressure, density, temperature and isospin asymmetry ($\delta = (\rho_n - \rho_p)/(\rho_n + \rho_p)$) described by the equation of state (EOS) of nuclear matter governs the compression achieved in supernova and neutron stars, as well as the structure and many other basic properties of nuclei. Thus, determination of the EOS has been one of the primary goals since first relativistic heavy ion beams started to operate in the beginning of the 80s of the last century [5–8]. The symmetry energy of nuclear matter is defined as the difference between the energy per nucleon in pure neutron matter and in symmetric nuclear matter, i.e., the energy related to the isospin asymmetry of the system. As one of the most important and unclear part of the EOS for isopin asymmetric nuclear matter, it plays an important role in the various fields, ranging from the structure of nuclei to gravitational collapse to neutron stars.

Within the parabolic approximation, the energy per nucleon for asymmetric nuclear matter $E(\rho, \delta)$ (known as EOS of cold nuclear matter) can be written as

$$
\begin{aligned}
E(\rho, \delta) &= E(\rho, \delta = 0) + E_{sym} \\
&= E(\rho, \delta = 0) + S(\rho)\delta^2 + \cdots .
\end{aligned}
\tag{83}
$$

The first term is the energy per nucleon for symmetric nuclear matter, $E_{sym}$ is the symmetry energy term. Here, one should note that the $S(\rho)$ in the second term describes the density dependence of the symmetry energy which was also named as symmetry energy in other literatures [62]. The approaches to describe the nuclear EOS have been performed by using the variety of effective interactions within mean-field theories, relativistic [374–383], non-relativistic [384–390] as well as the effective interaction based on chiral effective field theory [391–397]. In addition, the *ab initio* approaches based on high precision free space nucleon-nucleon interactions and the nuclear many-body problems being treated microscopically are also applied to study the nuclear EOS [398–413]. However, large uncertainties in predictions of the density dependence of the symmetry energy away from normal nuclear matter have been found. It stimulates lots of effort to reduce the uncertainties of the predicted density and momentum dependence of symmetry potential by using very neutron-rich HICs.

Quite a few important observables in HICs, such as the yield ratio of neutron to proton, the yield ratio of $\pi^-$ to $\pi^+$, and elliptic flow, etc., were predicted to be sensitive observables for sub- and supra-saturation densities based on the IBUU approach [63, 66, 69–71, 414]. Q.F. Li *et al.* also pointed out that comprehensive studies with multiple observables such as the yield ratios of free neutrons to protons, $\pi^-$ to $\pi^+$, $^3$H to $^3$He, etc can provide the density dependence of the symmetry energy at a wider density region based on the UrQMD model calculations as shown in Refs. [73–75]. Tsang *et al.* proposed and measured the isospin sensitive observables, double neutron to proton ratios [415], isospin diffusion [416], isospin transport ratios as a function of rapidity [417]. At present, the data for the isospin diffusion [416] and the neutron to proton yield ratio data [415] from NSCL/MSU, the mass asymmetry for largest and second largest residues in CHIMERA and the flow data from TAMU [418] and GSI [233], angular distribution of neutron to proton exceed ratios from Tsinghua/IMP [419], are important observables for constraining the symmetry energy around saturation density. The data of pion yield ratio, $\pi^-/\pi^+$ [78] and elliptic flow of neutron and proton [246, 420, 421] from GSI are available, which are supposed to be used to constrain the symmetry energy at supra-saturated densities. By comparing the data with transport model calculations, one indirectly obtains the density dependence of the symmetry energy. With the development of new generation of neutron-rich beam facilities in Institute of Modern Physics (IMP), National Superconducting Cyclotron Laboratory(NSCL), Texas A&M (TAMU), National Institute for Nuclear Physics (INFN), GSI, tight constraints of the density dependence of the symmetry energy will be possible in future.

### A. Symmetry energy at subsaturation density

The discussions in this section focus on the neutron to proton yield ratios, isospin diffusion as observables to constrain the density dependence of the symmetry energy from normal density to subnormal density. The ImQMD code is used to extract the information of the symmetry energy from heavy ion collision data. Here, one should note that the exact input of the transport model is the effective nucleon interaction, potential energy density functional or the single particle potential as mentioned in section II. Consequently, one can also obtain the EOS or symmetry energy at different density or temperature based on the interaction or nucleonic potential parameters that we used in the transport models. In the community, the EOS or symmetry energy at zero temperature, which are obtained with the interaction used in the transport models, are usually used to describe the constrained results on EOS or the density dependence of the symmetry energy $S(\rho)$.

#### 1. Constraints from n/p ratios and isospin diffusion

The single neutron to proton yield ratio was proposed to study the density dependence of the symmetry energy by Bao-An Li *et al.* in Ref. [63] in 1997 where the neutron proton yield ratios are related to the strength of symmetry energy. The first experimental data was published in 2006 for $^{112,124}$Sn+$^{112,124}$Sn by the NSCL/MSU group [415]. However, the different efficiencies for neutron and charged particles detectors cause the large errors for single n/p ratio. A double n/p ratio was proposed in order to reduce the uncertainties in the neutron detection efficiencies and the energy calibrations of neutrons and protons. The double ratio,

$$
\begin{aligned}
DR(Y(n)/Y(p)) &= \frac{R_{n/p}(A)}{R_{n/p}(B)} \\
&= \frac{(dM_n(A))/(dE_{c.m.})}{(dM_p(A))/(dE_{c.m.})} \\
&\quad \frac{(dM_p(B))/(dE_{c.m.})}{(dM_n(B))/(dE_{c.m.})},
\end{aligned}
\tag{84}
$$

is constructed from the ratios of energy spectra, $dM/dE_{c.m.}$ of neutrons and protons for two systems A and B characterized by different isospin asymmetries. The comparison between the ImQMD05 calculations and the experimental data is shown in Fig. 37. The star symbols in the left panel of Fig. 37 show the neutron-proton double ratios measured at $70^\circ < \theta_{c.m.} < 110^\circ$ as a function of center-of-mass (c.m.) energy of nucleons emitted from the central collisions of $^{124}$Sn+$^{124}$Sn and $^{112}$Sn +$^{112}$Sn [415]. Despite the large experimental uncertainties for higher energy data from [415], the comparisons between the data and ImQMD calculations definitely rule out both very soft ($\gamma_i=0.35$) and very stiff ($\gamma_i=2$) density-dependent symmetry terms. Here, $\gamma_i$ is the symmetry potential parameter as in Eq. (42) in section II.

The parameterization of the symmetry energy corresponding to the interactions we used in these calculations has the form as in Eq. (44), i.e., $S(\rho) = \frac{\hbar^2}{6m}(\frac{3\pi^2\rho}{2})^{2/3} + \frac{C_{s,p}}{2}(\frac{\rho}{\rho_0})^{\gamma_i}$. The potential parameters is $C_{s,p}=35.2$ MeV and the symmetry energy at saturation density, $S_0=30.1$ MeV. The right panel shows the dependence on $\gamma_i$ of the $\chi^2$ computed from the difference between predicted and measured double ratios. We determine, within a $2\sigma$ uncertainty, parameter values of $0.4\leq \gamma_i \leq 1.05$ corresponding to an increase in $\chi^2$ by 4 above its minimum near $\gamma_i \sim 0.7$.

The density dependence of the symmetry energy has also been probed in peripheral collisions between two nuclei with different isospin asymmetries by examining the diffusion of neutrons and protons across the neck that joins them. This "isospin diffusion" generally continues until the two nuclei separate or until the chemical potentials for neutrons and protons in both nuclei become equal. Thus, in theory, the changes of isospin asymmetry of projectile-like/target-like residues at their separation time reflect the isospin diffusion ability. We named this as isospin asymmetry of the emitting source. To isolate diffusion effects from other effects, such as pre-equilibrium emission, Coulomb effects and secondary decays, measurements of isospin diffusion compare "mixed" collisions, involving a neutron-rich nucleus $A$ and a neutron-deficient nucleus $B$, to the "symmetric" collisions involving $A+A$ and $B+B$. The degree of isospin equilibration in such collisions can be quantified by rescaling the isospin observable $X$ according to the definition of isospin transport ratio $R_i(X)$ [416],

$$R_i(X) = \frac{2X_{AB} - X_{AA} - X_{BB}}{X_{AA} - X_{BB}}, \qquad (85)$$

where $X$ is the isospin observable. In the absence of isospin diffusion, the ratios are $R = 1$ or $R = -1$. If isospin equilibrium is achieved, then the ratios $R = 0$ for the mixed systems. Equation (85) dictates that different observables, $X$, provide the same results if they are linearly related [416, 417]. The agreement of experimental isospin transport ratios obtained from isoscaling parameters, $\alpha$, [22] and from yield ratios of $A = 7$ mirror nuclei [417], $R_7 = R(X_7 = ln(Y(^7Li)/Y(^7Be))$ agree, i.e., $R_i(\alpha) \approx R_7$, reflects nearly linear relationships between $\alpha$, $X_7$, and the asymmetry $\delta$ of the emitting source [417].

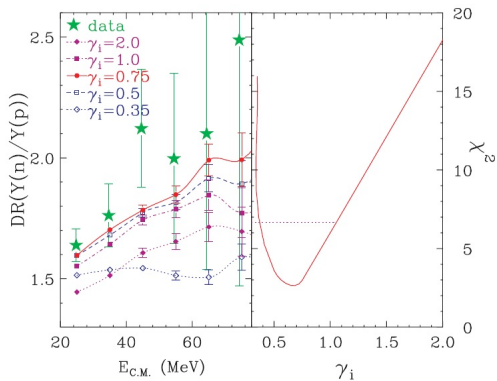

FIG. 37: (Color online) Left panel: Comparison of experimental double neutron-proton ratios (star symbols), as a function of nucleon center-of-mass energy, to ImQMD calculations (lines) with different density dependencies of the symmetry energy parameterized by $\gamma_i$ in Eq. (44). Right panel: A plot of $\chi^2$ as a function of $\gamma_i$. Taken from Ref. [422]

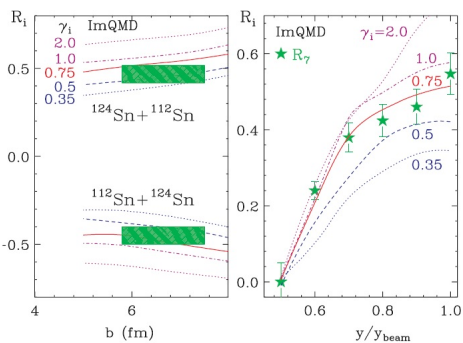

FIG. 38: (Color online) Left panel: Comparison of experimental isospin transport ratios (shaded regions) to ImQMD results (lines), as a function of impact parameter for different values of $\gamma_i$. Right panel: Comparison of experimental isospin transport ratios obtained from the yield ratios of $A = 7$ isotopes (star symbols), as a function of the rapidity to ImQMD calculations (lines) at $b = 6$ fm. Taken from Ref. [422].

In the ImQMD calculations, the isospin asymmetry $\delta$

of emitting source is calculated from the emitted fragments and free nucleons at the same velocity or rapidity region as in experiments [224, 416, 422]. Left panel of Fig. 38 shows the calculation results of $R_i(\delta)$ compared with the experimental isospin diffusion transport ratios, $R_i(\alpha)$, plotted as shaded regions. The ImQMD calculations are performed at impact parameters of $b=5, 6, 7$, and 8 fm for considering the impact parameter effects. The lines in the left panel of Fig. 38 show the predicted isospin transport ratio $R_i(\delta)$ as a function of impact parameter $b$ for $\gamma_i=0.35, 0.5, 0.75, 1$, and 2. Faster equilibration occurs for smaller $\gamma_i$ values which correspond to larger symmetry energies at subsaturation densities. Thus we see a monotonic decrease of the absolute value of $R_i(\delta)$ with decreasing $\gamma_i$. The $\chi^2$ analysis for both impact parameters $b=6$ and 7 fm, which corresponds to the impact parameter region in the measurements, are performed. Using the same $2\sigma$ criterion, the analysis brackets the regions $0.45 \leq \gamma_i \leq 1.0$ and $0.35 \leq \gamma_i \leq 0.8$ for $b=6$ and 7 fm, respectively. ImQMD calculations also provide predictions for fragment yields as a function of rapidity. The star symbols in the right panel of Fig. 38 represent measured values of $R_7$ obtained from the yield ratios of $^7Li$ and $^7Be$ [417] at $b=6$ fm, as shown by the lines. This first calculation of the shapes and magnitude of the rapidity dependence of the isospin transport ratios $R_7$ reproduces the trends accurately. The corresponding $\chi^2$ analysis with calculations at $b=6$ and 7 fm favors the region $0.45 \leq \gamma_i \leq 0.95$. The favorite $\gamma_i$ region obtained from double neutron to proton yield ratio and isospin diffusion observables are in coincidence.

Here, we should point out the constraints on the exponent $\gamma_i$ depends on the symmetry energy at saturation density, $S_0 = S(\rho_0)$. Increasing $S_0$ has the same effect on the isospin transport ratio as decreasing $\gamma_i$. To compare our results to constraints obtained from nuclear masses and nuclear structure, we expand $S(\rho)$ around the saturation density, $\rho_0$,

$$
\begin{aligned}
S(\rho) \;=\; & S_0 + \frac{L}{3}(\rho - \rho_0)/\rho_0 + \\
& \frac{K_{sym}}{18}((\rho - \rho_0)/\rho_0)^2 + \frac{Q_{sym}}{162}((\rho - \rho_0)/\rho_0)^3 ...
\end{aligned}
\qquad (86)
$$

where $L$, $K_{sym}$ and $Q_{sym}$ are the slope, curvature and skewness parameters of $S(\rho)$ at $\rho_0$. For realistic parameterization of $S(\rho)$, $K_{sym}$ is correlated to $L$ [423]. As the second term in Eq. (86) is much important than the third term, we believe $L$ can be determined more reliably than $K_{sym}$. Furthermore, the slope parameter, $L = 3\rho_0 dS(\rho)/d\rho|_{\rho_0}$ is related to $p_0$, the pressure from the symmetry energy for pure-neutron matter at saturation density. The symmetry pressure, $p_0$, provides the dominant baryonic contribution to the pressure in neutron stars at saturation density [424–426].

More calculations have been performed at $b = 6$ fm with different values of $\gamma_i$ and $S_0$ to locate the approximate boundaries in the $S_0$ and $L$ plane that satisfy the $2\sigma$ criterion in the $\chi^2$ analysis of the isospin diffusion data. The two diagonal lines in Fig.39 represent estimates in such an effort. Examination of the symmetry energy functional formed along these boundaries, where the diffusion rates are similar but $S_0$ and $L$ are different, reveal that diffusion rates predominantly reflect the symmetry energy at and somewhat below.

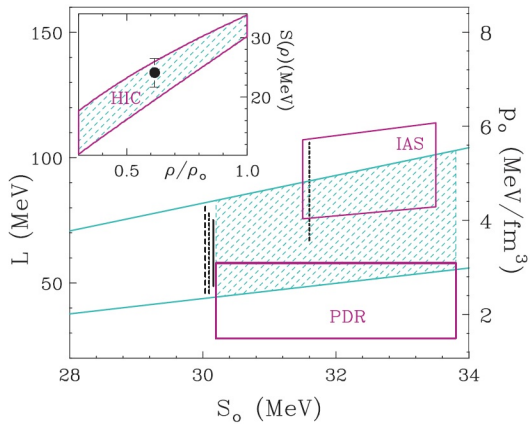

FIG. 39: (Color online). Representation of the constraints on parameters $S_0$ and $L$. The right axis corresponds to the neutron matter symmetry pressure at saturation density. The region bounded by the diagonal lines represents the constraints obtained in the present work. The vertical line at $S_0 \sim 31.6$ MeV is from Refs. [62, 431]. The lower and upper boxes are formed by the constraints from PDR data [432] and from symmetry energy analysis on nuclei [423], respectively. The inset shows the density dependence of the symmetry energy of the shaded region. The symbol in the inset represents the GDR results from Refs. [433, 434].

Up to now, there are a lot of efforts on the extraction of symmetry energy. They have been made by comparing measured isospin sensitive observables, such as isospin diffusion [427] at the beam energy 74 MeV/nucleon, angular distribution of neutron-excess for light charged particles at 35 MeV/nucleon [428], collective flows [246, 418] to various transport model calculations. Consensus on the symmetry energy coefficient and slope of symmetry energy has been obtained from nuclear structure and reaction studies, as partly summarized and shown in Fig. 40, where symmetry energy coefficient $S_0 = 30 - 32$MeV and slope of symmetry energy $L = 40 - 65$MeV [429, 430], but the uncertainties of symmetry energy constraints are still large.

Furthermore, the symmetry energy not only depends on the $S_0$ and $L$ but also on the higher terms, such as $K_{sym}$, $Q_{sym}$, ..., or depends on the effective mass $m^*$ and neutron proton effective mass splitting $\Delta m = (m_n^* - m_p^*)/m$. It naturally requires more accumulations of the data of the isospin sensitive observables and the development of transport models to distinguish those different physics.

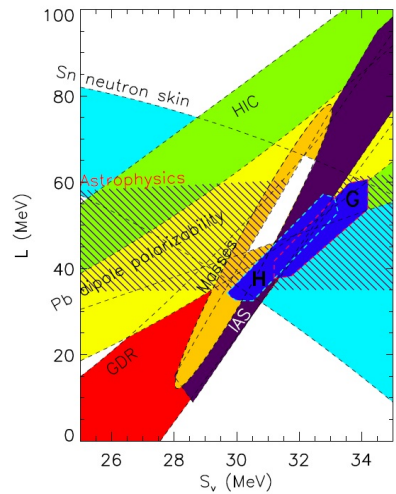

FIG. 40: (Color online) Constraints for symmetry energy parameters from different approaches. Taken from Ref. [430].

### 2. Novel probes of symmetry energy at subsaturation denisty

The new sensitive and clear observables to the symmetry potential are also welcome for obtaining accurate knowledge of the symmetry energy. For example, some types of the direct reaction, like the elastic or inelastic scattering as well as the direct projectile breakup, involve fewer degrees of freedom in the reaction process and may reduce the difficulties in modeling the collision and could be used to constrain the symmetry energy at the subsaturation density.

Li Ou *et al.* proposed a new sensitive probe that is the isovector reorientation of deuteron induced collisions on heavy nuclei based on the ImQMD model calculations [435]. It is demonstrated in [435] that in the deuteron induced peripheral collisions on heavy nuclei such as $^{124}$Sn, the loosely bound deuteron break up into neutron and proton, which moves with different directions relative to the incoming beam direction (named as reorientation effect), due to the isovector and Coulomb (only for proton) interaction with the target. It is found that the correlation angle determined by the relative momentum vector of the proton and the neutron originating from the breakup deuteron, which is experimentally detectable, exhibits significant dependence on the isovector nuclear potential but is robust against the variation of the isoscaler sector. Fig. 41 shows the distribution of correlation angle $\alpha$ defined as

$$cos\alpha = \frac{p_z^p - p_z^n}{|\mathbf{p}^p - \mathbf{p}^n|}, \quad (87)$$

for 100 MeV/nucleon (unpolarized and polarized) d + $^{124}$Sn. The calculation is performed with the ImQMD05 code and the expression of density functional is given in section II. The Skyrme density functionals SkA [436], SkT5, SkT1 [437], SkM$^*$ [229], Skz-1, Skz1, and Skz4

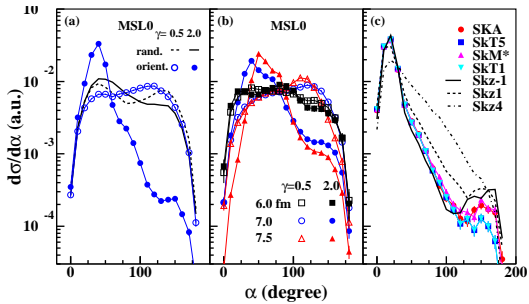

FIG. 41: (Color online) The distribution of the correlation angle $\alpha$ in 100 MeV/nucleon $\vec{d}+^{124}$Sn. Here, $\gamma$ is the $\gamma_i$ in Eq.(44). Taken from Ref. [435].

[438] are applied. The MSL0-like Skyrme interaction corresponding to the density dependent parameters $\gamma_i = 0.5$ and =2. From the results of ImQMD calculations, one can see that the correlation angle $\alpha$ is very sensitive to the density dependence of symmetry energy, and insignificantly depends on the variation of isoscalar potentials.

Left panel of Fig. 42 shows the logarithmic spectra of $cos\alpha$, i.e., $\ln\frac{d\sigma}{d\cos\alpha}$, for various $\gamma_i$ with impact parameters $b = 7.0$ fm for the reaction of $\vec{d}+^{124}$Sn. The slope of function $\ln\frac{d\sigma}{d\cos\alpha} = a_0 + a_1\cos\alpha$, which fitting the logarithmic spectra of $\cos\alpha$ in $|cos\alpha| \leq 0.2$, are shown in right panel of Fig. 42. Fig. 42 clearly demonstrates the sensitivity of the correlation angle distribution to the density dependence of the symmetry energy and thus can be taken as a more clear probe to the density dependence of symmetry energy at subsaturation densities. In terms of sensitivity and cleanness, the breakup reactions induced by the polarized deuteron beam at about 100 MeV/nucleon could provide a more stringent constraint to the symmetry energy at subsaturation densities.

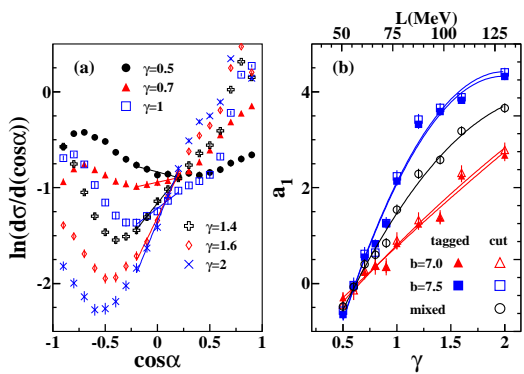

FIG. 42: (Color online) (a) The logarithmic spectra of $cos\alpha$ in $\vec{d} + ^{124}$Sn with $b = 7$ fm at 100 MeV/nucleon for various $\gamma$ and corresponding fitted quadratic functions, (b) the linear coefficient $a_1$ as a function of $\gamma$. Here, $\gamma$ is the $\gamma_i$ in Eq.(44). See text for detailed discussions. Taken from Ref. [435].

Another effort we have done is that we propose the angular distribution anisotropy of coalescence invariant neutron to proton yield ratio to probe the symmetry energy at subsaturation density. The mass asymmetry reaction system, i.e. $^{40}$Ar$+^{197}$Au, are analyzed, since this system has a gradient of Coulomb field during the reaction and the effects of symmetry potential at forward and backward region may different.

The angular distribution anisotropy of coalescence invariant neutron to proton yield ratio is defined as

$$CI\ n/p = \frac{dM_{n,CI}}{d\theta_{c.m.}}/\frac{dM_{p,CI}}{d\theta_{c.m.}}, \qquad (88)$$

where

$$\frac{dM_{n,CI}}{d\theta_{c.m.}} = \sum_{Z,N} N\frac{Y(N,Z)}{d\theta_{c.m.}}, \qquad (89)$$

$$\frac{dM_{p,CI}}{d\theta_{c.m.}} = \sum_{Z,N} Z\frac{Y(N,Z)}{d\theta_{c.m.}}. \qquad (90)$$

As shown in Fig. 43, the CIn/p ratios obtained with $\gamma_i = 0.5$ are greater than that obtained with $\gamma_i = 2.0$ for the beam energies we studied, because the symmetry energy for $\gamma_i = 0.5$ is stronger than that for $\gamma_i = 2.0$ at subsaturation density. Furthermore, the CIn/p ratios show a different $\theta_{c.m.}$ dependence for the different forms of symmetry potential. For $\gamma_i = 0.5$, the CI n/p ratios slightly increase as a function of $\theta_{c.m.}$. It is the result of the isospin asymmetry changing from $\delta_{proj} = 0.10$ at the projectile region to $\delta_{tar} = 0.19$ at the target region since the single particle potentials of neutrons and protons are close. However, for $\gamma_i = 2.0$, the CI n/p ratios obviously decrease with $\theta_{c.m.}$, and the CI n/p ratios at backward regions are smaller than that at forward regions. Our calculations show that this behavior also exists at beam energy for 50 and 100 MeV/nucleon.

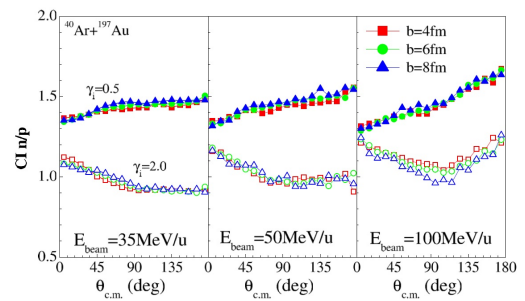

FIG. 43: (Color online) Left: calculated results for CI n/p ratios as a function of $\theta_{c.m.}$ for $b = 4$, 6, and 8 fm at $E_{beam} = 35$ MeV/nucleon. The solid symbols are for $\gamma_i = 0.5$ and open symbols are for $\gamma_i = 2.0$. Middle: results for 50 MeV/nucleon; right: results for 100 MeV/nucleon. Taken from Ref. [439].

## B. Symmetry energy at suprasaturation density

The symmetry energy and its constraints at suprasaturation density are very important for understanding the dense neutron-rich matter, such as the structure of neutron star and its cooling mechanism. In this section we will discuss the efforts on constraining the symmetry energy at suprasaturation densities based on the UrQMD model calculations. Some sensitive observables to the density dependence of the symmetry energy at suprasaturation densities are discussed and then the constraints on the density dependence of the symmetry energy at suprasaturation densities are deduced by comparing the UrQMD model calculations with the measurements of flow.

In Ref. [73, 74], a hard Skyrme-type EoS ( $K = 300$ MeV) without momentum dependence is used in the calculations, and it was demonstrated by the UrQMD model calculations that several observables, such as the transverse momentum distribution of free neutron to proton yield ratio, $^3$H to $^3$He ratio, $\pi^-$ to $\pi^+$, etc. in neutron-rich reaction systems like $^{208}$Pb + $^{208}$Pb, $^{132}$Sn+$^{124}$Sn, $^{96}$Zr +$^{96}$Zr are sensitive to the density dependence of symmetry potential energy. In the UrQMD calculations, a momentum independent symmetry potential is used, i.e., $v^q_{sym} = \frac{\partial(\rho\delta^2 E^{pot}_{sym})}{\partial\rho_q}$. Here, $E^{pot}_{sym}$ is the corresponding density dependence of the symmetry potential energy (in Eq. (43), it is the potential part of $S(\rho)$). It is written as

$$E^{pot}_{sym}(u) = (S_0 - \frac{\epsilon_F}{3})F(u),\qquad(91)$$

where $u = \rho/\rho_0$ is the reduced nuclear density, $S_0$ and $\epsilon_F$ are the symmetry energy coefficient and Fermi energy, respectively. The forms of $F(u)$ adopted in the calculations are (1) $u^\gamma$ with $\gamma=$ 1.5 (called F15 in the following text), (2) $u\frac{(a-u)}{a-1}$ with $a=$ 3 (Fa3), a is the so-called reduced critical density; (3) and (4) so-called DDH$\rho^*$ and DDH3$\rho\delta^*$ symmetry potential energies, which are inspired by the relativistic mean-field calculations of DDH$\rho^*$ and DDH3$\rho\delta^*$ [440]. Fig. 44 illustrates the form of density dependence of symmetry potential energy used in the UrQMD. Among them, F15 and DDH3$\rho\delta^*$ corresponds to the stiff symmetry potential energy and other two correspond to the soft one.

Fig. 45 shows the transverse momentum distribution of the neutron to proton yield ratios (n/p or Y(n)/Y(p)) of emitted free nucleons calculated with four different forms of the symmetry potential. The n/p ratio, especially in the low transverse momentum region, depends strongly on the choice of the symmetry potential. In the low transverse momentum region, the n/p ratio with DDH$\rho^*$ is the largest and that with F15 is the smallest. Obviously, nucleons with low transverse momenta are mainly emitted from the low density region. It follows that the deviation between Fa3 and DDH3$\rho\delta^*$ at densities $\rho < \rho_0$ is small. Correspondingly, the n/p ratios at low transverse momenta calculated with Fa3 and DDH3$\rho\delta^*$ are very close. For emitted nucleons with transverse momenta larger than $\sim$700 MeV/$c$, the n/p ratios in the Fa3 and DDH3$\rho\delta^*$ cases are close to those in the DDH$\rho^*$ and F15 cases, respectively. This is in correspondence with the fact that $E^{pot}_{sym}$ for Fa3 (DDH3$\rho\delta^*$) is close to DDH$\rho^*$ (F15) when $\rho > \rho_0$. Free nucleons with transverse momenta larger than $\sim$700MeV/$c$ are mainly squeezed out from higher densities [441]. The calculations performed in Ref. [73, 74] showed that those nucleons with high transverse momentum are emitted at early time when the density is high.

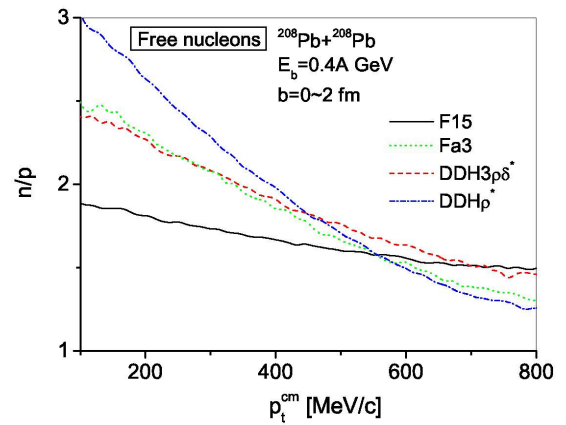

FIG. 45: Transverse momentum distributions of free neutron/proton ratios for different density dependent symmetry potentials. Taken from Ref. [73].

In addition, in Refs. [73, 75], the yields of $\pi^-$ and $\pi^+$ as well as the $\pi^-/\pi^+$ ratios were also investigated with the F15 and Fa3. Fig. 46 shows the calculation results of the yields of $\pi^-$ and $\pi^+$ as well as the $\pi^-/\pi^+$ ratios for central collision in $^{208}$Pb +$^{208}$Pb at 0.4 GeV/nucleon. One can find from the figure that the sequence of the relative differences between calculation results of the n/p ratios at transverse momenta larger than $\sim$700 MeV/$c$ with 4 different symmetry potentials is the same as that of the $\pi^-/\pi^+$ ratios at higher $p_t$. It is because that emitted nucleons with high $p_t$ are also from high density region as pions do. But, the UrQMD calculations show that the relative difference between the calculated results of

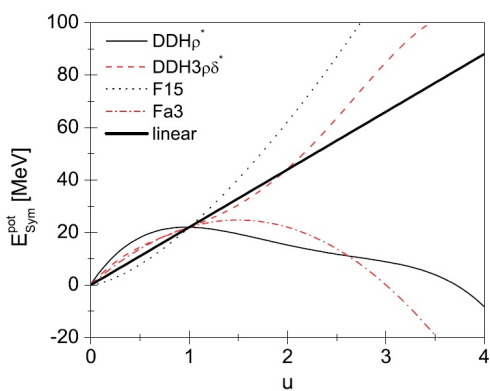

FIG. 44: (Color online) Parameterizations of the nuclear symmetry potential energy DDH$\rho^*$, DDH3$\rho\delta^*$, F15, Fa3, and the linear one as a function of the reduced density $u$. Taken from Ref. [73].

$\pi^-/\pi^+$ ratio from different symmetry energies is weaker than n/p ratio.

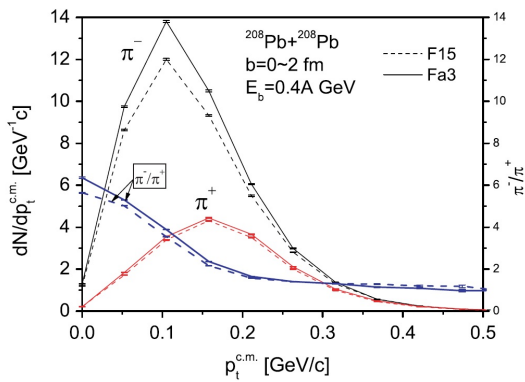

FIG. 46: (Color online) Transverse-momentum distributions of $\pi^-$ and $\pi^+$ from central $^{208}$Pb + $^{208}$Pb collisions at $E_{beam}$ = 0.4 GeV/nucleon for different symmetry potentials. The $\pi^-/\pi^+$ ratios are also shown as functions of transverse-momentum. Taken from Ref. [73].

Up to now a lot of effort have been made to extract the symmetry energy at suprasaturation densities by reproducing the FOPI $\pi^-/\pi^+$ ratio. An apparent systematic discrepancies between the extracting symmetry energy from extreme soft to extreme stiff appeared by different transport models calculations [138, 442–447]. Also, the calculations of Hong and Danielewicz [447] shows the $\pi^-/\pi^+$ ratios are independent of the form of density dependence of symmetry energy after including the strong pion interaction. Further work will thus be required before pion yields and yield ratios can be reliably applied to the investigation of the high-density symmetry energy.

Flow observables have been proposed by several groups as probes for the symmetry energy at high densities [69, 81]. In Ref. [421], the UrQMD model calculations were performed with the SM EOS (soft EOS and with momentum dependent interaction) in isoscalar part, and power law form of symmetry energy as in Eq.(44) in isovector part. The results showed that the elliptic-flow ratio of neutrons with respect to protons or light complex particles in reactions of neutron-rich systems at relativistic energies was sensitive to the strength of the symmetry term at suprasaturation densities. The comparison of existing data of the ratio between the elliptic flow of neutrons and hydrogen isotopes $v_2^n/v_2^H$ from the FOPI/LAND experiment [82, 83] with calculations performed with the UrQMD model suggested a moderately soft to linear symmetry term, characterized by a coefficient $\gamma_i = 0.9 \pm 0.4$ in a power law form of density dependence of the symmetry energy as in Eq.(44).

In Ref. [252], the updated version of UrQMD model, where the Skyrme potential energy density functional was adopted in the mean field part, was applied to study the flow effect and to constrain the density dependence of the symmetry energy from the FOPI/LAND elliptic flow data. Fig. 47 shows the calculation results of $v_2^n/v_2^H$ for $^{197}$Au+$^{197}$Au at 400 MeV/nucleon with impact parameter $b < 7.5$ fm. Left penal shows the comparison between the calculations with different Skyrme potential energy density functional with FOPI/LAND data. The calculation was performed with the same range of laboratory angles accepted by LAND. The slope parameter of $L = 89 \pm 45$MeV for the density dependence of the symmetry energy ($2\sigma$ uncertainty) was extracted by fitting the FOPI/LAND data. The right panel shows the calculation results for intermediate rapidity window $0.25 \le |y_0| \le 0.75$, for the same impact parameter and rapidity interval but without the gate on laboratory angles. It is clearly seen that the differences of the various predictions steadily grow as one moves to the region of low transverse momentum. The $v_2^n/v_2^H$ ratio in the rapidity window $0.25 \le |y_0| \le 0.75$ seems considerably more sensitive to the density dependent symmetry energy than in the mid-rapidity interval $|y_0| \le 0.5$, thus offering interesting opportunities for future experiments.

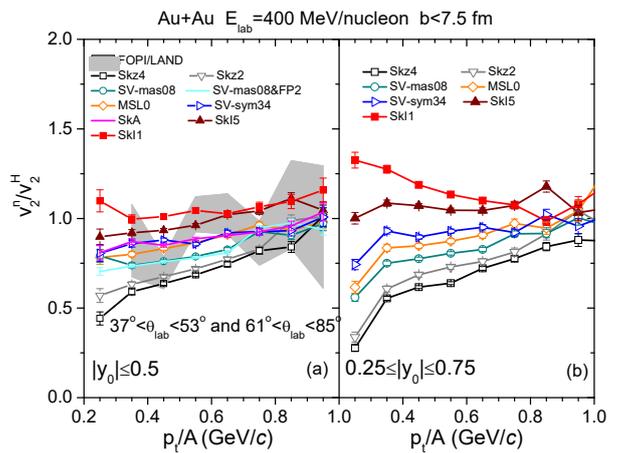

FIG. 47: (Color online) (a) Elliptic flow ratio of neutrons vs. hydrogen isotopes ($Z = 1$) as a function of the transverse momentum $p_t$/A, calculated with the indicated 9 Skyrme forces for central ($b < 7.5$ fm) $^{197}$Au+$^{197}$Au collisions at $E_{beam}$ = 400 MeV/nucleon in the mid-rapidity interval $|y_0| \le 0.5$ in comparison with the FOPI/LAND data (shaded area) reported in Ref. [421]; (b) the same quantity calculated with the indicated 7 Skyrme forces for the intermediate rapidity interval $0.25 \le |y_0| \le 0.75$. Taken from Ref. [319].

Furthermore, the newly measured directed and elliptic flows of neutrons and light-charged particles for the reaction $^{197}$Au+$^{197}$Au at 400 MeV/nucleon incident energy within the ASY-EOS experimental campaign at the GSI laboratory were applied to extract the density dependence of the symmetry energy at supra-saturation densities [246]. Fig. 48 shows the ASY-EOS/GSI data and the UrQMD predictions for the elliptic flow ratio of neutrons over all charged particles $v_2^n/v_2^{ch}$ for central ($b < 7.5$ fm) collisions of $^{197}$Au+$^{197}$Au at 400 MeV/nucleon as a function of the transverse momentum per nucleon $p_t$/A . The deduced symmetry-term coefficient $\gamma_i = 0.75 \pm 0.10$.

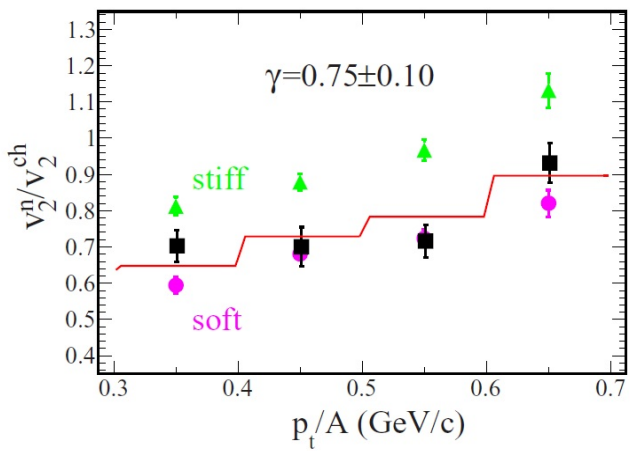

FIG. 48: (Color online) Elliptic flow ratio of neutrons over all charged particles for central collisions of $^{197}$Au+$^{197}$Au at 400 MeV/nucleon as a function of the transverse momentum per nucleon $p_t/A$. The black squares represent the experimental data [246]; the green triangles and purple circles represent the UrQMD predictions for stiff ($\gamma = 1.5$) and soft ($\gamma = 0.5$) power-law exponents of the potential term, respectively. $\gamma$ in this figure means $\gamma_i$ in Eq.(44). Taken from Ref. [246].

It is interesting to see that all the deduced density dependence of the symmetry potential energy at suprasaturation densities from the comparison with the elliptic flow data of FOPI/LAND and ASY-EOS are in coincidence, which suggests a moderate and linear density dependence of the symmetry potential energy.

### C. EOS and symmetry energy at finite temperature

For HICs, the compressed nuclear matter are excited and the temperature is not zero. How does the EOS or the symmetry energy changes at finite temperature is very important for understanding the HICs observables and extracting the EOS or symmetric energy. In Ref. [448] the temperature and density dependence of the symmetry energy was studied. The energy per nucleon at density $\rho$ and temperature $T$ for pure neutron and symmetry matter are calculate by using Skyrme interactions within mean-field approach and the temperature dependence arises from the modification of the zero temperature step-like momentum distributions, which becomes Fermi-Dirac distributions.

The Skyrme density functional reads

$$
\begin{aligned}
H =\ & \frac{\hbar^2}{2m}[\tau_n + \tau_p] \\
& + \frac{1}{4}t_0[(2+x_0)\rho^2 - (2x_0+1)(\rho_n^2 + \rho_p^2)] \\
& + \frac{1}{24}t_3\rho^\alpha[(2+x_3)\rho^2 - (2x_3+1)(\rho_p^2 + \rho_n^2)] \\
& + \frac{1}{8}[t_1(2+x_1) + t_2(2+x_2)]\tau\rho \\
& + \frac{1}{8}[t_2(2x_2+1) - t_1(2x_1+1)](\tau_n\rho_n + \tau_p\rho_p),
\end{aligned}
\tag{92}
$$

where $\rho = \rho_n + \rho_p$, and $\tau = \tau_n + \tau_p$. The $\rho_q$ and $\tau_q$ are calculated by $\rho_q = \frac{1}{(2\pi\hbar)^3}2\int_0^\infty n_q(p)d^3p$ and $\tau_q = 2\int_0^\infty (n_q(p)p^2)/\hbar^2 d^3p/\hbar^3$. The occupation number distribution for species $q$, obeys Fermi-Dirac distribution.

$$
n_q(p) = \frac{1}{1 + exp[\beta(\epsilon_q - \mu_q)]},
\tag{93}
$$

$\epsilon_q$ and $\mu_q$ are the single particle energy and chemical potential for species $q$, and $\beta = 1/K_B T$.

By solving above equations iteratively at any pair of $\mu_n$ and $\mu_p$ the proton and neutron density $\rho_p$ and $\rho_n$ at temperature $T$, the energy per nucleon in neutron matter (NM) and symmetric matter (SM) can be obtained. Then the symmetry energy at finite temperature can be calculated. Fig. 49 shows the density and temperature dependence of the symmetry energy in nuclear matter at temperatures $T = 0$, 5, 10 and 20 MeV calculated with 9 different Skyrme interactions. The 9 subfigures in Fig. 49 are ordered according to the magnitude of the ratio of effective mass $R_m^0$ for the applied Skyrme interactions. The quantity $R_m^0$ is the value of $R_m = m_1^*/m_0^*$ at normal density, i.e., $R_m^0 = R_m(\rho_0)$, where the subscripts 0 and 1 indicate the isospin asymmetry $\delta = 0$ (for SM) and 1 (for NM), respectively. Obviously, $R_m$ is proportional to the neutron and proton effective mass splitting in asymmetric matter, which can characterize the strength of the splitting of the neutron and proton effective mass for the Skyrme interaction. Fig. 49 shows that there is no obvious correlation between the trend of the density dependence of the symmetry energy in cold matter and the magnitude of the $R_m^0$ of the corresponding Skyrme interactions. Concerning the temperature dependence of the symmetry energy, one sees that the nuclear symmetry energy decreases with increasing temperature at all densities for Skyrme interactions with small $R_m^0$, such as SLy7, Skz3, Skt5 and SkMP, etc., in agreement with the results given in Ref. [449]. However, for SkM, SKXm, SkP and v070 with large $R_m^0$, the symmetry energy decreases with increasing temperature at low density and increases with temperature when the density is higher than a certain density. We call this phenomenon as the transition of the temperature dependence of the symmetry energy (TrTDSE). It shows that the Skyrme interactions, such as SkM, SKXm, SkP, and v070 for which the TrTDSE phenomenon occurs all satisfy $m_n^* > m_p^*$.

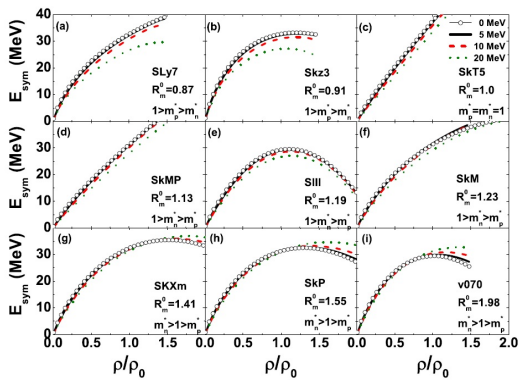

FIG. 49: (Color online) Density dependence of symmetry energy at T=0, 5, 10, 20 MeV calculated with different Skyrme interactions. The ranges of corresponding neutron and proton effective masses for the neutron matter and symmetric matter at saturation density are also presented in each sub-figure. Taken from Ref. [448].

The density for the onset of the TrTDSE depends on the magnitude of the $R_m^0$ of the corresponding Skyrme interactions. The larger $R_m^0$ is, the lower the density for the TrTDSE onset is. For the Skyrme interactions not satisfying $m_n^* > m_p^*$, the TrTDSE phenomenon will not occur.

### D. Uncertainties in symmetry energy constraints

#### 1. Uncertainties of the symmetry energy associated parameters

With the progress of the study of the density dependence of the symmetry energy, tight constraints become urgently requisite, which need the efforts from experimental measurements, improvements of transport models, and understanding the physics parameter correlations. For example, we know that there are not only $S_0$ and $L$ in the Taylor expansion of $S(\rho)$ as in Eq. (86), but also the high order terms, such as curvature $K_{sym}$ and skewness $Q_{sym}$ of $S(\rho)$. Obviously, the uncertainties of $K_{sym}$ and $Q_{sym}$ can influence the constraints on the $S_0$ and $L$, or the density dependence of the symmetry energy. There are also many efforts to constrain the $K_{sym}$ and $Q_{sym}$ from neutron skin and neutron star [450–453].

Margueron et al.'s calculations show that the simple Taylor expansion of the EOS cannot be used to reproduce the EOS well at the whole density region as well as for the symmetry energy, and they proposed a meta-EOS model to describe it [451–453]. Another method to well describe the Skyrme EOS and symmetry energy is to use the nuclear matter parameters, such as $\rho_0$, $E_0$, $K_0$, $S_0$, $L$, $m_s^*$, $m_v^*$, with two additional coefficients $g_{sur}$ and $g_{sur,iso}$ [159, 454, 455]. Here, $\rho_0$ is the normal density,

$K_0 = 9\rho_0 \frac{\partial^2 \epsilon/\rho}{\partial \rho^2}|_{\rho_0}$ is the incompressibility of symmetric nuclear matter, $m_s^*/m = (1 + \frac{2m}{\hbar^2} \frac{\partial}{\partial \tau} \frac{E}{A})|_{\rho_0}$ is the isoscalar effective mass, $m_v^* = \frac{1}{1+\kappa}$ is the isovector effective mass where $\kappa$ is the enhancement of a factor of the Thomas-Reich-Kuhn sum rule. $g_{sur}$, $g_{sur,iso}$ are the coefficients related to density gradient terms.

A lot of theoretical works have evidenced that all of them are related to the symmetry energy. For example, in the Skyrme-Hartree-Fock approaches, the density dependence of symmetry energy is written as,

$$S(\rho) = \frac{1}{3} \frac{\hbar^2}{2m}(\frac{3\pi^2}{2}\rho)^{2/3} \qquad (94)$$
$$+ (A_{sym}u + B_{sym}u^\eta + C_{sym}(m_s^*, m_v^*)u^{5/3}),$$

where $u$ is the reduced density, i.e., $\rho/\rho_0$. A recent theoretical study by Mondal et al. also provide evidence that the $S(\rho)$ depends on the effective mass [456]. In the standard Skyrme interaction, one also observed that the $m_s^*$ is also related to the $K_0$ based on the formula of Skyrme Hartree-Fock (SHF) as pointed out in Ref. [386],

$$K_0 = B + C\sigma + D(1 - \frac{3}{2}\sigma)\frac{8\hbar^2}{m\rho_0}(\frac{m}{m_s^*} - 1), \qquad (95)$$

with $B = -9E_0 + \frac{3}{5}\epsilon_F$, $C = -9E_0 + \frac{9}{5}\epsilon_F$ and $D = \frac{3}{20}\rho_0 k_F^2$. If the $E_0$ and $\rho_0$ are well known, the $K_0$ depends on the $m_s^*$ and $\sigma$. Focusing on the correlation between $m_s^*$ and $K_0$, one can say $K_0$ is independent of $m_s^*$ if $\sigma = 2/3$, but $K_0$ linearly depends on the inverse of $m_s^*$ if $\sigma \neq \frac{2}{3}$. Thus, one can expect that the constraint of $S(\rho)$ with less biased uncertainty should depend on the values of $\rho_0$, $E_0$, $K_0$, $S_0$, $L$, $m_s^*$, $m_v^*$ rather than only on the uncertainties of $S_0$ and $L$.

Fig. 50 shows the values of nuclear matter parameters, $K_0$, $S_0$, $L$, $m_s^*/m$, and $f_I$ calculated from 224 effective Skyrme interactions published from the years 1970-2015 [457]. Here, $f_I = \frac{1}{2\delta}(\frac{m}{m_n^*} - \frac{m}{m_p^*}) = \frac{m}{m_s^*} - \frac{m}{m_v^*}$; it can be analytically incorporated into the transport model and its sign reflects the $m_n^* > m_p^*$ or $m_n^* < m_p^*$. The nuclear matter incompressibility from Skyrme parameter sets converges to the region of 200-280 MeV after the year ∼1990, except for the parameter from the original quark meson coupling (QMC) method [458] (red circles in upper panels of Fig. 50) which were readjusted in 2006. In Ref. [459], they show the value of $K_0$ is higher than generally accepted by a considerable margin, i.e. $K_0 = 240 \pm 20$ MeV, based on the most precise and up-to-date data on GMR energies of Sn and Cd isotopes, together with a selected set of data from $^{56}$Ni to $^{208}$Pb. This result is $250 < K_0 < 315$ MeV which has been obtained without any microscopic model assumptions, except (marginally) the Coulomb effect, and revealed the essential role of surface properties in vibrating nuclei.

For other nuclear matter parameters, such as $S_0$, $L$, $m_s^*$, and $f_I$, most of their values fall into the regions of $S_0 = 25 - 35$ MeV, $L = 30 - 120$ MeV, $m_s^*/m = 0.6 - 1.0$, $f_I = -0.5 - 0.4$. The very recent results on the estimated nuclear matter parameters [451] are shown as black squares in Fig. 50.

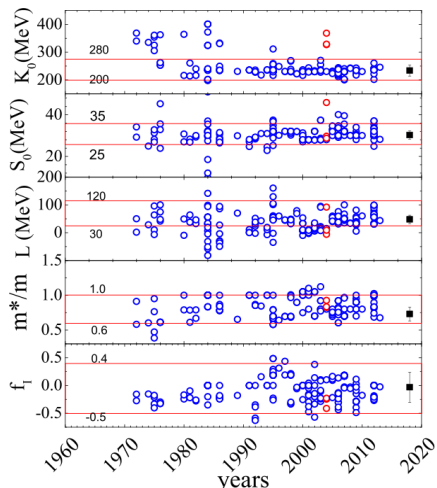

FIG. 50: (Color online) Extracted values of nuclear matter parameters, $K_0$, $S_0$, $L$, $m_s^*/m$, and $f_I$ as a function of published year. The values are obtained from the compiled Skyrme parameter sets by Dutra *et al.* [457]. The black points are the results obtained in Ref. [451]. Taken from Ref.[460].

In the simulations of neutron-rich HICs, the calculations results of the isospin sensitive observables could be influenced by those nuclear matter parameters, such as effective mass splitting and isoscalar effective mass, which makes the effect of the symmetry energy in a certain extent [159, 461–463]. Thus, the effect of different $K_0$, $m_s^*$ and different effective mass splitting should also be investigated and make unentanglement with the effect of the symmetry potential on the isospin sensitive observables in HICs.

### 2. Influence of effective mass splitting on $R_i$, $R_i(y)$, and n/p ratios

In the investigations with the standard Skyrme interaction in HICs, four Skyrme interaction parameter sets, SLy4, SkI2, SkM* and Gs [229, 386, 464, 465] which have similar incompressibility ($K_0$), symmetry energy coefficient ($S_0$) i.e., $K_0 = 230 \pm 20$ MeV, $S_0 = 32 \pm 2$ MeV, but different isoscalar effective mass $m^*/m = 0.7 \pm 0.1$ and different effective mass splitting, are adopted in the ImQMD-Sky calculations. The SLy4 and SkI2 [386, 464] have similar neutron/proton effective mass splitting with $m_n^* < m_p^*$, but very different slopes of symmetry energy $L$ values, 46 MeV for SLy4, and 104 MeV for SkI2. The other two Skyrme interaction parameter sets with $m_n^* > m_p^*$ and $m^*/m \sim 0.78$ also have different $L$ values, 46 MeV for SkM* [229], and 93 MeV for Gs [465]. The saturation properties of nuclear matter for these four Skymre interactions are listed in Table. IV. By analyzing the results calculated with these interactions, the sensitivities of the isospin observables on the different nuclear matter parameters can be investigated.

TABLE IV: Corresponding saturation properties of nuclear matter in, SLy4, SkI2, SkM*, and Gs Skyrme parameters. All entries are in MeV, except for $\rho_0$ in $fm^{-3}$ and the dimensionless effective mass ratios for nucleon, neutron and proton. The effective mass for neutron and proton are obtained for isospin asymmetric nuclear matter with $\delta = 0.2$. Taken from Ref. [224].

| Para. | $\rho_0$ | $E_0$ | $K_0$ | $S_0$ | $L$ | $K_{sym}$ | $m^*/m$ | $m_n^*/m$ | $m_p^*/m$ |
|---|---|---|---|---|---|---|---|---|---|
| SLy4 | 0.160 | -15.97 | 230 | 32 | 46 | -120 | 0.69 | 0.68 | 0.71 |
| SkI2 | 0.158 | -15.78 | 241 | 33 | 104 | 71 | 0.68 | 0.66 | 0.71 |
| SkM* | 0.160 | -15.77 | 217 | 30 | 46 | -156 | 0.79 | 0.82 | 0.76 |
| Gs | 0.158 | -15.59 | 237 | 31 | 93 | 14 | 0.78 | 0.81 | 0.76 |

We simulated the collisions of $^{124}Sn+^{124}Sn$, $^{124}Sn+^{112}Sn$, $^{112}Sn+^{124}Sn$, and $^{112}Sn+^{112}Sn$ reactions at beam energy of 50 MeV/nucleon using the ImQMD-Sky code. 64,000 events were performed for each reaction at each impact parameter. Previous theoretical studies [158, 466] and recent experimental studies [467] suggest that there is no strong dependence of transverse emitted neutron to proton yield ratios on the impact parameter. In the left panel of Fig. 51, we plot the isospin transport ratios obtained with SLy4, SkI2, SkM* and Gs interactions at $b = 6$ fm. As in previous studies [158, 422], we analyze the amount of isospin diffusion by constructing a tracer, $X = \delta$, from the isospin asymmetry of emitting source which includes all emitted nucleons (N) and fragments (frag) with the velocity cut ($v_z^{N,frag} > 0.5v_{beam}^{c.m.}$). The shaded region is experimental data obtained by constructing the isospin transport ratio using isoscaling parameter $X = \alpha$, near the projectile rapidity regions [416]. Our results show that the isospin transport ratio $R_i$ (see Eq. (85)) values for SLy4 (solid circle) and SkM* (solid squares), both with $L = 46$ MeV, lie within the experimental uncertainties while the $R_i$ values for SkI2 (open circle, $L = 104$ MeV) and Gs (open square, $L = 93$ MeV) are above the data range. Even though the isospin diffusion process is accelerated at subsaturation densities with the stronger Lane potential, the overall effect of mass splitting on isospin diffusion is small. This conclusion is similar with previous results even from the IBUU and SMF models [70, 461]. Since the isospin diffusion process is strongly related to the difference of local isospin concentration, the strong repulsive momentum-dependent isoscalar potential reduces the effect of isovector potential on the reaction dynamics, thus, there is no clear pattern that $R_i$ values decrease significantly with the strength of Lane potentials.

We also compare results of the calculations to $R_i$ as a function of the scaled rapidity $y/y_{beam}^{c.m.}$ as shown in the right panel of Fig. 51. The star symbols in the right panel are experimental data, $R_i(X_7) = R_7$, obtained in Ref. [417]. This transport ratio was generated using the isospin tracer $X = ln[Y(^7Li)/Y(^7Be)]$, where

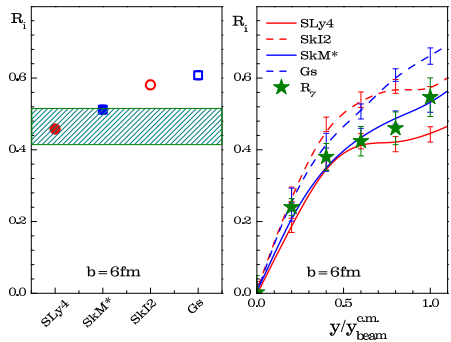
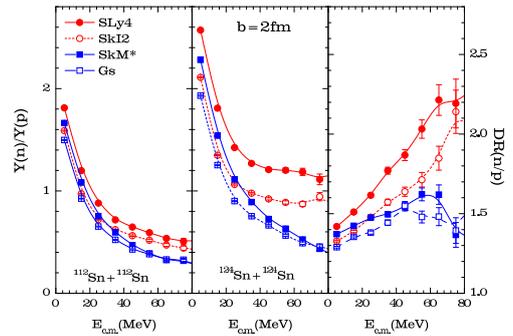

FIG. 51: (Color online) Left panel: isospin diffusion transport ratios obtained with SLy4, SkI2, SkM* and Gs. The shaded region corresponds to the data from [416]. Right panel: isospin transport ratios as a function of rapidity for SLy4, SkI2, SkM* and Gs. The star symbols are data from [417]. Taken from Ref. [224].

FIG. 52: (Color online) Left panel: Y(n)/Y(p) as a function of kinetic energy for $^{112}$Sn+$^{112}$Sn at $b = 2$ fm with angular cuts $70° < \theta_{c.m.} < 110°$; Middle panel is the Y(n)/Y(p) for $^{124}$Sn+$^{124}$Sn. Right panel: DR(n/p) ratios as a function of kinetic energy. The calculated results are for SLy4 (solid circles), SkI2( open circles), SkM* (solid squares) and Gs(open squares). Taken from Ref. [224].

$Y(^7Li)/Y(^7Be)$ is the yield ratio of the mirror nuclei, $^7Li$ and $^7Be$ [417]. For comparison, the ImQMD-Sky calculations of $R_i$ are plotted as lines for $b = 6$ fm. The interactions with smaller $L$ values, SLy4 and SkM* (solid lines) agree with the data better especially in the high rapidity region. However, $\chi^2$ analysis suggests that the quality of fit with isospin diffusion data is not good enough to draw definite conclusions about mass splitting effect with confidence. We need a more sensitive and reliable observable to extract quantitative information about the nucleon effective mass splitting.

The calculated results on single ratio R(n/p) and double ratio DR(n/p) are shown in Fig. 52. We plot the Y(n)/Y(p) ratios as a function of kinetic energy of lowest emitted nucleon in center of mass frame, $E_{c.m.}$, for $^{112}$Sn+$^{112}$Sn (left panel) and $^{124}$Sn+$^{124}$Sn (middle panel) at $b = 2$ fm with angular gate $70° < \theta_{c.m.} < 110°$. The lines connecting the circles correspond to $m_n^* < m_p^*$ case, and the lines connecting the squares correspond to $m_n^* > m_p^*$ case. Not surprisingly, the Y(n)/Y(p) ratios are larger for the neutron rich system, $^{124}$Sn+$^{124}$Sn, in the middle panel. Consistent with Refs. [461, 462], the differences in the Y(n)/Y(p) ratios between them $m_n^* < m_p^*$ (circles) and $m_n^* > m_p^*$ (squares) increase with nucleons kinetic energy. At high nucleon energies, the stronger Lane potentials with $m_n^* < m_p^*$ enhance neutron emissions, leading to flatter Y(n)/Y(p) dependence on the nucleon kinetic energy. The calculations with SLy4 ($L = 46MeV, m_n^* < m_p^*$) are consistent with the double ratios data from Ref. [415] which was published in 2006, especially at high kinetic energy region.

Recently, remeasurements of the neutron to proton yield data has been finished and was published in Ref. [468]. The measured coalescence invariant spectral double ratios DR(n/p) for both beam energies, 50 and 120 MeV/nucleon, were analyzed. There are systematic uncertainties in DR(n/p) of about 10% at $E_{beam} = 50$

MeV/nucleon and 15% at $E_{beam} = 120$ MeV/nucleon stemming from the dependence of the neutron detection efficiencies on the charged particle and neutron-scattering backgrounds in LANA. The current data at $E_{beam} = 50$ MeV/nucleon have a factor of 4 smaller uncertainties and extend over a wider energy range than those of Ref. [415]. The two data overlap within statistical and systematic uncertainties, except for the lowest two energy data points. The new precision data show there is no agreement between the data and calculations due to the inadequate description of cluster formation mechanism at the beam energy of 50 MeV/nucleon. At higher incident energy, $E_{beam} = 120$ MeV/nucleon, the calculation describes the nucleon spectra fairly well. Furthermore, at high kinetic energy of emitted nucleons ($E_{c.m.} > 60$ MeV), the ImQMD-Sky calculations of Ref. [224] showed the greatest sensitivity to the effective mass splitting, and calculations with SLy4 interaction lie close to the data whereas the calculations with SkM* lie below the data.

Further theoretical analyses have indicated that n/p yield ratios (as well as other observables) are also somewhat sensitive to the other nuclear matter parameters. For example, in Ref. [159], we discussed the linear correlation coefficient $C_{AB}$ between the nuclear matter parameter and the isospin sensitive observable in the sampled parameter space. The linear-correlation coefficient $C_{AB}$ between variable $A$ and observable $B$ is calculated

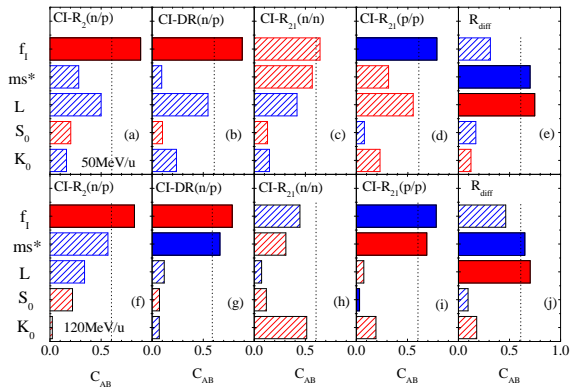

FIG. 53: (Color online) Correlations of five observables, CI-R$_2$(n/p)(a), CI-DR(n/p)(b), CI-R$_{21}$(n/n)(c), CI-R$_{21}$(p/p)(d), $R_{diff}$ (e) with five force parameter, $K_0$, $S_0$, $L$, $m_s^*$ and $f_I$. Up panels are the results for 50 MeV/nucleon, and bottom panels are for 120 MeV/nucleon. Taken from Ref. [159].

as follows[469]:

$$C_{AB} = \frac{cov(A, B)}{\sigma(A)\sigma(B)} \quad (96)$$

$$cov(A, B) = \frac{1}{N-1}\sum_i (A_i - <A>)(B_i - <B>) \quad (97)$$

$$\sigma(X) = \sqrt{\frac{1}{N-1}\sum_i (X_i - <X>)^2}, X = A, B \quad (98)$$

$$<X> = \frac{1}{N}\sum_i X_i, i = 1, N. \quad (99)$$

$cov(A, B)$ is the covariance, $\sigma(X)$ is the variance. $C_{AB} = \pm 1$ means there is a linear dependence between $A$ and $B$, and $C_{AB} = 0$ means no correlations. As shown in the panel (e) and (j) of Fig. 53, isospin diffusion also related to the isoscalar effective mass [159], which may have to be better constrained in order to accurately determine the effective mass splitting and slope of symmetry energy. Those studies also stimulate the further statistical analysis or Bayesian analysis in the multi-dimension parameter surface with respect to the neutron to proton yield data [470].

### 3. Influence of $K_0$, $S_0$, $L$, $m_s^*$, and $f_I$ on isospin diffusion

In order to investigate the impact of other nuclear matter parameters on the isospin diffusion observable, we calculate it in five-dimensional (5D) parameter space, such as $K_0$, $S_0$, $L$, $m_s^*$, $f_I$, with ImQMD-Sky. We sampled 120 points in the range which we listed in Table V under the condition that $\eta \geq 1.1$. $\eta \geq 1.1$ is used for guaranteeing the reasonable three-body force in the transport model calculations. The ranges of these nuclear matter parameters are chosen based on the *prior* information of Skyrme parameters as shown in Fig. 50. As an example, the 120 sampled points are presented as open and solid circles in two-dimensional projection in Fig. 54. The points of parameter sets uniformly distribute in two-dimensional projection except for the plots of $K_0$ and $m_s^*/m$ due to the restriction of $\eta \geq 1.1$. We perform the calculations for isospin transport diffusion at 35 MeV/nucleon and 50 MeV/nucleon at $b = 5 - 8$ fm with the impact parameter smearing [471] for $^{112,124}$Sn+$^{112,124}$Sn. 10,000 events are calculated for each point in the parameter space and simulations are stopped at 400 fm/$c$. The calculations are performed on TianHe-1 (A), the National Supercomputer Center in Tianjin.

TABLE V: Model parameter space used in the codes for the simulation of $^{112,124}$Sn+$^{112,124}$Sn reaction. 120 parameter sets are sampled in this space by using the Latin Hyper-cuber Sampling method. Taken from Ref.[460]

| Para. Name | Values | Description |
|---|---|---|
| $K_0$ (MeV) | [200,280] | Incompressibility |
| $S_0$ (MeV) | [25,35] | Symmetry energy coefficient |
| $L$ (MeV) | [30,120] | Slope of symmetry energy |
| $m_s^*/m_0$ | [0.6,1.0] | Isoscalar effective mass |
| $f_I = (\frac{m_0}{m_s^*} - \frac{m_0}{m_v^*})$ | [-0.5,0.4] | $f_I = \frac{1}{2\delta}(\frac{m_0}{m_n^*} - \frac{m_0}{m_p^*})$ |

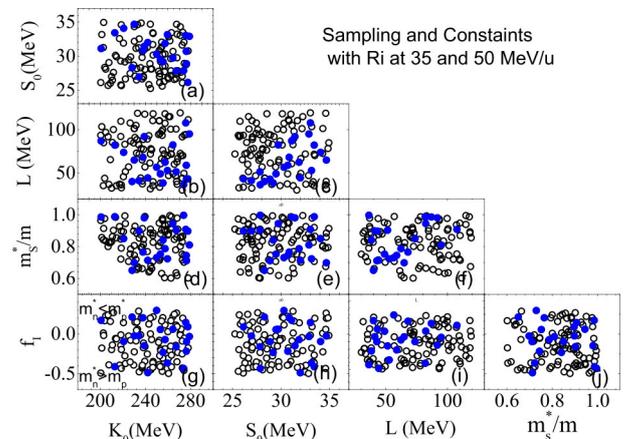

FIG. 54: (Color online) Sampled points in 5D parameter space, blue solid points are the sets which can reproduce two isospin diffusion data. Taken from Ref.[460].

In Fig. 55, the lines represent the calculated results of the isospin transport ratio $R_i$ with 120 parameter sets. Two stars are the experimental data [416, 417, 472] which is constructed from the isoscaling parameter $X = \alpha_{iso}$ at 50 MeV/nucleon [416] and the ratio of $X = ln(Y(^7Li)/Y(^7Be))$ [417, 472] at the beam energy of 35 MeV/nucleon, which was assumed and evidenced to linearly related to the isospin asymmetry of emitting source [417]. And thus, one can compare the $R_i(\delta)$ to $R_i(\alpha)$ or $R_i(ln(Y(^7Li)/Y(^7Be))$. As shown in Fig. 55, the calculated results show a large spread around the experimental data. By comparing the calculations to the data, we find 22 parameter sets that can reproduce the isospin diffusion data within experimental errors. The extracted 22 parameter sets are listed in Table VI. We highlight those points that can reproduce the experimental data within experimental errors with blue colors in Fig. 54. Generally, one can observe that the $L$ increases with the $S_0$. The constrained points distribute in the bottom-right corner in the $S_0$-$L$ plot [panel (c)], and the large $L$ with small $S_0$ are ruled out.

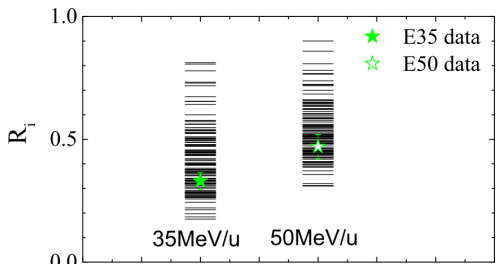

FIG. 55: (Color online) Stars are the isospin diffusion data at 35 MeV/nucleon and 50 MeV/nucleon [417, 472], lines are the calculated isospin transport ratios with 120 parameter sets. Taken from Ref.[460].

The results in panel (j) of Fig. 54 show that isospin diffusion data is not sensitive to the effective mass and its splitting. Based on the Fig. 54, one can learn that there is no obvious correlations between $R_i$ and $S_0$, $L$, $m_s^*/m$, and $f_I$ in 5D parameter space. It is because the $R_i$ is not only correlated to $L$ but also correlated to $m_s^*/m$ [159], which broke the $R_i$ dependence of $L$ when we randomly chose the values of $K_0$, $S_0$, $L$, $m_s^*/m$, and $f_I$. If we fix the values of $K_0$, $S_0$, $m_s^*/m$, and $f_I$, the positive correlation between $R_i$ and $L$ can be found.

## VI. DISCUSSION AND PROSPECT

The transport model has been widely and successfully applied in the study of heavy ion collisions from low energies to relativistic energies, for understanding the nuclear phenomena, such as, collective flow, particle emission, fussion-fission, multifragmentation mechanism and the properties of nucleonic equation of state (EOS) in the laboratory.

However, the model uncertainties still need to be understood for reliable extracting the physics information of interest. In last 30 years, the transport code comparison project has been performed in the community to seek the reasons [226, 372, 473, 474] and reduce the model uncertainties. Despite a lot of efforts devoted to the improvement of the model, we still have to face some fundamental problems for the demands of more exact theoretical description and accurate calculation of heavy ion collisions.

Because of the complexity of transport equations, and in particular their dimensionality, the collision term and mean field potential term are treated separately and it causes the model uncertainties in solving the transport equation [474] and extracting the properties of nuclear system. Ideally, the final but very difficult goal for transport model is to develop a precise and accurate enough transport model, which takes enough quantum effects and treat the mean field and collision self-consistently for the quantum $N$-body system.

The quantum effects in transport model play more important roles for heavy ion reaction at low energy than that at high energy, but this is not well considered in the current models. One of the reason is that the coarse-grain method is adopted to derive the transport equation. It leads to a semi-classical transport equation, where the phase space distribution $f$ will finally evolve to the Boltzmann distribution. It is mimicked by the Pauli blocking, but it is always underestimated in the current transport models, especially for low energy heavy ion reaction [372]. It seems to us that there is a long standing task to improve the treatment of the quantum effects, such as the shell effect which plays important role in some cases as well as the Pauli blocking effect for the better description of low energy heavy ion collision by transport model. However, it is also needed to go beyond the mean field for heavy ion reaction at low energy.

Fluctuations in transport theory are the main venue to go beyond dissipative mean field dynamics, and it plays an important role in the large amplitude motions of nuclear systems, such as nuclear fission, fusion, MNT, ternary break, and multifragmentation reactions as one observed in experiments of heavy ion collisions. As an example, in Ref. [475], the models of MNT reaction was tested by reaction $^{136}$Xe+$^{198}$Pt and it turned out that the ImQMD calculations provide better agreement with experimental data, and it seems to us that it is due to the stronger fluctuation and dissipation being considered dynamically in the ImQMD model. To understand and properly treating the fluctuation and dissipation in transport models is of importance for a better description of the fusion, MNT as well as the multifragmentation,etc., in heavy ion collisions from both theoretical study and practical applications.

The cluster formation mechanism is another important issue which needs to be solved in the transport models, since the heavy ion collision observables used to extract the physics information of interest are obtained from the measurement of the emitted light particles or fragments. In physics, the cluster formation is related to the fluctuation when the system enters into the spinodal region as well as the $N$-body correlations. But both mechanisms are hard to accurately deal with and to incorporate in the calculations for describing the light particles productions in the mean field level. For heavy ion reactions at intermediate energy, one has clearly observed the multifragmentation phenomena, and there are lot of light charged particle formed. But in the transport model, ex-cept for the AMD considering the NN correlation to form light particles [476], the calculations always overpredict the $Z = 1$ particles and underestimate the $Z = 2$ particles. It naturally requires including a reasonable cluster formation mechanism besides the current Hamiltonian dynamics we used.

TABLE VI: Extracted 22 nuclear matter parameter sets and the corresponding standard Skyrme parameters. $t_0$ in $MeV fm^3$, $t_1$ and $t_2$ in MeVfm$^5$, $t_3$ in MeVfm$^{3\sigma+3}$, $x_0$ to $x_3$ is dimensionless quantities. Taken from Ref.[460].

| No. | $K_0$ | $S_0$ | $L$ | $m_s^*/m$ | $f_I$ | $t_0$ | $t_1$ | $t_2$ | $t_3$ | $x_0$ | $x_1$ | $x_2$ | $x_3$ | $\sigma$ |
|---|---|---|---|---|---|---|---|---|---|---|---|---|---|---|
| 1 | 234.391 | 26.936 | 41.147 | 0.898 | -0.024 | -1890.80 | 427.97 | -490.81 | 12571.72 | 0.10669 | -0.19396 | -0.7161 | 0.15416 | 0.29804 |
| 2 | 277.553 | 26.124 | 43.235 | 0.897 | 0.089 | -1374.17 | 428.19 | -607.42 | 10814.29 | 0.04292 | -0.26258 | -0.81939 | 0.24329 | 0.51892 |
| 3 | 259.484 | 33.146 | 52.855 | 0.723 | -0.366 | -1569.42 | 474.60 | 3.93 | 9415.46 | 0.21035 | -0.03708 | -41.13867 | -0.02844 | 0.37265 |
| 4 | 257.436 | 31.863 | 62.418 | 0.787 | -0.072 | -1572.00 | 455.14 | -359.50 | 10186.44 | 0.10568 | -0.18487 | -0.69112 | 0.07323 | 0.38608 |
| 5 | 249.937 | 30.298 | 56.647 | 0.73 | 0.295 | -1714.97 | 472.30 | -688.83 | 10110.07 | 0.34791 | -0.39789 | -1.01437 | 0.97341 | 0.31666 |
| 6 | 267.291 | 27.828 | 51.482 | 0.903 | -0.16 | -1452.20 | 426.91 | -352.89 | 10979.89 | -0.02416 | -0.11056 | -0.50064 | -0.25793 | 0.46733 |
| 7 | 276.418 | 28.86 | 42.831 | 0.711 | -0.097 | -1395.03 | 478.63 | -263.07 | 8737.27 | 0.20269 | -0.18678 | -0.68719 | 0.48667 | 0.47509 |
| 8 | 200.821 | 31.098 | 87.039 | 0.986 | 0.171 | -3048.33 | 410.78 | -744.73 | 19381.38 | -0.28089 | -0.3043 | -0.8462 | -0.35056 | 0.16036 |
| 9 | 228.2 | 28.292 | 40.048 | 0.65 | 0.212 | -3312.92 | 501.46 | -515.21 | 17988.52 | 1.00059 | -0.36089 | -1.06232 | 1.48966 | 0.10376 |
| 10 | 253.203 | 29.474 | 49.084 | 0.752 | 0.055 | -1644.99 | 465.37 | -460.59 | 10070.75 | 0.24038 | -0.26259 | -0.86375 | 0.55912 | 0.34745 |
| 11 | 242.098 | 31.985 | 44.36 | 0.713 | -0.488 | -1914.52 | 477.95 | 140.60 | 10865.66 | 0.15117 | 0.02588 | -2.31398 | -0.12133 | 0.25704 |
| 12 | 239.014 | 31.441 | 91.905 | 0.981 | -0.148 | -1766.26 | 411.68 | -411.04 | 12629.01 | -0.43493 | -0.10372 | -0.52328 | -0.93988 | 0.34248 |
| 13 | 230.13 | 34.676 | 64.931 | 0.698 | -0.026 | -2480.04 | 483.17 | -323.15 | 13757.39 | 0.39189 | -0.22784 | -0.82337 | 0.54526 | 0.16807 |
| 14 | 220.763 | 34.081 | 73.762 | 0.85 | -0.096 | -2359.49 | 438.85 | -383.47 | 14591.08 | -0.02704 | -0.15899 | -0.63047 | -0.17633 | 0.20869 |
| 15 | 237.836 | 30.837 | 68.072 | 0.765 | 0.203 | -1945.23 | 461.46 | -625.89 | 11613.88 | 0.15946 | -0.34378 | -0.95171 | 0.41995 | 0.26249 |
| 16 | 276.165 | 30.705 | 58.846 | 0.744 | -0.218 | -1393.55 | 467.84 | -169.89 | 9181.01 | 0.06398 | -0.11247 | -0.2356 | -0.13504 | 0.48318 |
| 17 | 212.881 | 33.425 | 82.13 | 0.988 | -0.413 | -2406.93 | 410.43 | -139.80 | 15831.81 | -0.50854 | 0.06498 | 0.90667 | -1.02398 | 0.21879 |
| 18 | 273.816 | 27.854 | 36.382 | 0.997 | -0.435 | -1396.68 | 408.85 | -121.72 | 11646.34 | 0.0986 | 0.08113 | 1.25635 | -0.43832 | 0.51157 |
| 19 | 278.918 | 32.888 | 95.046 | 0.81 | -0.033 | -1368.43 | 448.90 | -418.69 | 9938.92 | -0.21753 | -0.20303 | -0.73659 | -0.6341 | 0.51343 |
| 20 | 255.597 | 29.184 | 38.419 | 0.841 | -0.233 | -1579.20 | 441.03 | -234.77 | 10767.52 | 0.19335 | -0.08009 | -0.25897 | 0.09028 | 0.39256 |
| 21 | 275.783 | 33.03 | 107.768 | 0.908 | 0.143 | -1386.09 | 425.85 | -670.47 | 10922.75 | -0.33682 | -0.29417 | -0.85204 | -0.68126 | 0.51127 |
| 22 | 264.335 | 29.718 | 82.428 | 0.945 | 0.059 | -1474.21 | 418.40 | -605.68 | 11375.98 | -0.25086 | -0.23842 | -0.78177 | -0.4952 | 0.45917 |
| | $\bar{K}_0$ | $\bar{S}_0$ | $\bar{L}$ | $\bar{m_s^*}/m$ | $\bar{f}_I$ | $\bar{t}_0$ | $\bar{t}_1$ | $\bar{t}_2$ | $\bar{t}_3$ | $\bar{x}_0$ | $\bar{x}_1$ | $\bar{x}_2$ | $\bar{x}_3$ | $\bar{\sigma}$ |
| Average | 250.54 | 30.62 | 62.31 | 0.83 | -0.072 | -1838.43 | 447.08 | -383.78 | 11971.69 | 0.05613 | -0.17691 | -2.4674 | -0.0112 | 0.3534 |
| error | (22.87) | (2.39) | (21.01) | (0.11) | (0.22) | (553.90) | (28.05) | (223.12) | (2783.59) | (0.3255) | (0.133) | (8.66) | (0.608) | (0.130) |

Baryons and mesons production and propagation mechanism near the threshold energy are also important issue for extracting the information of EOS at high density. At this energy region, the issue related to a covariant dynamics model in the propagation and collisions part, in-medium cross sections of $NN \leftrightarrow NR/NN \leftrightarrow RR$, energy conservation in the process of $NN \leftrightarrow NR/NN \leftrightarrow RR/R \leftrightarrow N\pi$ ($R$ is a resonance particles) and the potential for mesons, should be well investigated.

## Acknowledgments

Yingxun Zhang acknowledges the supports in part by National Science Foundation of China Nos.11875323, 11875125, 11475262, 10675172, 11075215, 11475262, 11790323, 11790324, and 11790325, 11961141003, the National Key R&D Program of China under Grant No. 2018 YFA0404404 and the Continuous Basic Scientific Research Project (No. WDJC-2019-13). Ning Wang acknowledges the supports in part by National Natural Science Foundation of China Nos. U1867212, 11422548, the Guangxi Natural Science Foundation Nos.2015GXNSFDA139004, 2017GXNSFGA198001. Qingfeng Li acknowledges the supports in part by National Science Foundation of China Nos. 11875125, 11847315, 11375062, 11505057, 11947410, and 11747312, and the Zhejiang Provincial Natural Science Foundation of China (No. LY18A050002), and the "Ten-Thousand Talent Program" of Zhejiang province. Junlong Tian acknowledges the supports in part by National Science Foundation of China Nos.11961131010, 11475004. Li Ou acknowledges the supports in part by National Science Foundation of China No.11965004, and Natural Science Foundation of Guangxi province (2016GXNSFFA380001), Foundation of Guangxi innovative team and distinguished scholar in institutions of higher education. Min Liu, acknowledges the supports in part by National Science Foundation of China Nos.11875323. Kai Zhao, acknowledges the supports in part by National Science Foundation of China Nos. 11675266, 11005155, 11475262, 11275052,

11375062, 11547312, 11275068 and the National Key Basic Research Development Program of China under Grant Nos. 2007CB209900, 2013CB834404. Xizhen Wu, acknowledges the supports in part by National Science Foundation of China No. 10235020, 10979023, 11005155, 11365004, 11475004, 11675266. Zhuxia Li, acknowledges the supports in part by National Science Foundation of China Nos. 19975073, 10175093, 10175089, 10235030, 11275052, 11375062, 11475262, 11475004, 11875323, 11875125 and the National Key Basic Research Development Program of China Nos. G20000774, 2007CB209900). .

---